\documentclass[%
superscriptaddress,
 amsmath,amssymb,
 aps,
prb,
]{revtex4-1}

\usepackage{graphicx}% Include figure files
\usepackage{dcolumn}% Align table columns on decimal point
\usepackage{bm}% bold math
\usepackage{xcolor,colortbl}
\usepackage{csquotes}
\usepackage{SupRefs}

\renewcommand{\figurename}{\textbf{Fig.}}
\def\plotFigs{}

\makeatletter
\def\fnum@figure{\figurename\nobreakspace\textbf{\thefigure}}
\makeatother

\definecolor{Blue}{rgb}{0.8,1,0.8}
\definecolor{Red}{rgb}{1,0.8,0.8}
\definecolor{Green}{rgb}{0.8,0.8,1}
\newcolumntype{a}{>{\columncolor{Green}}c}
\newcolumntype{b}{>{\columncolor{Blue}}c}
\newcolumntype{d}{>{\columncolor{Red}}c}

\begin{document}

%%%%%%%%%%%%
% Commands %
%%%%%%%%%%%%
\newcommand{\uofa}{Department of Physics, University of Alberta, Edmonton, Alberta T6G 2E1, Canada}
\newcommand{\tum}{Department of Chemistry, Technical University of Munich, Garching bei M\"unchen 85748, Germany}

\preprint{APS/123-QED}

\title{Ultrafast photoconductivity and terahertz vibrational dynamics \linebreak in double-helix SnIP nanowires}

\author{David N. Purschke}
\email[Corresponding Author: ]{purschke@ualberta.ca}

\affiliation{\uofa}
\author{Markus R. P. Pielmeier}
\affiliation{\tum}
\author{Ebru \"Uzer}
\affiliation{\tum}
\author{Claudia Ott}
\affiliation{\tum}
\author{Charles Jensen}
\affiliation{\uofa}
\author{Annabelle Degg}
\affiliation{\tum}
\author{Anna Vogel}
\affiliation{\tum}
\author{Naaman Amer}
\affiliation{\uofa}
\author{Tom Nilges}
\affiliation{\tum}
\author{Frank A. Hegmann}
\affiliation{\uofa}%

\date{\today}% It is always \today, today,

\begin{abstract}
Tin iodide phosphide (SnIP), an inorganic double-helix material, is a quasi-1D van der Waals semiconductor that shows promise in photocatalysis and flexible electronics. However, our understanding of the fundamental photophysics and charge transport dynamics of this new material is limited. Here, we use time-resolved terahertz (THz) spectroscopy to probe the transient photoconductivity of SnIP nanowire films and, with insight into the highly anisotropic electronic structure from quantum chemical calculations, measure an electron mobility as high as 280 $cm^2\,V^{-1}s^{-1}$. Additionally, the THz vibrational spectrum reveals a photoexcitation-induced charge redistribution that reduces the amplitude of a twisting mode of the outer SnI helix on picosecond timescales. Finally, we show that the carrier lifetime and mobility are limited by a trap density greater than 10$^{18}\,cm^{-3}$. Our results provide insight into the optical excitation and relaxation pathways of SnIP and demonstrate a remarkably high carrier mobility for such a soft and flexible material.
\end{abstract}

%Use showkeys class option if keyword
\maketitle

\newpage
The importance of the double helix structure in biology has led to significant fundamental interest in abiotic and inorganic analogues \cite{soghomonian_inorganic_1993,su_inorganic_2011,ivanov_inorganic_2012,zhao_emerging_2014,haldar_metal-free_2009}. SnIP recently surprised the inorganic materials community as the first carbonless atomic-scale double helix \cite{pfister_inorganic_2016}. Subsequently, several compounds were predicted to form with the SnIP structure, indicating that SnIP could be the first of a new class of inorganic double-helix materials \cite{baumgartner_inorganic_2017,li_landscape_2017,bijoy_atomic_2020}. With strong intra-helix covalent bonds and weak inter-helix dispersion forces, SnIP belongs to the group of newly emerging 1D van der Waals (vdW) materials with potential applications in nanoelectronics and photonics \cite{ott_flexible_2019,xiang_one-dimensional_2020, qin_raman_2020, burdanova_ultrafast_2020}. In contrast to the DNA structure, which consists of two equal radius helices, SnIP forms with an outer [SnI]$^+$ helix wrapping around an inner [P]$^-$ helix, as pictured in Fig. \ref{intro}a. SnIP crystallizes monoclinically with a unit cell containing two opposite-handed double helices so that there is no net chirality. It is composed of abundant and non-toxic elements and can grow uninhibited to cm-length needles with a low-temperature synthesis\cite{pfister_inorganic_2016,pielmeier_formation_2020} (\supls\noteMorph{} and \noteDist) or in nanotubes using vapor deposition \cite{uzer_vapor_2019, pielmeier_toward_2020}. Its 1.86 eV band gap, as determined by band structure calculations (Fig. \ref{intro}b, c) and verified experimentally\cite{pfister_inorganic_2016}, is well situated for solar absorption and photocatalytic water splitting \cite{li_landscape_2017,ott_flexible_2019,uzer_vapor_2019}. SnIP is also an extremely soft and flexible semiconductor and is a promising material for applications in flexible electronics\cite{pfister_inorganic_2016,ott_flexible_2019}. It is predicted to have a high carrier mobility \cite{li_landscape_2017}, however, as-grown SnIP is highly resistive so that the current lack of doped samples has made it difficult to explore its electronic properties \cite{pfister_inorganic_2016}. Moreover, despite the exciting properites and unique structure of SnIP, there have been no investigations probing its ultrafast photophysical properties.
\par

Here, we use time-resolved THz spectroscopy (TRTS), a powerful non-contact ultrafast probe \cite{jepsen2011}, to study picosecond charge carrier dynamics in SnIP nanowire films, as shown in Fig. \ref{intro}d. From analysis of the photoconductivity spectra, along with insight into the highly anisotropic energy landscape from density functional theory (DFT), we make the first measurement of the carrier mobility in SnIP. We find a maximum mobility of 280 $cm^2V^{-1}s^{-1}$ along the double-helix axis, an extraordinarily high mobility for a material as soft and flexible as SnIP. Using terahertz time-domain spectroscopy (THz-TDS), we also make the first observation of optically-active vibrational modes in SnIP, finding two resonances in the THz range that we assign to rocking and twisting modes of the outer [SnI]$^+$ helix by comparison to quantum-chemical calculations. Interestingly, we see a suppression of the photoconductivity near a strong resonance at 1.495 THz, which indicates a reduction in oscillator strength after photoexcitation due to a photoexcitation-induced charge redistribution\cite{koeberg_simultaneous_2007,ulbricht_carrier_2011,butler_ultrafast_2016,zhao_monitoring_2019}. Tracking the time evolution of the oscillator parameters, we see that the charge redistribution occurs on a timescale comparable to the lifetime of the mode. Finally, we show that the carrier lifetime is limited by ultrafast trapping and estimate the trap density to be greater than 10$^{18}cm^{-3}$. This suggests that the carrier lifetime and mobility can be improved by optimized synthesis or passivation of traps, which could dramatically enhance photocatalytic activity\cite{parkinson_carrier_2009,ozawa_correlation_2018}. Our results show that SnIP could be an important new material for flexible electronics due to it's unique combination of electronic and mechanical properties.
\par
\vfill

\ifdefined\plotFigs
\begin{figure}[htb]
	\begin{center}
		\includegraphics[scale=0.9]{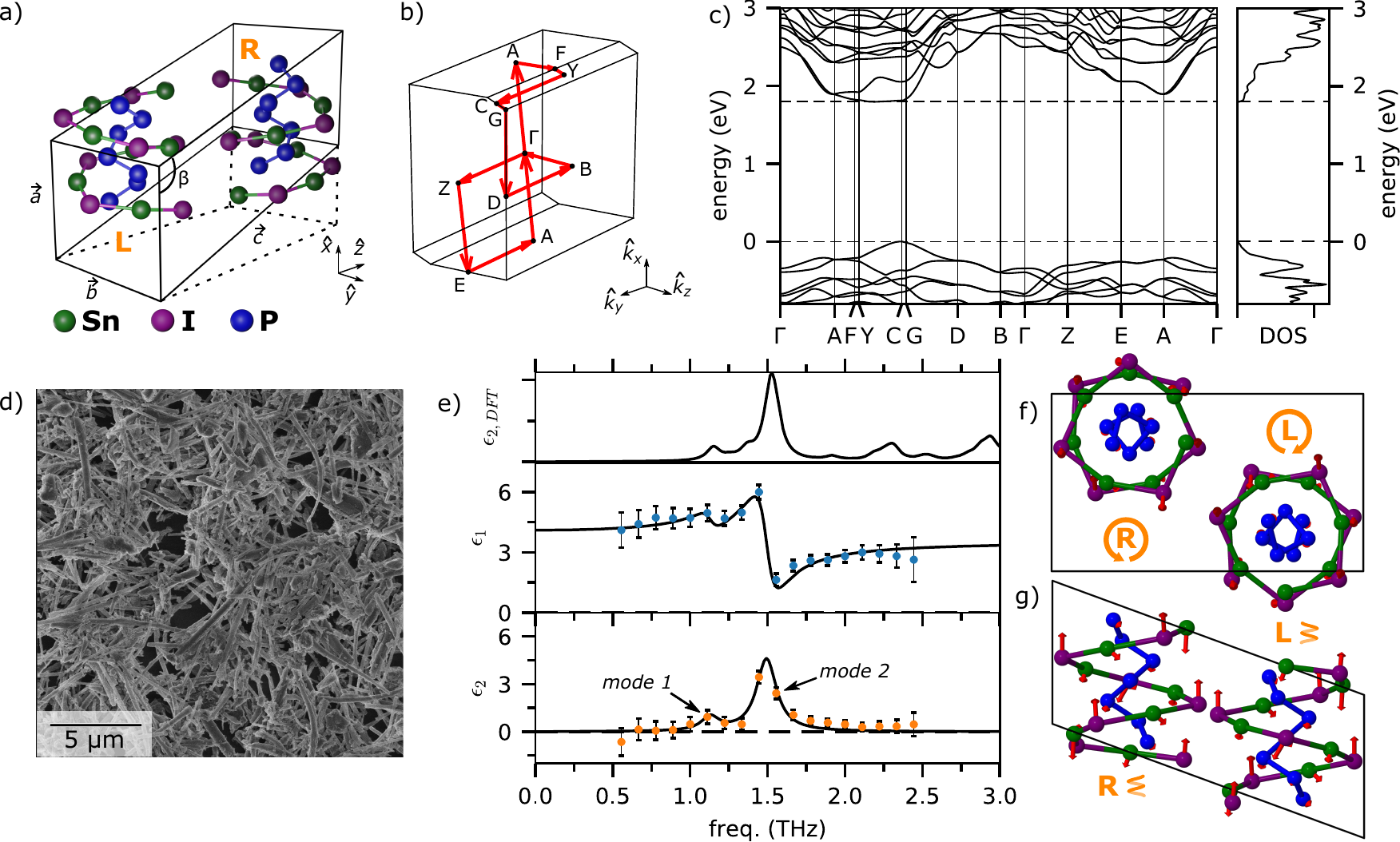}
		\caption{ \textbf{SnIP crystal and electronic structure, thin-film morphology, and THz vibrational spectrum. a} SnIP structure with two opposite-helicity strands in the monoclinic unit cell. $\vec{a}$, $\vec{b}$, and $\vec{c}$ indicate the lattice vectors while $\hat{x}$, $\hat{y}$, and $\hat{z}$ are the Cartesian axes. \textbf{b} The associated Brillouin zone with $\hat{k}_x$, $\hat{k}_y$, and $\hat{k}_z$ indicating the Cartesian axes.  \textbf{c} Calculated band structure and density of states (DOS). \textbf{d} Helium-ion microscope image of the ultrasonicated drop-cast nanowire film on a quartz substrate. \textbf{e} Measured real, $\epsilon_1$, and imaginary, $\epsilon_2$, effective dielectric function of the thin film in the THz range (blue and orange circles, respectively). The solid lines are a fit with two Lorentz oscillators of center frequency $\omega_0/2\pi=1.13\,(1.495)$ THz, amplitude $A_{osc}=5.3\,(42)$ THz$^2$, and damping of $\gamma/2\pi=0.14\,(0.15)$ THz for mode 1 (mode 2), and a high-frequency dielectric constant of $\epsilon_{\infty}=$3.52. The weighted average (see \enquote{Methods}) of the calculated imaginary dielectric constant assuming a 0.1 THz linewidth, $\epsilon_{2,DFT}$ is plotted above. Calculated normal mode displacement vectors of the 1.52 THz  oscillation projected along the \textbf{f} $\vec{a}$-axis and \textbf{g} $\vec{b}$-axis.}
		\label{intro}
	\end{center}
\end{figure}
\fi

\section{Results}
\noindent
\textbf{Terahertz vibrational modes.} We measured the effective dielectric function, $\tilde{\epsilon}_r=\epsilon_1+i\epsilon_2$, of the SnIP nanowire film using THz-TDS in the frequency range from 0.5 to 2.5 THz (see \enquote{Methods} for a description of the THz-spectroscopy setup and data analysis). As shown in Fig. \ref{intro}e, a weak vibrational resonance at 1.13 THz and stronger resonance at 1.495 THz are observed (referred to here as mode 1 and mode 2, respectively). The solid lines are a simultaneous fit to the real and imaginary components using two Lorentz oscillators,
\begin{equation}
\tilde{\epsilon}_r(\omega)=\epsilon_\infty+\sum_{i=1}^2\frac{A_{osc,i}}{\omega_{0,i}^2-\omega^2-i\omega\gamma_i},%+\frac{\omega_{p_2}^2}{\omega_2^2-\omega^2+i\omega\gamma_2},
\label{eq:osc}
\end{equation}
where $\omega$ is the angular frequency and $\omega_{0,i}$, $A_{osc,i}$, and $\gamma_i$, are the resonant angular frequency, amplitude, and damping of the $i^{th}$ mode, respectively, and $\epsilon_\infty$ is the constant offset due to higher frequency vibrational modes and electronic transitions. To date, the vibrational modes in SnIP have been measured by Raman spectroscopy, which is insensitive to the infrared active (IR) modes, and inelastic neutron scattering, which can only resolve features in this region with careful background subtraction and do not distinguish between IR active and inactive modes \cite{pfister_inorganic_2016,ott_flexible_2019}. We therefore make the first identification these modes based on comparison to density functional theory (DFT) calculations using the Crystal software package (see \enquote{Methods}). 
\par

While the random orientation of nanowires in our thin films obscures the underlying symmetry of the modes, making it more difficult to precisely assign peaks in the experimentally observed spectrum to individual modes \cite{schubert_phonon_2019}, the calculated average imaginary dielectric function shows good agreement with data, as seen in Fig. \ref{intro}e. In general, the calculations suggest that motion in the 0.5-2.5 THz range is associated with oscillations of the [SnI]$^+$ (outer) helix. Two high intensity modes near mode 2 (1.495 THz) are revealed by DFT calculations at 1.52 THz and at 1.55 THz. Both peaks can be classified as primarily twisting modes of the outer [SnI]$^+$ helix. Figure \ref{intro}f, g shows the vibrational conformation of the 1.52 THz mode, the highest intensity mode in the 0.5-2.5 THz range. In this mode, the Sn$^{2+}$ and I$^-$ atoms show the largest displacement with motion predominantly along the double-helix axis so that the mode is longitudinally polarized with respect to the nanowires. A smaller peak near mode 1 (1.13 THz) is revealed by the calculated spectra at 1.15 THz, corresponding to rocking motion of the outer [SnI]$^+$ helix. For further discussion of the calculations, see \supls\noteEpsDFT{} and \noteVibVecs, which includes plots of the polarization-dependent dielectric function and animations of several high intensity modes. 
\par
\ifdefined\plotFigs
\begin{figure}[htb]
	\includegraphics[scale=1]{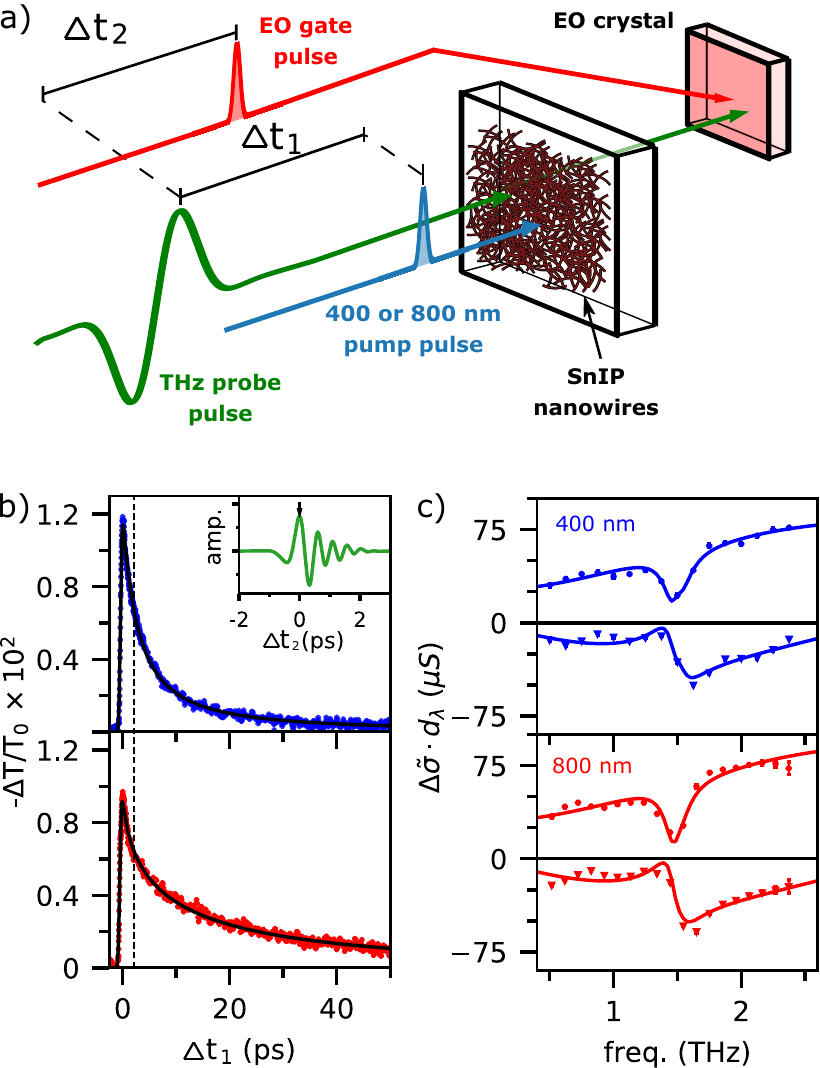}
	\caption{\textbf{Ultrafast transient photoconductivity and relaxation dynamics. a} Schematic of the TRTS experiment. A THz pulse is coincident on a drop-cast SnIP thin film along with a time-delayed ($\Delta t_1$) 400 or 800 nm photoexcitation (pump) pulse. A third pulse is used as a gating beam at time delay $\Delta t_2$ for field-resolved electro-optic sampling of the transmitted THz pulse to monitor the pump-induced change in transmission. \textbf{b} Pump-induced differential THz transmission as a function of pump-probe time delay with $\Delta t_2=0\,ps$  (indicated by the black arrow in the inset, which shows the THz-pulse waveform) for 400 nm (top) and 800 nm (bottom) pump wavelengths at fluences of 190$\,\mu J\,cm^{-2}$ and 850$\,\mu J\,cm^{-2}$, respectively. The solid lines are bi-exponential fits (see \enquote{Methods})with $A_f$=0.76(0.54), $A_s$=0.24(0.46), $\tau_f$=2.6 (1.4) ps, and $\tau_s$=20 (35) ps for 400 (800) nm excitation. The vertical dashed black lines indicate a time delay of 2.2 ps, which was chosen for the photoconductivity measurement in \textbf{c}. \textbf{c} Extracted real (circles) and imaginary (triangles) areal conductivity, $d_\lambda\cdot\Delta\tilde{\sigma}$, for 400 nm (top) and 800 nm (bottom) pump wavelengths. The solid lines are a fit to the model described in the text.}
	\label{800_400comp}
\end{figure}
\fi
\noindent
\textbf{Photocarrier lifetime.} To study the ultrafast carrier transport, we measure the carrier lifetime and photoconductivity for above-gap and below-gap excitation using TRTS, as shown schematically in Fig. \ref{800_400comp}a. Briefly, a 400 nm or 800 nm photoexcitation laser pulse at a fixed time delay, $\Delta t_1$, with respect to the THz pulse modulates the THz transmission by generating free carriers in the SnIP film (see \enquote{Methods} for further details). The transmitted THz field is electro-optically sampled using a third optical pulse with time delay $\Delta t_2=0$ and monitored for changes induced by photoexcitation. In Fig. \ref{800_400comp}b, we see the time-dependent pump-induced differential THz transmission, $-\Delta T/T_0$, with 400 nm and 800 nm pump wavelengths and 190 $\mu J\,cm^{-2}$ and 850 $\mu J\,cm^{-2}$ excitation fluences, respectively. These fluences were chosen to yield approximately equal amplitude differential signals. The differential THz transmission is proportional to the change in conductivity, $\Delta\tilde\sigma=\Delta\sigma_1+i\Delta\sigma_2$. To parameterize the relaxation, we fit the decay curves with a bi-exponential function (see \enquote{Methods}). We find a slower intial decay for 400 nm than 800 nm excitation, with $\tau_f$=2.6 ps and 1.4 ps, respectively. On the other hand, the long-lived component is shorter for 400 nm than 800 nm, with $\tau_s$=20 ps and 35 ps, respectively. Due to the high surface-to-volume ratio, it is likely that the lifetime is limited by the surface recombination velocity, as seen in some semiconductor nanowires \cite{joyce_electronic_2013}. This conclusion is supported by an estimate of the diffusion timescale in our nanowires (see \supl\noteDifTime) and is consistent with the longer $\tau_s$ for 800 nm excitation, where carriers are excited deeper within the material due to the weaker absorption as compared to above-gap excitation. 
\par

The observation of photoconductivity with sub-gap excitation is interesting on its own, as it implies free carriers must be generated either by two-photon or band/defect-tail absorption \cite{fox2001optical}. The linearity of 800 nm transmission (\supl\noteOptLin) is inconsistent with two-photon absorption while the photon energy of 1.55 eV is too far below the band gap for band-tail absorption, suggesting transitions from deep-level states are the dominant excitation channel. The large amplitude of the signal, with similar photoconductivity at only 4.5 times the fluence (9 times the photon flux), is then somewhat surprising. This can be attributed in part to the increased thickness of the photoexcited layer due to the smaller absorption coefficient and partly to a high density of mid-gap states. It is also possible that weak light-trapping effects due to the nanowire-film morphology increase the effective interaction length of the 800 nm pulse\cite{pathirane_hybrid_2015}.
\par

\noindent
\textbf{Transient photoconductivity spectra.} The corresponding photoconductivity spectra for 400 nm and 800 nm excitation, acquired 2.2 ps after the peak of the transient photoconductivity, are shown in Fig. \ref{800_400comp}c. Due to the geometry of the SnIP nanowires and needles, quantitative measurement of the optical penetration depth, $d_\lambda$, for a given excitation wavelength is difficult. As a result, we study the areal conductivity, $d_\lambda\cdot\Delta\tilde{\sigma}$, which is directly accessible in TRTS of thin films via the Tinkham formula (see \enquote{Methods}). In TRTS, the distinction between areal and volume conductivity does not affect the measurement of carrier mobility, which is derived from the scattering time and is therefore a property of the dispersion rather than the amplitude of the conductivity.
\par

We can learn a great deal about the nature of photoconductivity by studying the qualitative properties of the spectra, which are quite similar for 400 and 800 nm despite the difference in excitation channels. Immediately apparent is the reduced photoconductivity near 1.5 THz. Resonance-like features in the conductivity spectrum are often attributed to resonant modulation of the conductivity due to polaronic effects from polar longitudinal-optical (LO) modes\cite{ziwritsch_direct_2016,yang_time-resolved_2018, cinquanta_ultrafast_2019}, however, we estimate the LO mode frequency to be at a significantly higher frequency (1.73 THz, see \enquote{Methods}). The close proximity of the resonance-like feature in the photoconductivity spectra to the transverse-optical (TO) mode at 1.495 THz, which does not couple strongly through the polar-optical mechanism responsible for strong electron-phonon coupling\cite{yu2010fundamentals}, instead suggests that the lineshape of mode 2 is modified by photoexcitation\cite{koeberg_simultaneous_2007,sim_ultrafast_2014,butler_ultrafast_2016,zhao_monitoring_2019}.
\par

Additionally, we see that the imaginary conductivity is negative and the real conductivity is suppressed at low frequencies. Several models are commonly applied to understand this behavior\cite{jepsen2011,ulbricht_carrier_2011,joyce_review_2016,kuzel_terahertz_2020}. Bruggeman effective-medium theory has been used to describe the conductivity in inhomogeneous systems\cite{baxter_conductivity_2006,walther_terahertz_2007,ulbricht_carrier_2011,joyce_review_2016}, however, we were not able to obtain satisfactory fits to our experimental data using this model. A Drude-Lorentz model has been used to describe surface plasmon resonances in nanowires\cite{parkinson_carrier_2009,joyce_electronic_2013,joyce_review_2016,boland_high_2018} or to describe hydrogen-like transitions in excitonic systems \cite{poellmann_resonant_2015,luo_ultrafast_2017}. However, the plasmon model predicts a specific scaling of the resonant frequency with increasing excitation density, which we do not observe in our data (see \supl\noteModDep{} for fluence-dependent spectra and fits). Moreover, in SnIP there are no excitonic signatures in either the absorbance or photoluminescence spectra \cite{pfister_inorganic_2016,ott_flexible_2019}.
\par

Alternatively, modified forms of Drude conductivity, such as the Drude-Smith model\cite{smith_classical_2001,nemec_far-infrared_2009,cocker_microscopic_2017}, are commonly used to describe nanomaterial conductivity when carrier localization arises from nanoscale morphology \cite{baxter_conductivity_2006,walther_terahertz_2007, cooke_ultrabroadband_2012,jensen_ultrafast_2013,laforge_conductivity_2014,evers_high_2015,liu_ultrahigh_2016,ziwritsch_direct_2016,titova_ultrafast_2016,luo_ultrafast_2017,zhao_monitoring_2019,li_dynamical_2020}. The Drude-Smith model has been shown to fit photoconductivity spectra over a broad frequency range\cite{cooke_ultrabroadband_2012}, yield comparable conductivity to standard transport measurements\cite{laforge_conductivity_2014}, and provide qualitative information about carrier localization\cite{walther_terahertz_2007}. The functional form is given by,
\begin{equation}
\label{eq:DS}
\tilde{\sigma}_{DS}(\omega)=\frac{Ne^2\tau_{DS}/m}{1-i\omega\tau_{DS}}\left(1+\frac{c}{1-i\omega\tau_{DS}}\right),
\end{equation}
where $m$ is the effective mass, $N$ is the density of electron-hole pairs, $\tau_{DS}$ is the scattering time, $e$ is the elementary charge, and $c$ is a phenomenological localization parameter that ranges from 0 to -1 for free and localized carriers, respectively. For the Drude-Smith fits, we use $m=\bar{m}^*$, the average of the direction-dependent effective mass (see \enquote{Methods}), because our thin films consist of an ensemble of randomly-oriented nanowires. It is clear that to quantitatively interpret the spectra we must understand the effective mass anisotropy in SnIP, which we obtain from quantum-chemical calculations.
\par
\ifdefined\plotFigs
\begin{figure}[htb]
	\includegraphics[scale=1]{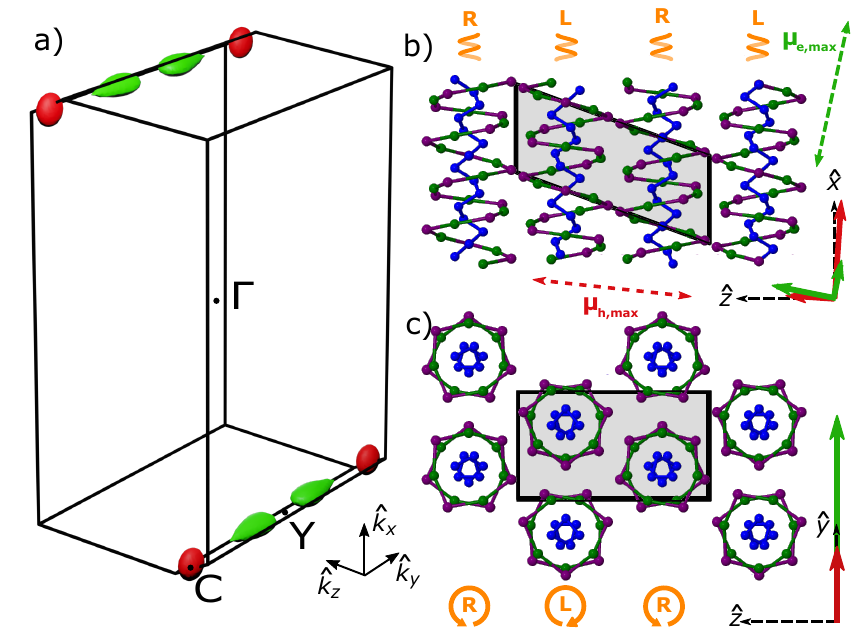}
	\caption{\textbf{Geometry of constant-energy surfaces. a} Constant energy surfaces 5 meV below the valence band maximum (red) and above the conduction band minimum (green). Immediately apparent is the extreme anisotropy of the conduction band. The conduction (valence) band effective masses along the principle axes of the ellipsoidal constant energy surfaces were found to be 0.28 (0.71)$\,m_e$, 2.0 (0.66)$\,m_e$, and 0.51 (0.33)$\,m_e$ in the $\hat{x}'$, $\hat{y}$, and $\hat{z}'$ directions, respectively. Projections of the crystal structure to the \textbf{b} \textit{xz}-plane and \textbf{c} \textit{yz}-plane with the vectors of the principle axes of the effective mass tensor for the conduction band (green) and valence band (red) shown in the bottom right corners. The directions of the colored vectors in \textbf{b}, which are rotated with respect to the Cartesian axes ($\hat{x}$, $\hat{y}$, $\hat{z}$), define the directions of $\hat{x}'$ and $\hat{z}'$ for the conduction band (green) and valence band (red). The preferred transport directions (directions of highest mobility) are indicated by the dashed green arrow for the conduction band (labelled $\mu_{e,max}$), which is almost parallel to the double-helix axis, and the dashed red arrow for the valence band (labelled $\mu_{h,max}$), which is almost perpendicular to the double-helix axis in the $\hat{z}$ direction. Additionally, the helicity of adjacent planes of double helices is indicated schematically. }
	\label{CES}
\end{figure}
\fi

\noindent
\textbf{Anisotropic electronic structure and effective mass.} We can visualize the electronic structure by plotting the constant energy surfaces for the conduction (green) and valence (red) bands with a 5 meV offset from the extrema, as shown in Fig. \ref{CES}a. We can see that the band gap is indirect with the valence band maximum centered at the C point and two conduction band minima slightly offset from the Y point in the direction of the valence band (see Fig. \ref{intro}b, c for the full Brillouin zone and bandstructure as well as \supl\noteEffM{} for more discussion). To extract the effective mass, the band structure was calculated on a uniform grid around the conduction and valence band extrema and the effective mass tensor was calculated from the inverse Hessian matrix (see \enquote{Methods} and \supl\noteHess{}).
\par

Due to the low symmetry of the crystal and anisotropy of bonding, with strong covalent interactions parallel to the double helices ($\hat{x}$ direction) and weak vdW interactions perpendicular to the double helix ($\hat{y}$ and $\hat{z}$ directions), we expect anisotropy in the $\hat{x}$-direction effective mass in comparison to the $\hat{y}$ and $\hat{z}$ directions. This intuition holds for the conduction band, where $m_{cx'}=0.28\,m_e$, where $m_e$ is the free-space electron mass, is the preferred transport direction, while the $\hat{y}$ and $\hat{z}'$ masses are both significantly heavier. Here, we use the notation $\hat{x}'$ and $\hat{z}'$ because the principle axes of the effective mass tensor in the $xz$-plane are rotated by a small amount (10.4$^\circ$/3.8$^\circ$ for the conduction/valence bands) with respect to the Cartesian axes, as seen in Fig. \ref{CES}b. We see that in the valence band the preferred transport direction is $m_{vz'}=0.33\,m_e$, and that in general that the masses perpendicular to the double-helix axis are significantly lower in the valence band than the conduction band. This can be understood based on the results of Li \textit{et al.}, who showed that the valence band states are more localized around the outer SnI helices\cite{li_landscape_2017}, which could facilitate inter-helix hopping in comparison to the conduction band states that are more localized on the inner P helices.
\par

The calculation also reveals large anisotropy in the effective masses in the plane perpendicular to the double-helix axis as $m_{z'}\,<\,m_y$ for both the conduction band and valence band. This anisotropy is more surprising, however, we highlight an additional asymmetry in the $\hat{y}$ and $\hat{z}$ directions that affects the inter-helix interaction strength. As seen in Fig. \ref{CES}b, c, which shows projections of the crystal structure onto the $xz$ and $yz$ planes, stacking in the $\hat{z}$ direction consists of alternating planes of right and left handed helices. Conversely, in the $\hat{y}$ direction, the stacking planes themselves are composed of alternating left and right handed helices with a buckled structure. This asymmetry, which results in anisotropic van der Waals interactions, has also been shown to directly affect the anisotropic mechanical properties of SnIP\cite{ott_flexible_2019}.
\par

\ifdefined\plotFigs
\begin{table}[htb]
	\begin{center}
		\caption{Fit parameters obtained from the model described in the text for the differential conductivity, where $N$ is the volume charge density, and $\tau_{DS}$ and $c$ are the Drude-Smith scattering time and localization parameter, respectively. $\Delta\omega_{0,2}$, $\Delta\gamma_2$, and $\Delta A_2$ are the photoexcitation-induced changes in oscillator resonant frequency, damping, and amplitude, respectively (see \enquote{Methods}). $d_\lambda$ is the photoexcited film thickness, so that $N\cdot d_\lambda$ is the areal charge density.}
		\def\arraystretch{1.5}%
		\setlength{\tabcolsep}{10pt}
		\begin{tabular}{|c|c|c|c|c|c|c|c|c|}
			\hline\label{tab:fitParam}
			$\lambda_{pump} $ & $ N\cdot d_\lambda $  &$\tau_{DS} $ & $c$  & $\Delta A_{osc,2}\cdot d_\lambda$     & $\Delta \omega_{0,2}/2\pi\cdot d_\lambda$ & $\Delta\gamma_2/2\pi\cdot d_\lambda$\\
			(nm)              &  ($cm^{-2}$)  &    $(fs)$   &      & (THz$^2\cdot\mu m$)     &  (THz $\cdot\mu m$)    & (THz$\cdot\mu m$)    \\
			\hline
			400           &  $3.5\times 10^{12}$ &  45      & -0.84 &        -6.7            &   0.9$\times 10^{-3}$  &  -9.1$\times 10^{-3}$ \\
			800           &  $3.8\times 10^{12}$ &  42      & -0.84 &        -6.4            &   1.6$\times 10^{-3}$  &  -6.2$\times 10^{-3}$ \\
			\hline
		\end{tabular}
	\end{center}
\end{table}
\fi

\noindent
\textbf{Carrier mobility and vibrational mode suppression.} To fit the conductivity spectra over the full range, we must also consider the photoexcitation-induced modification to mode 2, the high-amplitude mode at 1.495 THz. To do so, we include a term in our fit function proportional to the differential lineshape using the peak-shift model of Zhao \textit{et al.}\cite{zhao_monitoring_2019},
\begin{equation}
\Delta\tilde{\sigma}=\tilde{\sigma}_{DS}+\tilde{\sigma}_{PS}
\end{equation}
where $\tilde{\sigma}_{PS}$ is the peak-shift conductivity (see \enquote{Methods}, eq. \ref{PSmodel}, and \supl\notePS{} for a discussion of the lineshape and range of validity of this model). As discussed previously, we do not know the exact photoexcited film thickness, $d_\lambda$, which we address by fitting with $d_\lambda\Delta\tilde{\sigma}$ and using as fit parameters the scattering time, localization parameter, areal charge density, $N\cdot d_\lambda$, and grouping the thickness with the differential oscillator parameters (see \enquote{Methods}). A simultaneous fit to the measured real and imaginary components of the conductivity in Fig. \ref{800_400comp}c was performed. The fit parameters are summarized in Table \ref{tab:fitParam}. 
\par

From the Drude-Smith portion of the fit, we can extract the areal charge density, localization parameter, and scattering time. Good fits were obtained for three different fluences ranging from 40 to 470 $\mu J\,cm^{-1}$ with a scattering time that decreases with fluence and a similar localization parameter, indicating that the Drude-Smith model effectively describes the photoconductivity spectra over an order of magnitude range of fluences (see \supl\noteModDep). The measured localization parameter of -0.82, which is a similar to that observed in 1D systems such as graphene nanoribbons and carbon nanotubes, indicates that carriers are highly localized \cite{jensen_ultrafast_2013}. We note that implicit in our model is the approximation that the Drude-Smith scattering time, measured to be $\tau_{DS}=$45 fs and 42 fs for 400 nm  and 800 nm excitation, respectively, is isotropic and the same for electrons and holes. In combination with the calculated effective mass, the measured scattering time allows us to extract the carrier mobility. In the Drude-Smith model (eq. \ref{eq:DS}), the mobility is defined as,
\begin{equation}
\mu=\frac{e\tau_{DS}}{m},
\label{DSmobility}
\end{equation}
where we can study the anisotropic carrier mobility by using the direction-dependent mass instead of $\bar{m}^*$ as in the Drude-Smith fits. Along the $\hat{x}'$ direction (almost parallel to the double-helix axis), with $m_{cx'}=0.28\,m_e$, we find a carrier mobility of $280\,cm^2\,V^{-1}s^{-1}$ while along the $\hat{y}$ direction, with $m_{cy}=2.0\,m_e$,  we find a much smaller carrier mobility of $39\,cm^2\,V^{-1}s^{-1}$ (see \supl\noteEffM{} for a complete summary of the anisotropic mobility). We can also see that the long-range mobility, $\mu_{lr}=\mu(1+c)$, is suppressed by localization due to the nanoscale morphology. For example, in the $\hat{x}'$ direction we find $\mu_{lr,x'}=45\,cm^2\,V^{-1}s^{-1}$. A detailed analysis (see \supl\noteDifTime) indicates that the characteristic length scale defining long-range transport in our samples is on the order of 20 nm, which is several times smaller than the average nanowire diameter. This indicates that, \textit{e.g.}, grain boundaries within nanowires are the origin of localization\cite{laforge_conductivity_2014,titova_ultrafast_2016}
\par

In addition to the free-carrier response, the parameters extracted from the peak-shift model allow us to quantify the effect of photoexcitation on the vibrational mode. The differential oscillator parameters, which are also summarized in Table \ref{tab:fitParam}, indicate a large photoexcitation-induced reduction in the amplitude along with a small reduction in linewidth and blue shift of the resonant frequency (for an estimation of the absolute change in lineshape, see \supl\notePS). This behavior indicates that screening from the photoexcited carriers induces a charge redistribution that stiffens the potential-energy landscape of this mode and reduces its effective charge, \textit{i.e.}, reduces its dipole moment\cite{koeberg_simultaneous_2007,ulbricht_carrier_2011,butler_ultrafast_2016, zhao_monitoring_2019}. Interestingly, the differential linewidth is negative, indicating a line narrowing. We speculate on two possible origins of this effect. First, the charge redistribution could lead to a lattice reorganization that reduces strain inhomogeneity,\cite{zhou_giant_2016,zheng_ultrafast_2020} which would reduce the amount of inhomogeneous broadening\cite{neumann_raman_2015}. Second, it could result from a reduction of lattice anharmonicity, which is sensitive to dispersion forces\cite{marcondes_importance_2018} that can in turn be modified by screening\cite{li_faraday_2018}.
\par

\ifdefined\plotFigs
\begin{figure}[htb]
	\includegraphics[scale=0.95]{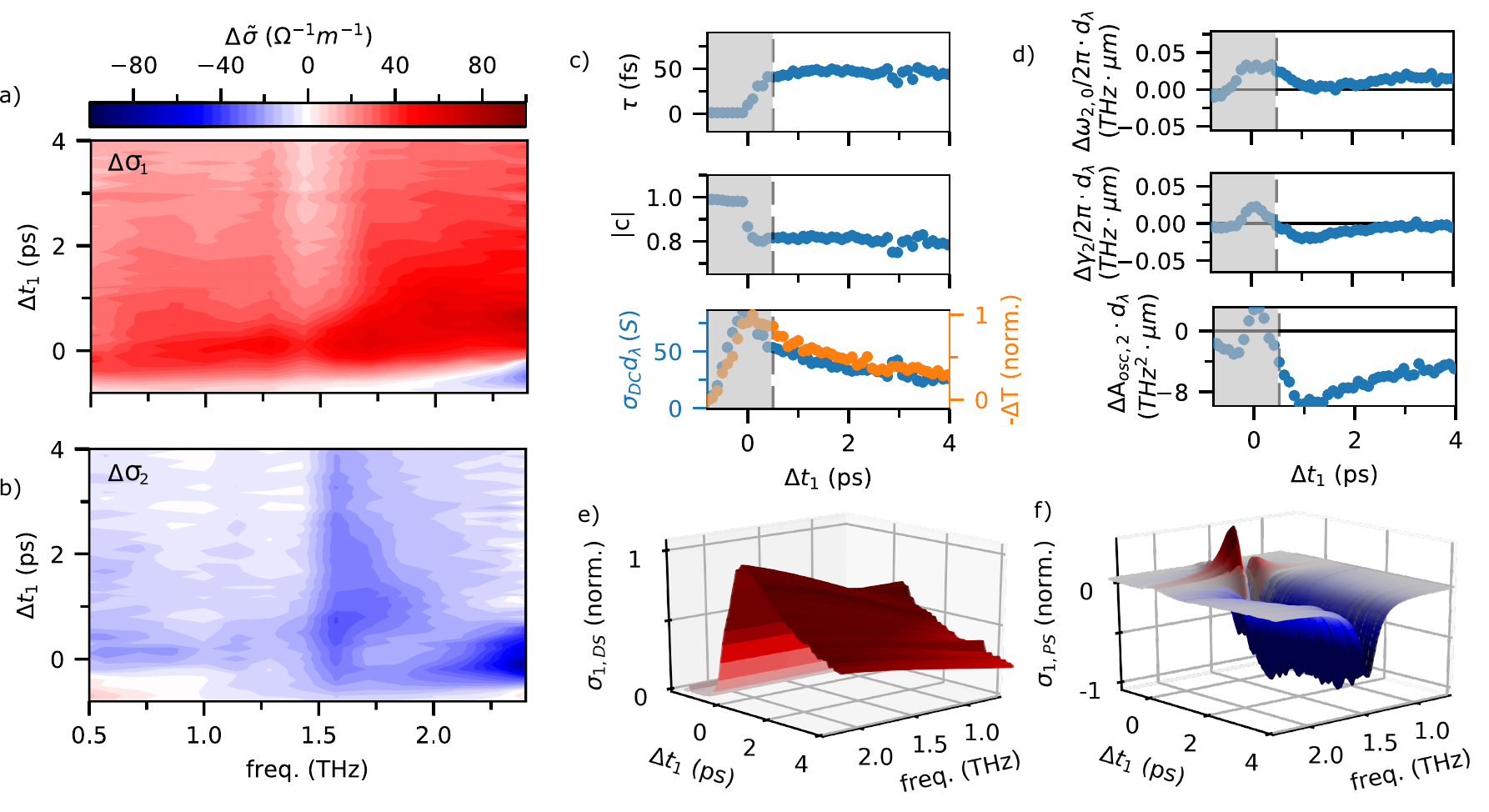}
	\caption{\textbf{Picosecond time-evolution of the transient photoconductivity. a} Real part and \textbf{b} imaginary part of the time evolution of the photoconductivity for 400 nm excitation at a fluence of 190$\,\mu J\,cm^{-2}$. \textbf{c} Time-dependence of the Drude-Smith fit parameters: the scattering time $\tau_{DS}$ (top), localization parameter c (middle), and low frequency limit of the Drude-Smith fit, $\sigma_{DC}\cdot d$ with the time-domain differential transmission superimposed (bottom). \textbf{d}  Time-dependence of the differential oscillator parameters of mode 2 extracted from fits to the the peak-shift model: the center frequency (top) linewidth (middle) and amplitudes (bottom), as a function of time delay. The shaded areas for time-delays less than 0.5 ps indicate the region where early-time transient artifacts could potentially distort the extracted spectra. Evolution of the real part of the \textbf{e} Drude-Smith and \textbf{f} differential oscillator conductivity.}
	\label{2dScan}
\end{figure}
\fi

\noindent
\textbf{Picosecond charge redistribution dynamics.} It is also interesting to study the time evolution of the real and imaginary parts of the complex conductivity after photoexcitation with a 2D scan, as seen in Fig. \ref{2dScan}a, b, respectively. This allows us to see how the Drude-Smith and differential oscillator parameters change as a function of time after femtosecond excitation. The behavior of the vibrational mode is most striking in $\Delta\sigma_2$, suggesting some sort of transient behavior in the differential oscillator parameters at early times. It is, however, important to identify and rule out early-time artifacts that arise due to the system response \cite{beard_transient_2000,larsen_finite-difference_2011}. Our calculations (not shown) suggest that for the low amplitude of modulation observed here the extracted conductivity accurately represents the real conductivity for delay times larger than 0.5 ps after the peak of the transient photoconductivity (chosen as $\Delta t_1=0\,ps$), indicated by the dashed horizontal line in Fig. \ref{2dScan}a, b.
\par

Shown in Fig. \ref{2dScan}c are the time-dependent Drude-Smith fit parameters extracted from fitting the 2D conductivity to the model described in the previous section. The scattering time and localization parameter are stable throughout the entire window, which means the mobility is essentially constant over this range of time delays. In contrast, the peak-shift fit parameters, as seen in Fig. \ref{2dScan}d, show more interesting behavior. The change in resonant frequency, $\Delta\omega_0/2\pi$, shows evidence of oscillatory behavior as a function of time after excitation and the amplitude of negative differential amplitude and linewidth continue to grow until approximately 1 ps. We note the similarity between this 1 ps timescale and the oscillator lifetime, $\gamma_2^{-1}\approx 1\,ps$. We can also see that the differential linewidth, which approaches zero by 3 ps, shows a faster recovery than the reduction in amplitude, which decays on a timescale similar to the extrapolated DC limit of the Drude-smith fit function in eq. \ref{eq:DS}, $\sigma_{DC}$, which is proportional to the carrier density. Finally, using the extracted fit parameters, we can decouple the Drude-Smith and peak-shift portions of the differential signal to visualize the time evolution of each separately, as seen in Fig. \ref{2dScan}e, f, respectively.
\par

\ifdefined\plotFigs
\begin{figure}[htb]
	\includegraphics[scale=1]{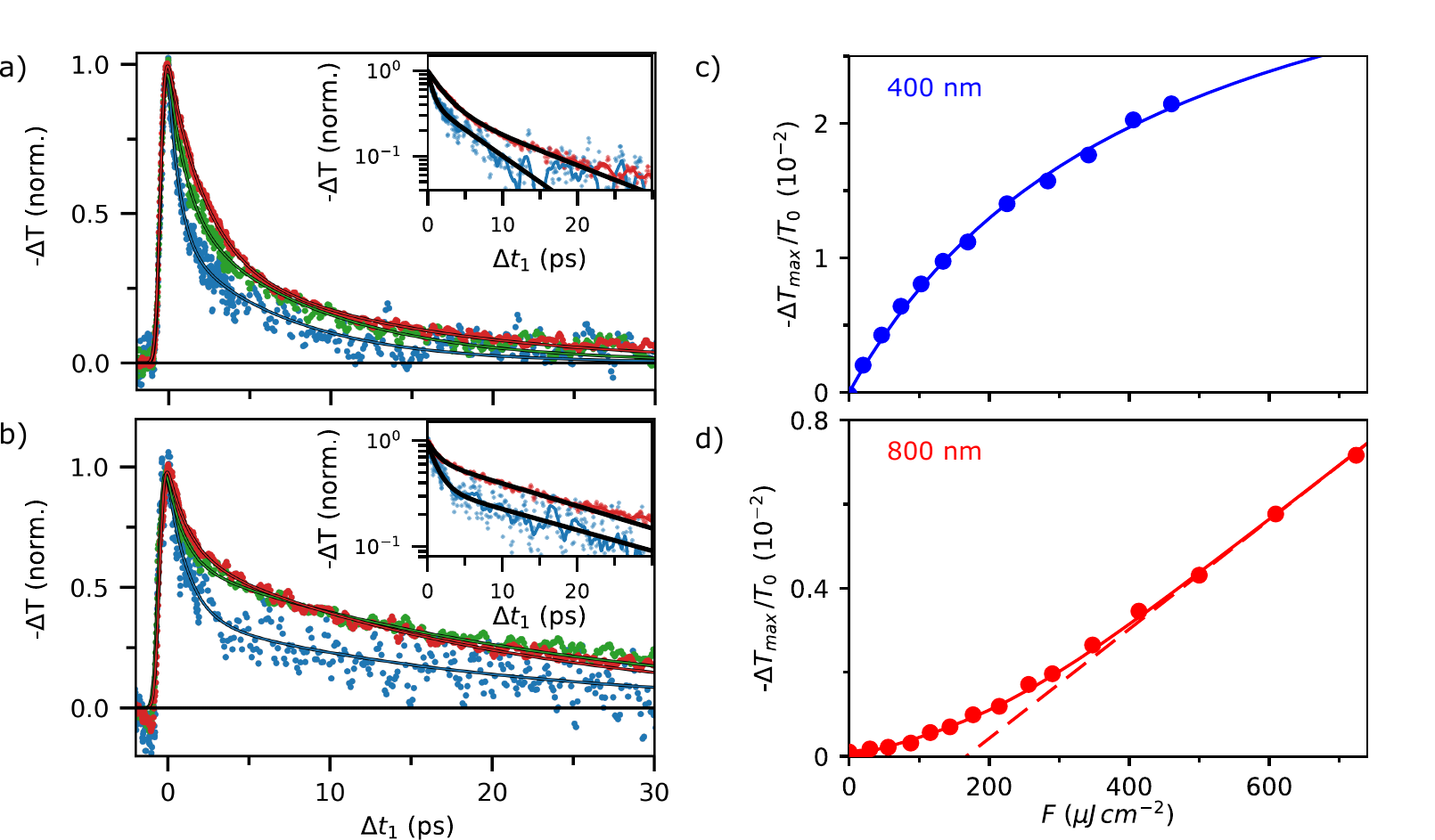}
	\caption{\textbf{Ultrafast trap-filling dynamics. a} Normalized differential THz transmission as a function of pump-probetime-delay for 20, 100, and 450 $\mu J\,cm^{-2}$ excitation fluences (blue, green, and red circles, respectively) at a pump wavelength of 400 nm. Inset: semi-log plot of the 20 and 450 $\mu J\,cm^{-2}$ decays. \textbf{b} Normalized differential THz transmission as a function of time-delay for 170, 570, and 860 $\mu J\,cm^{-2}$ excitation fluences (blue, green, and red circles, respectively) at a pump wavelength of 800 nm. The solid lines are bi-exponential fits in both \textbf{a} and \textbf{b}. Inset: semi-log plot of the 170 and 860 $\mu J\,cm^{-2}$ decays.  Peak change in negative differential transmission as a function of photoexcitation fluence, $F$, for \textbf{c} 400 nm and \textbf{d} 800 nm excitation, respectively. The 400 nm curve was fit using a phenomenological saturation model while the 800 nm curve was fit using a rate equation model with a saturable trap state convoluted with a Gaussian system response (see \supl\noteRateMod).}
	\label{powDep}
\end{figure}
\fi

\noindent
\textbf{Trap filling dynamics and defect density.} From the 2D scan in Fig. \ref{2dScan} we also see that the differential transmission, $\Delta T$, and extrapolated DC conductivity, $\sigma_{DC}$, have very similar time dependence so that we can monitor the relaxation of photoconductivity with a 1D scan along $\Delta t_1$. Fig. \ref{powDep}a and b show the fluence-dependent normalized differential transmission with 400 nm and 800 nm excitation, respectively, revealing an increasing lifetime with fluence in both cases (see \supl\noteBiExp{} for the corresponding bi-exponential fit parameters). This response is characteristic of trap-filling dynamics \cite{uhd_jepsen_ultrafast_2001,parkinson_carrier_2009}. Even at the highest fluence, the lifetime is still quite short, which implies that the surface states are non-saturable. We therefore propose that at low fluence ultrafast trapping in the bulk dominates relaxation, while at high fluence bulk traps are saturated and the lifetime is limited by surface recombination velocity.
\par

We also study how the peak photoconductivity changes as a function of pump fluence, as seen in Fig. \ref{powDep}c, d for 400 nm and 800 nm excitation, respectively. Here we see dramatically different behavior for 800 nm versus 400 nm excitation: the peak differential transmission increases sublinearly with 400 nm and superlinearly with 800 nm excitation. As we have already argued that two-photon absorption is not the dominant excitation channel for 800 nm excitation (see \supl\noteOptLin), we instead attribute the super-linear behavior at low fluence to trapping on timescales faster than the system response time\cite{sahota_many-body_2019}, which is approximately 0.4 ps. 
\par

To fit the curve in Fig. \ref{powDep}d, we use a rate equation model with one saturable and one non-saturable relaxation pathway combined with a Gaussian system response function (see \supl\noteRateMod), which gives excellent quantitative agreement with data. From the $x$-intercept extrapolated linearly from the high fluence data (dashed line in Fig. \ref{powDep}d) we find a saturation fluence of 150 $\mu J\,cm^{-2}$ (photon flux of 6$\times 10^{14}cm^{-2}$), with which we can estimate the bulk trap density to be 6$\times 10^{18}cm^{-3}$ by assuming a penetration depth of 1$\,\mu m$ (similar to the film thickness). Furthermore, an estimate based on the fluence-dependent lifetime with 400 nm excitation yields a similar trap density.
\par

To understand the saturation behavior with 400 nm excitation, we can rule out the introduction of new recombination channels as the lifetime does not decrease with increasing fluence \cite{sahota_many-body_2019}. We can also rule out optical nonlinearities (see \supl\noteOptLin) and reduction of mobility at high fluence (see \supl\noteModDep). At high fluences the carrier density is greater than $10^{18}cm^{-3}$ (see \supl\noteModDep{} or \enquote{Methods}), which is high enough that the quasi-Fermi level is pushed into the conduction band where the dispersion is highly non-parabolic. The electronic dispersion in these higher lying states becomes essentially flat in the y-direction and the system becomes effectively 2D in a large range of the Brillouin zone (see \supl\noteEffM). The sub-linearity at high fluence could therefore result from a combination of non-parabolicity in the $\hat{x}'$ and $\hat{z}'$ directions and a partial freeze out of transport in the $\hat{y}$-direction.
\par

\ifdefined\plotFigs
\begin{figure}[htb]
	\includegraphics[scale=1]{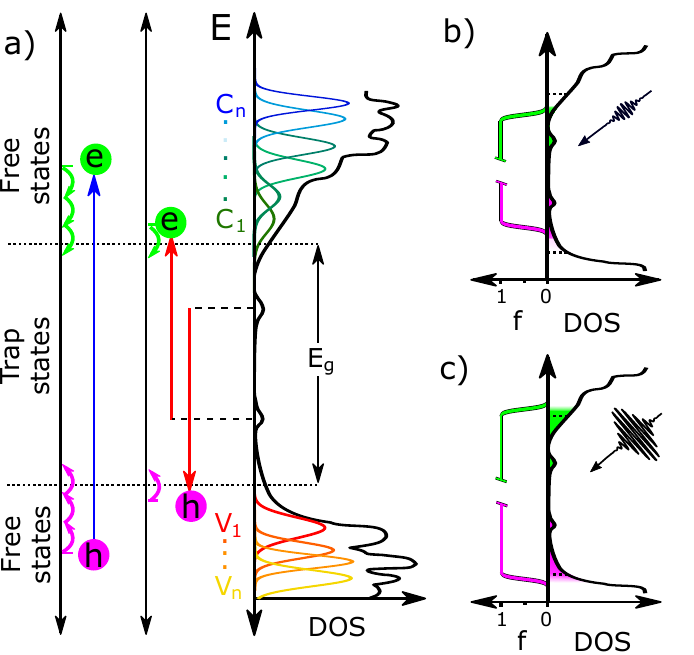}
	\caption{\textbf{Excitation and relaxation pathways. a} Right: Schematic illustration of the DOS (black curve) with contributing conduction (green/blue) and valence (red/orange) bands. The DOS below/above the conduction/valence bands consist of traps due to band-tail states and deep-level defects. Left: Illustration of the excitation pathways for 400 nm (blue) and 800 nm (red) pump wavelengths. The 400 nm pump produces electron-hole pairs with significant excess energy while the 800 nm pump, which has sub-gap photon energy, excites either holes or electrons out of mid-gap trap states. Illustration of the occupancy of electrons (green) and holes (purple) in \textbf{b} the low-fluence and \textbf{c} high-fluences cases. On picosecond timescales the distribution can be approximated as thermal with a quasi-Fermi level for electrons and holes. In the low fluence case, trap states near the band edge are not filled and carriers are rapidly trapped. Alternatively, in the high-fluence case, traps are filled and the lifetime is enhanced. The distribution of trap states can consist of band-tail states (exponential tails below/above the dotted line in the conduction/valence band DOS) and localized defect and impurity states within the gap (peaks near the middle of the gap). }
	\label{exChan}
\end{figure}
\fi

\noindent
\textbf{Ultrafast excitation and relaxation pathways.} Using the results described in the previous sections, we can develop a qualitative picture of excitation and relaxation channels in SnIP, as illustrated in Fig. \ref{exChan}. The system consists of free states, corresponding to the delocalized conduction and valence bands, and trap states, which includes band-tail states and deep levels. With 400 nm excitation (above gap), electron-hole pairs are excited in band-like regions with a high joint density of states. Alternatively, with 800 nm excitation (sub gap), electrons are excited to or from regions with a high density of states in the gap to create free holes or electrons, respectively. 
\par

With 400 nm excitation (blue line in Fig. \ref{exChan}a), carriers are generated with significant average excess energy, which must be dissipated before trapping, leading to a slower initial decay, $\tau_f$, than with 800 nm excitation. This relaxation can involve both intraband and interband relaxation through multiple conduction/valence bands (indicated by C$_i$/V$_i$ in Fig. \ref{exChan}). Alternatively, with 800 nm excitation (red line in Fig. \ref{exChan}a) carriers are generated with less excess energy and they can be trapped on timescales faster than the response time of the system, leading to the superlinear behavior seen in Fig. \ref{powDep}d. Importantly, trapping on such fast timescales must occur via bulk trap states, as there is insufficient time for diffusion to the surface. 
\par

Finally, in Fig. \ref{exChan}b, c, we illustrate the trap filling-behavior that leads to the lifetime enhancement observed at high fluence. With low fluence excitation (Fig. \ref{exChan}b), the quasi-Fermi levels of electrons and holes are still within the gap and a high density of unoccupied traps is present. Alternatively, with high fluence excitation (Fig. \ref{exChan}c), the quasi-Fermi levels have reached the band-like states and all traps are filled, resulting in an enhanced lifetime. This description applies to bulk traps and the surface trap density is likely high enough to pin the quasi-Fermi levels to the middle of the gap even at the highest
fluences.
\par

\section{Discussion}
Our results provide significant insight into the optoelectronic properties of SnIP. Using the Drude-Smith model, we find that the microscopic mobility in our SnIP nanowire films is as high as $280\,cm^2\,V^{-1}s^{-1}$, which is similar to other vdW materials\cite{yu_analyzing_2017,qin_raman_2020} and many times higher than mobilities in organic semiconductors with comparable mechanical properties typically used in flexible electronics\cite{gao_high_2015}. In a survey of a wide variety of materials, we find the combination of high mobility and low bulk modulus seen in SnIP is quite rare. Additionally, its extreme flexibility sets it apart from contemporary materials, giving SnIP a unique combination of electronic and mechanical properties (see \supl\noteMatSum{}).
\par

We also make the first measurement of optically-active vibrational modes in SnIP and observe a reduction in oscillator amplitude of a strong 1.495 THz mode after photoexcitation. The fingerprints of low-frequency vibrational dynamics on TRTS spectra has been the focus of several recent studies that have elaborated on a variety of interesting mechanisms\cite{huber_femtosecond_2005,kubler_coherent_2007-1,gaal_internal_2007,koeberg_simultaneous_2007,sim_ultrafast_2014,ziwritsch_direct_2016,butler_ultrafast_2016, yang_time-resolved_2018,cinquanta_ultrafast_2019,lan_ultrafast_2019, zhao_monitoring_2019}. In this work, a simple estimate yields a Debye length on the order of 3 nm (see \enquote{Methods}), which is several times larger than the interatomic spacing, indicating that we are in an excitation regime where modulation of optically-active modes is not expected in prototypical covalent semiconductors\cite{varga_coupling_1965,huber_femtosecond_2005,turchinovich_femtosecond_2017}. However, in other material systems, changes to the vibrational spectrum have been linked to a photoinduced charge redistribution \cite{kubler_coherent_2007-1,koeberg_simultaneous_2007,ulbricht_carrier_2011,butler_ultrafast_2016, zhao_monitoring_2019}. In SnIP, where long-range dispersion forces account for a great deal of the cohesive energy, the structure and charge distribution could be especially sensitive to screening\cite{li_faraday_2018}. The delayed reduction in oscillator amplitude revealed by the 2D scan is also interesting as it implies that there is a characteristic timescale associated with the reduction in oscillator amplitude. Further exploration of this behavior in the early time regime or at low temperature could yield more insight into the dynamics of the charge redistribution\cite{lan_coherent_2019,lan_ultrafast_2019}.
\par

Importantly, we construct a picture of optical excitation and relaxation pathways in SnIP nanowires, identifying bulk trapping as the dominant source of recombination at low fluence. From a practical standpoint, the high trap density of 6$\times 10^{18}\,cm^{-3}$, which results in trapping timescales of several hundred femtoseconds or less in the bulk, is a barrier for photocatalysis and optoelectronics as a large fraction of carriers are trapped before they can diffuse to the surface or interface. This suggests that a significant enhancement in photocatalytic activity could be achieved by passivation or annealing to remove bulk defects \cite{parkinson_carrier_2009,chang_electrical_2012,ozawa_correlation_2018}. The high trap density is also a likely source of scattering centers and it is likely that eliminating bulk defects would significantly enhance the carrier mobility. We note that some of the traps appear to be thermally active, as the lifetime appears to increase with temperature (see \supl\noteTemp). Future studies exploring this behavior in more detail may help with understanding the nature of these states.
\par

In summary, we have performed the first ultrafast spectroscopy and made the first measurement of carrier mobility of SnIP, a fundamentally exciting new inorganic double-helix material with extreme flexibility and quasi-1D nature. THz spectroscopy, supplemented by quantum chemical calculations that reveal a highly anisotropic electronic structure, indicates that SnIP has a carrier mobility as high as $280\,cm^2\,V^{-1}s^{-1}$, which is remarkably high for a material as soft and flexible as SnIP. Tracking photoexcitation-induced modification of a strong vibrational resonance reveals that photoexcitation results in a redistribution of charge that occurs on a timescale comparable to the lifetime of the resonance. Finally, we explored the excitation and relaxation channels and suggested that reducing the bulk trap density could yield a considerable enhancement in carrier mobility and lifetime in this promising new material. 
\par

\section{Methods}
\noindent
\textbf{Sample Preparation}.
SnIP was grown by heating and annealing a stoichiometric mixture of SnI$_4$, Sn, and P$_{red}$ in a vacuum-sealed argon-purged ampoule\cite{pfister_inorganic_2016}. The sample was cleaned in toluene to remove any unreacted SnI$_4$, and ultrasonicated to produce a suspension with a distribution of nanowires with, on average, 190 nm in diameter and 0.5-10 $\mu m$ in length (\supl\noteDist). The suspended SnIP was drop cast on a z-cut quartz wafer to produce an approximately uniform sample of 1.5 $\mu m$ thickness suitable for THz spectroscopy (Fig. \ref{intro}d and \supl\noteMorph). Raman microscopy of a variety of ultrasonicated nanowires as well as powder x-ray diffraction (XRD) show only a small line-broadening on average compared to bulk SnIP needles, indicating high crystal quality of the nanowires (\supls\noteXRD{} and \noteRam). In addition to SnIP, the XRD reveals that the thin film contains a small amount of the clathrate compound $Sn_{24}P_{19.3}I_8$ \cite{shatruk_first_1999}. 
\par

\noindent
\textbf{Laser source and THz spectroscopy system}. We used an ultrafast laser surce consisting of a 1 kHz, 100 fs transform limit, 700 $\mu J$, Ti:Sapphire oscillator/amplifier system operating at 800 nm. The beam was split into three paths, one to generate THz pulses using optical rectification in a 1 mm ZnTe crystal, one to electro-optically sample the THz pulse in another 1 mm ZnTe crystal, and one used for photoexcitation with either its fundamental frequency or second harmonic. The THz beam path was enclosed in a high vacuum environment ($<10^{-5}mbar$). To extract the conductivity from THz spectra, we used the Tinkham formula,
\begin{equation}
\tilde{\sigma}=\frac{\tilde{n}_{sub}+1}{Z_0d}\left(\frac{1}{\tilde{t}}-1\right),
\label{tinkham}
\end{equation}
where $\tilde{n}_{sub}$ is the complex substrate index, $Z_0$ is the impedance of free space, and $\tilde{t}$ is the complex transmission of the THz pulse. For TDS measurements, we used $\tilde{t}=\tilde{E}_{samp}/\tilde{E}_{ref}$, where $\tilde{E}_{samp}$ and $\tilde{E}_{ref}$ are the complex amplitudes extracted from the Fourier transform of the waveform transmitted through the substrate and film \textit{vs} through the substrate alone, respectively, with no photoexcitation pulse. The conductivity then was related to the dielectric function using,
\begin{equation}
\tilde{\epsilon}_r(\omega)=1+i\frac{\tilde{\sigma}(\omega)}{\omega\,\epsilon_0},
\label{eq:eps_sig}
\end{equation}
where $\epsilon_0$ is the permittivity of free space. 
\par

TRTS was performed using the same system with an additional photoexcitation pulse of 400 (800) nm wavelength to photoexcite the sample above (below) the band gap. 1D scans of the time delay along $\Delta t_1$, with $\Delta t_2=0\,ps$ (see inset of Fig. \ref{800_400comp}b) were used to probe the carrier lifetime and 1D scans of the THz time delay with $\Delta t_1-\Delta t_2$ fixed were used to extract reference and photoexcited spectra, which are related to the conductivity a fixed time after photoexcitation using eq. \ref{eq:eps_sig}\cite{beard_transient_2000,jepsen2011}. The reference and photoexcited spectra were measured simultaneously using a two-chopper scheme to achieve a high signal-to-noise ratio\cite{iwaszczuk_simultaneous_2009,li_dynamical_2020}. The photoconductivity spectra were extracted using a modified version of the Tinkham formula that took into account the dispersion of the unexcited film, which only had a small affect on the extracted $\Delta\tilde{\sigma}$ (see \supl\noteTinkMod). The 2D scan was performed with 400 nm excitation by collecting separate 1D scans along $\Delta t_1-\Delta t_2$ for various $\Delta t_1$. 
\par

\noindent
\textbf{Bi-Exponential Fit} The relaxation dynamics were fit with a bi-exponential decay given by the equation,
\begin{equation}
-\Delta T=A_fe^{-\Delta t_1/\tau_f}+A_se^{-\Delta t_1/\tau_s},
\label{eq:biexp}
\end{equation}
where $A_f$ and $A_s$ are the fast and slow decay amplitudes, respectively, and $\tau_f$ and $\tau_s$ are the fast and slow decay lifetimes, respectively. The average lifetime varied by up to several picoseconds depending on the location of excitation on the film. 
\par

\noindent
\textbf{Oscillator differential lineshape.} To study the oscillator parameters in the non-equilibrium state we must include a term proportional to the difference oscillator lineshape in the photoexcited state, $\tilde{\sigma}_{osc,2}'$, and the static state, $\tilde{\sigma}_{osc,2}$,
\begin{equation}
\Delta\tilde{\sigma}_{osc,2}(\omega)=\tilde{\sigma}_{osc,2}'-\tilde{\sigma}_{osc,2}.
\end{equation}
Using eqs. \ref{eq:osc} and \ref{eq:eps_sig}, this can be written as,
\begin{equation}
\label{eq:oscDif}
\frac{i\Delta\tilde{\sigma}_{osc,2}(\omega)}{\omega\epsilon_0}=\frac{A'_{osc,2}}{{\omega'}_{0,2}^2-\omega^2+i\omega\gamma_2'}-\frac{A_{osc,2}}{\omega_{0,2}^2-\omega^2+i\omega\gamma_2},
\end{equation}
where the prime superscript indicates the value of each parameter in the photoexcited state. We see that a change in the center frequency or damping results in a non-Lorentzian differential lineshape. Unfortunately, we cannot group the photoexcited film thickness, which is not well known for SnIP, into one of the fit parameters as is possible with the Drude-Smith portion of the fit. To overcome this issue, we employ the peak-shift (PS) model of Zhao \textit{et al.}\cite{zhao_monitoring_2019}, where the differential lineshape is given by,
\begin{equation}
\tilde{\sigma}_{PS}(\omega)=\sum_{i=1}^{3}\frac{\partial\tilde{\sigma}_{osc,2}}{\partial x_i}\Delta x_i
\label{PSmodel}
\end{equation}
where the $x_i$'s represent $\gamma_2$, $\omega_{0,2}$, and $A_{osc,2}$ and the $\Delta x_i$'s are the corresponding photoexcitation induced changes. With this functional form, we can use the products $d_\lambda\cdot\Delta\gamma_2$, $d_\lambda\cdot\Delta A_{osc,2}$, and $d_\lambda\cdot\Delta\omega_{0,2}$ as the free parameters. We note that the validity of this differential form is restricted to small differential changes in the center frequency and damping constants while it is exact for the differential amplitude. For further discussion see \supl\notePS.
\par

\noindent
\textbf{Quantum-chemical calculations and effective mass.} The band structure and density of states, shown in Fig. \ref{intro}c, was calculated in the framework of DFT with the HSE06 functional using the Crystal17 sofware package. Six Monkhorst-Pack-type $k$-points were used for sampling the reciprocal space (a double-density Gilat net of 12 points was used in the calculation of the Fermi energy). The calculations were carried out using Grimme's DFT-D3 dispersion correction scheme as implemented in Crystal17. The vibrational spectrum was calculated, again with Crystal17, using the coupled-perturbed Kohn-Sham method. The imaginary dielectric tensor was calculated using the Born-effective charges (see \supls\noteEpsDFT{} and \noteVibVecs) assuming a Lorentzian lineshape and 0.1 THz damping constant. The weighted average, plotted in Fig. \ref{intro}e, was calculated as $\epsilon_{2,DFT}=0.5\epsilon_{2,xx}+0.25\epsilon_{2,yy}+0.25\epsilon_{2,zz}$.
\par

To calculate the effective masses, the band structure was calculated on a uniform grid in $k$-space near the conduction-band minimum and valence-band maximum. The energies of the $i^{th}$ band in $k$-space can be expanded around the extrema as,
\begin{equation}
E_{\vec{k}}^i=E^i|_{\vec{k}_0}+\frac{1}{2!}H^i_{mn}|_{\vec{k}_0}\vec{k}_m\vec{k}_n+\frac{1}{3!}T^i_{mnl}|_{\vec{k}_0}\vec{k}_m\vec{k}_n\vec{k}_l+\cdots,
\label{effMassExp}
\end{equation}
where $H^i_{mn}=\partial_{km}\partial_{kn}E_{i\vec{k}}$ and $T^i_{mnl}=\partial_{km}\partial_{kn}\partial_{kl}E_{i\vec{k}}$ are the second (Hessian) and third rank partial derivative tensors, the subscripts m, n, and l are the Cartesian indices, and an implicit sum over like indices is assumed. Including only the parabolic term yields the effective-mass approximation and the effective-mass tensor, $M^i_{mn}$, can be calculated from the inverse of the Hessian as,
\begin{equation}
M^i_{mn}=\left(\hbar^2H^i_{mn}\right)^{-1}.
\end{equation}
Diagonalizing the effective-mass tensor yields the principle axes (eigenvectors) and masses (eigenvalues), which are shown in Fig. \ref{CES}b, c. Additionally, \supls\noteHess{} and \noteEffM{} contain a detailed description of the principle axes and a discussion of the range of validity of the effective-mass approximation, respectively. We find the average effective mass by integrating the inverse of the direction-dependent effective mass for a given band, $m_i(\theta,\phi)$, over the distribution of nanowire directions,
\begin{equation}
\bar{m}_i^{-1}=\frac{1}{2\pi^2}\int_0^{2\pi}d\theta\int_0^{\pi}d\phi\, m_i^{-1}(\theta,\phi),
\end{equation}
where $\theta$ is the azimuthal angle and $\phi$ is the polar angle with respect to the double-helix axis and the factor of $(2\pi^2)^{-1}$ is a normalization constant. Note that due to the nanowire tendency to lie flat, the integration assumes a constant weighting as a function of $\theta$ and $\phi$ as opposed to a constant weight as a function of solid angle. The direction-dependent effective mass can be found directly from the definition of the effective mass approximation in eq. \ref{effMassExp}. The result of the integration is,
\begin{equation}
\frac{1}{\bar{m}_i}=\frac{a}{m_{ix'}}+\frac{0.25}{m_{iy}}+\frac{b}{m_{iz'}},
\end{equation}
where $a$ and $b$ are constants that depend on the principle axes of the effective mass tensor and are equal to $a$=0.485 (0.499) and $b$=0.245 (0.250) for the conduction (valence) band. We find $\bar{m}_c$=0.42$\,m_e$ and $\bar{m}_v$=0.54$\,m_e$. Finally, assuming an equal concentration of electrons and holes we calculate the average reduced mass,
\begin{equation}
\frac{1}{\bar{m}^*}=\frac{1}{\bar{m}_v}+\frac{1}{\bar{m}_c},
\end{equation}
which is the mass used to determine the areal charge density from the Drude-Smith fits. We find $\bar{m}^*=0.24\,m_e$.
\par

\noindent
\textbf{Carrier density and screening length.} In order to estimate the screening length and plasma frequency, we need to know photoexcited volume charge density instead of the areal charge density that is measured by THz spectroscopy so we must estimate the penetration depth. In typical semiconductors, with excitation well above the band gap the penetration depth is typically less than 100 nm. With $N\cdot d_\lambda=7\times 10^{12}\,cm^{-2}$ and an estimate of 100 nm for the penetration depth with 400 nm excitation, we find a carrier density of $N=7\times 10^{17}cm^{3}$.
\par

In the low-density high-temperature limit, we can use the Debye length as a simple estimate of the screening length,
\begin{equation}
\lambda_{D}=\left(\frac{\epsilon_{st}k_BT}{Ne^2}\right)^{1/2},
\end{equation}
where $\epsilon_{st}$, $k_B$, $T$, and $N$ are the static (zero frequency) permittivity, Boltzmann constant, carrier temperature, and free carrier density, respectively. For a temperature of 300$\,$K and density of $7\times 10^{17}\,cm^{-3}$ we find a screening length of 3 nm, which is several times larger than the dimensions of the unit cell and significantly larger than the bond lengths.
\par

\noindent
\textbf{LO frequency and Fr\"ohlich constant.} The Lyddane-Sachs-Teller (LST) relation is often used to calculate the LO phonon frequency from the TO phonon frequency along with the static (zero frequency) and infinite (high frequency) dielectric constants,
\begin{equation}
\frac{f_{LO}^2}{f_{TO}^2}=\frac{\epsilon_{st}}{\epsilon_{\infty}}.
\label{LST}
\end{equation}
To find the correct dielectric constants to use in eq. \ref{LST}, we must account for the anisotropy of the dielectric constant and the filling fraction of the film. For a given filling fraction, $f$, we assume the experimental dielectric function, $\epsilon_{exp}$, is given by the linear combination,
\begin{equation}
f(0.5\epsilon_\parallel+0.5\epsilon\bot)+(1-f)\epsilon_{vac}=\epsilon_{exp},
\end{equation}
where $\epsilon_\parallel$ and $\epsilon_\bot$ are the dielectric constants parallel and perpendicular to the double-helix axis, respectively, $\epsilon_{vac}=1$ is the vacuum dielectric constant, and $f$ is the filling fraction, which we take to be 0.35 (see \supl\noteDist). We make the additional assumption that the parallel and perpendicular high-frequency dielectric constants are equal, \textit{i.e.}, $\epsilon_{\infty\parallel}=\epsilon_{\infty\bot}$, which is shown to be approximately true based on DFT calculations in \supl\noteEpsDFT. From the experimentally measured values, $\epsilon_{st}=4.1$ and $\epsilon_\infty=3.5$, we find a static dielectric constant of $\epsilon_{st\parallel}=11.5$ and infinite dielectric constant of $\epsilon_{\infty\parallel}=\epsilon_{\infty\bot}=8.17$. 
\par

Finally, we find the contribution of mode 2 to the static dielectric constant, where $\Delta\epsilon_i$ is the contribution of mode $i$, using the relation,
\begin{equation}
\epsilon_{st\parallel}=\Delta\epsilon_{1\parallel}+\Delta\epsilon_{2\parallel}+\epsilon_{\infty\parallel}=\frac{A_{1\parallel}}{\omega_{0,1}^2}+\frac{A_{2\parallel}}{\omega_{0,2}^2}+\epsilon_{\infty\parallel},
\end{equation}
where $A_{i\parallel}$ is the amplitude of the $i^{th}$ oscillator parallel to the double-helix axis, and assuming the ratio $A_{1\parallel}/A_{2\parallel}=A_{osc,1}/A_{osc,2}$, we find an effective static permittivity for mode 2 of $\epsilon_{2st\parallel}=\Delta\epsilon_{2\parallel}+\epsilon_{\infty\parallel}=10.9$. 
\par

To calculate the LO frequency, we use eq. \ref{LST}, with $\epsilon_{\infty}=\epsilon_{\infty\parallel}$ and $\epsilon_{st}=\epsilon_{2,st\parallel}$. We note that the both the assumptions and the uncertainty in film thickness contribute to uncertainty in our calculation of the LO frequency. Taking this into consideration, we estimate $f_{LO}=1.73\pm 0.10$ THz. We further justify the application of the LST relation in this manner in \supl\noteLST.
\par

The Fr\"ohlich constant, $\alpha$, sets the strength of the polar-optical electron-phonon coupling and can be estimated using\cite{yu2010fundamentals},
\begin{equation}
\alpha=\frac{1}{137}\left(\frac{m^*c^2}{2hf_{LO}}\right)^{1/2}\left(\frac{1}{\epsilon_{\infty}}-\frac{1}{\epsilon_{st}}\right),
\end{equation}
where $h$ is the Planck constant, $c$ is the speed of light and $m^*$ is the effective mass. This simple estimate does not take into account effective mass anisotropy or mode polarization. With $m_{cx'}$=0.28, $\epsilon_{st\parallel}=10.9$, $\epsilon_{\infty\parallel}=8.2$, and $f_{LO}=1.73\,THz$, we estimate $\alpha=0.7\pm 0.4$, which is comparable to the moderately polar group of II-VI semiconductors. The Fr\"ohlich constant should also renormalize the effective masses\cite{lan_ultrafast_2019}, which would result in a small reduction in mobility. However, because it does not result in large corrections to the effective mass and as the uncertainty is rather large due to the approximations made in its calculation, in this work we use bare mass from DFT to calculate the density and mobility.
\par

\section{Data Availability}
The data that support the findings of this study are available from the corresponding author upon reasonable request.

\bibliographystyle{naturemag}
\bibliography{SnIP}% Produces the bibliography via BibTeX.

\begin{thebibliography}{10}
\expandafter\ifx\csname url\endcsname\relax
  \def\url#1{\texttt{#1}}\fi
\expandafter\ifx\csname urlprefix\endcsname\relax\def\urlprefix{URL }\fi
\providecommand{\bibinfo}[2]{#2}
\providecommand{\eprint}[2][]{\url{#2}}

\bibitem{soghomonian_inorganic_1993}
\bibinfo{author}{Soghomonian, V.}, \bibinfo{author}{Chen, Q.},
  \bibinfo{author}{Haushalter, R.~C.}, \bibinfo{author}{Zubieta, J.} \&
  \bibinfo{author}{O'Connor, C.~J.}
\newblock \bibinfo{title}{An {Inorganic} {Double} {Helix}: {Hydrothermal}
  {Synthesis}, {Structure}, and {Magnetism} of {Chiral}
  [({CH$_3$}){$_2$NH$_2$}]{K$_4$}[{V$_{10}$O$_{10}$}({H$_2$O})$_2$({OH})$_4$({PO$_4$})$_7$]$\cdot${4H$_2$O}}.
\newblock \emph{\bibinfo{journal}{Science}} \textbf{\bibinfo{volume}{259}},
  \bibinfo{pages}{1596--1599} (\bibinfo{year}{1993}).
\newblock
  \urlprefix\url{https://www.sciencemag.org/lookup/doi/10.1126/science.259.5101.1596}.

\bibitem{su_inorganic_2011}
\bibinfo{author}{Su, D.~S.}
\newblock \bibinfo{title}{Inorganic {Materials} with {Double}-{Helix}
  {Structures}}.
\newblock \emph{\bibinfo{journal}{Angewandte Chemie International Edition}}
  \textbf{\bibinfo{volume}{50}}, \bibinfo{pages}{4747--4750}
  (\bibinfo{year}{2011}).
\newblock \urlprefix\url{http://doi.wiley.com/10.1002/anie.201007147}.

\bibitem{ivanov_inorganic_2012}
\bibinfo{author}{Ivanov, A.~S.}, \bibinfo{author}{Morris, A.~J.},
  \bibinfo{author}{Bozhenko, K.~V.}, \bibinfo{author}{Pickard, C.~J.} \&
  \bibinfo{author}{Boldyrev, A.~I.}
\newblock \bibinfo{title}{Inorganic {Double}-{Helix} {Structures} of
  {Unusually} {Simple} {Lithium}-{Phosphorus} {Species}}.
\newblock \emph{\bibinfo{journal}{Angewandte Chemie International Edition}}
  \textbf{\bibinfo{volume}{51}}, \bibinfo{pages}{8330--8333}
  (\bibinfo{year}{2012}).
\newblock \urlprefix\url{http://doi.wiley.com/10.1002/anie.201201843}.

\bibitem{zhao_emerging_2014}
\bibinfo{author}{Zhao, M.-Q.}, \bibinfo{author}{Zhang, Q.},
  \bibinfo{author}{Tian, G.-L.} \& \bibinfo{author}{Wei, F.}
\newblock \bibinfo{title}{Emerging double helical nanostructures}.
\newblock \emph{\bibinfo{journal}{Nanoscale}} \textbf{\bibinfo{volume}{6}},
  \bibinfo{pages}{9339--9354} (\bibinfo{year}{2014}).
\newblock \urlprefix\url{http://xlink.rsc.org/?DOI=C4NR00271G}.

\bibitem{haldar_metal-free_2009}
\bibinfo{author}{Haldar, D.} \& \bibinfo{author}{Schmuck, C.}
\newblock \bibinfo{title}{Metal-free double helices from abiotic backbones}.
\newblock \emph{\bibinfo{journal}{Chem. Soc. Rev.}}
  \textbf{\bibinfo{volume}{38}}, \bibinfo{pages}{363--371}
  (\bibinfo{year}{2009}).
\newblock \urlprefix\url{http://xlink.rsc.org/?DOI=B803553A}.

\bibitem{pfister_inorganic_2016}
\bibinfo{author}{Pfister, D.} \emph{et~al.}
\newblock \bibinfo{title}{Inorganic {Double} {Helices} in {Semiconducting}
  {SnIP}}.
\newblock \emph{\bibinfo{journal}{Advanced Materials}}
  \textbf{\bibinfo{volume}{28}}, \bibinfo{pages}{9783--9791}
  (\bibinfo{year}{2016}).
\newblock \urlprefix\url{http://doi.wiley.com/10.1002/adma.201603135}.

\bibitem{baumgartner_inorganic_2017}
\bibinfo{author}{Baumgartner, M.}, \bibinfo{author}{Weihrich, R.} \&
  \bibinfo{author}{Nilges, T.}
\newblock \bibinfo{title}{Inorganic {SnIP}-{Type} {Double} {Helices} in
  {Main}-{Group} {Chemistry}}.
\newblock \emph{\bibinfo{journal}{Chemistry - A European Journal}}
  \textbf{\bibinfo{volume}{23}}, \bibinfo{pages}{6452--6457}
  (\bibinfo{year}{2017}).
\newblock \urlprefix\url{http://doi.wiley.com/10.1002/chem.201700929}.

\bibitem{li_landscape_2017}
\bibinfo{author}{Li, X.} \emph{et~al.}
\newblock \bibinfo{title}{Landscape of {DNA}-like inorganic metal free double
  helical semiconductors and potential applications in photocatalytic water
  splitting}.
\newblock \emph{\bibinfo{journal}{Journal of Materials Chemistry A}}
  \textbf{\bibinfo{volume}{5}}, \bibinfo{pages}{8484--8492}
  (\bibinfo{year}{2017}).
\newblock \urlprefix\url{http://xlink.rsc.org/?DOI=C7TA01349C}.

\bibitem{bijoy_atomic_2020}
\bibinfo{author}{Bijoy, T.~K.}, \bibinfo{author}{Murugan, P.} \&
  \bibinfo{author}{Kumar, V.}
\newblock \bibinfo{title}{Atomic and electronic structure of solids of
  {Ge}$_{\textrm{2}}${Br}$_{\textrm{2}}${PN},
  {Ge}$_{\textrm{2}}${I}$_{\textrm{2}}${PN},
  {Sn}$_{\textrm{2}}${Cl}$_{\textrm{2}}${PN},
  {Sn}$_{\textrm{2}}${Br}$_{\textrm{2}}${PN} and
  {Sn}$_{\textrm{2}}${I}$_{\textrm{2}}${PN} inorganic double helices: a first
  principles study}.
\newblock \emph{\bibinfo{journal}{RSC Advances}} \textbf{\bibinfo{volume}{10}},
  \bibinfo{pages}{14714--14719} (\bibinfo{year}{2020}).
\newblock \urlprefix\url{http://xlink.rsc.org/?DOI=D0RA02007A}.

\bibitem{ott_flexible_2019}
\bibinfo{author}{Ott, C.} \emph{et~al.}
\newblock \bibinfo{title}{Flexible and {Ultrasoft} {Inorganic} 1d
  {Semiconductor} and {Heterostructure} {Systems} {Based} on {SnIP}}.
\newblock \emph{\bibinfo{journal}{Advanced Functional Materials}}
  \textbf{\bibinfo{volume}{29}}, \bibinfo{pages}{1900233}
  (\bibinfo{year}{2019}).
\newblock
  \urlprefix\url{https://onlinelibrary.wiley.com/doi/abs/10.1002/adfm.201900233}.

\bibitem{xiang_one-dimensional_2020}
\bibinfo{author}{Xiang, R.} \emph{et~al.}
\newblock \bibinfo{title}{One-dimensional van der {Waals} heterostructures}.
\newblock \emph{\bibinfo{journal}{Science}} \textbf{\bibinfo{volume}{367}},
  \bibinfo{pages}{537--542} (\bibinfo{year}{2020}).
\newblock
  \urlprefix\url{https://www.sciencemag.org/lookup/doi/10.1126/science.aaz2570}.

\bibitem{qin_raman_2020}
\bibinfo{author}{Qin, J.-K.} \emph{et~al.}
\newblock \bibinfo{title}{Raman response and transport properties of tellurium
  atomic chains encapsulated in nanotubes}.
\newblock \emph{\bibinfo{journal}{Nature Electronics}}
  \textbf{\bibinfo{volume}{3}}, \bibinfo{pages}{141--147}
  (\bibinfo{year}{2020}).
\newblock \urlprefix\url{https://doi.org/10.1038/s41928-020-0365-4}.

\bibitem{burdanova_ultrafast_2020}
\bibinfo{author}{Burdanova, M.~G.} \emph{et~al.}
\newblock \bibinfo{title}{Ultrafast {Optoelectronic} {Processes} in {1D}
  {Radial} van der {Waals} {Heterostructures}: {Carbon}, {Boron} {Nitride}, and
  {MoS}$_{\textrm{2}}$ {Nanotubes} with {Coexisting} {Excitons} and {Highly}
  {Mobile} {Charges}}.
\newblock \emph{\bibinfo{journal}{Nano Letters}} \textbf{\bibinfo{volume}{20}},
  \bibinfo{pages}{3560--3567} (\bibinfo{year}{2020}).
\newblock
  \urlprefix\url{https://pubs.acs.org/doi/10.1021/acs.nanolett.0c00504}.

\bibitem{pielmeier_formation_2020}
\bibinfo{author}{Pielmeier, M. R.~P.} \& \bibinfo{author}{Nilges, T.}
\newblock \bibinfo{title}{Formation {Mechanisms} for {Phosphorene} and {SnIP}}.
\newblock \emph{\bibinfo{journal}{Angewandte Chemie International Edition}}
  \bibinfo{pages}{Accepted}.

\bibitem{uzer_vapor_2019}
\bibinfo{author}{Üzer, E.} \emph{et~al.}
\newblock \bibinfo{title}{Vapor growth of binary and ternary phosphorus-based
  semiconductors into {TiO}$_{\textrm{2}}$ nanotube arrays and application in
  visible light driven water splitting}.
\newblock \emph{\bibinfo{journal}{Nanoscale Advances}}
  \textbf{\bibinfo{volume}{1}}, \bibinfo{pages}{2881--2890}
  (\bibinfo{year}{2019}).
\newblock \urlprefix\url{http://xlink.rsc.org/?DOI=C9NA00084D}.

\bibitem{pielmeier_toward_2020}
\bibinfo{author}{Pielmeier, M. R.~P.}, \bibinfo{author}{Karttunen, A.~J.} \&
  \bibinfo{author}{Nilges, T.}
\newblock \bibinfo{title}{Toward {Atomic}-{Scale} {Inorganic} {Double}
  {Helices} via {Carbon} {Nanotube} {Matrices}—{Induction} of {Chirality} to
  {Carbon} {Nanotubes}}.
\newblock \emph{\bibinfo{journal}{The Journal of Physical Chemistry C}}
  \textbf{\bibinfo{volume}{124}}, \bibinfo{pages}{13338--13347}
  (\bibinfo{year}{2020}).
\newblock \urlprefix\url{https://pubs.acs.org/doi/10.1021/acs.jpcc.0c02079}.

\bibitem{jepsen2011}
\bibinfo{author}{Jepsen, P.}, \bibinfo{author}{Cooke, D.} \&
  \bibinfo{author}{Koch, M.}
\newblock \bibinfo{title}{Terahertz spectroscopy and imaging – {Modern}
  techniques and applications}.
\newblock \emph{\bibinfo{journal}{Laser \& Photonics Reviews}}
  \textbf{\bibinfo{volume}{5}}, \bibinfo{pages}{124--166}
  (\bibinfo{year}{2011}).
\newblock
  \urlprefix\url{http://onlinelibrary.wiley.com/doi/10.1002/lpor.201000011/abstract}.

\bibitem{koeberg_simultaneous_2007}
\bibinfo{author}{Koeberg, M.} \emph{et~al.}
\newblock \bibinfo{title}{Simultaneous ultrafast probing of intramolecular
  vibrations and photoinduced charge carriers in rubrene using broadband
  time-domain {THz} spectroscopy}.
\newblock \emph{\bibinfo{journal}{Physical Review B}}
  \textbf{\bibinfo{volume}{75}}, \bibinfo{pages}{195216}
  (\bibinfo{year}{2007}).
\newblock \urlprefix\url{https://link.aps.org/doi/10.1103/PhysRevB.75.195216}.

\bibitem{ulbricht_carrier_2011}
\bibinfo{author}{Ulbricht, R.}, \bibinfo{author}{Hendry, E.},
  \bibinfo{author}{Shan, J.}, \bibinfo{author}{Heinz, T.~F.} \&
  \bibinfo{author}{Bonn, M.}
\newblock \bibinfo{title}{Carrier dynamics in semiconductors studied with
  time-resolved terahertz spectroscopy}.
\newblock \emph{\bibinfo{journal}{Reviews of Modern Physics}}
  \textbf{\bibinfo{volume}{83}}, \bibinfo{pages}{543--586}
  (\bibinfo{year}{2011}).
\newblock \urlprefix\url{https://link.aps.org/doi/10.1103/RevModPhys.83.543}.

\bibitem{butler_ultrafast_2016}
\bibinfo{author}{Butler, K.~T.} \emph{et~al.}
\newblock \bibinfo{title}{Ultrafast carrier dynamics in {BiVO}$_4$ thin film
  photoanode material: interplay between free carriers, trapped carriers and
  low-frequency lattice vibrations}.
\newblock \emph{\bibinfo{journal}{Journal of Materials Chemistry A}}
  \textbf{\bibinfo{volume}{4}}, \bibinfo{pages}{18516--18523}
  (\bibinfo{year}{2016}).
\newblock \urlprefix\url{http://xlink.rsc.org/?DOI=C6TA07177E}.

\bibitem{zhao_monitoring_2019}
\bibinfo{author}{Zhao, D.} \emph{et~al.}
\newblock \bibinfo{title}{Monitoring {Electron}–{Phonon} {Interactions} in
  {Lead} {Halide} {Perovskites} {Using} {Time}-{Resolved} {THz}
  {Spectroscopy}}.
\newblock \emph{\bibinfo{journal}{ACS Nano}} \textbf{\bibinfo{volume}{13}},
  \bibinfo{pages}{8826--8835} (\bibinfo{year}{2019}).
\newblock \urlprefix\url{https://pubs.acs.org/doi/10.1021/acsnano.9b02049}.

\bibitem{parkinson_carrier_2009}
\bibinfo{author}{Parkinson, P.} \emph{et~al.}
\newblock \bibinfo{title}{Carrier {Lifetime} and {Mobility} {Enhancement} in
  {Nearly} {Defect}-{Free} {Core}-{Shell} {Nanowires} {Measured} {Using}
  {Time}-{Resolved} {Terahertz} {Spectroscopy}}.
\newblock \emph{\bibinfo{journal}{Nano Letters}} \textbf{\bibinfo{volume}{9}},
  \bibinfo{pages}{3349--3353} (\bibinfo{year}{2009}).
\newblock \urlprefix\url{https://doi.org/10.1021/nl9016336}.

\bibitem{ozawa_correlation_2018}
\bibinfo{author}{Ozawa, K.} \emph{et~al.}
\newblock \bibinfo{title}{Correlation between {Photocatalytic} {Activity} and
  {Carrier} {Lifetime}: {Acetic} {Acid} on {Single}-{Crystal} {Surfaces} of
  {Anatase} and {Rutile} {TiO}$_{\textrm{2}}$}.
\newblock \emph{\bibinfo{journal}{The Journal of Physical Chemistry C}}
  \textbf{\bibinfo{volume}{122}}, \bibinfo{pages}{9562--9569}
  (\bibinfo{year}{2018}).
\newblock \urlprefix\url{https://pubs.acs.org/doi/10.1021/acs.jpcc.8b02259}.

\bibitem{schubert_phonon_2019}
\bibinfo{author}{Schubert, M.}, \bibinfo{author}{Mock, A.},
  \bibinfo{author}{Korlacki, R.} \& \bibinfo{author}{Darakchieva, V.}
\newblock \bibinfo{title}{Phonon order and reststrahlen bands of polar
  vibrations in crystals with monoclinic symmetry}.
\newblock \emph{\bibinfo{journal}{Physical Review B}}
  \textbf{\bibinfo{volume}{99}}, \bibinfo{pages}{041201}
  (\bibinfo{year}{2019}).
\newblock \urlprefix\url{https://link.aps.org/doi/10.1103/PhysRevB.99.041201}.

\bibitem{joyce_electronic_2013}
\bibinfo{author}{Joyce, H.~J.} \emph{et~al.}
\newblock \bibinfo{title}{Electronic properties of {GaAs}, {InAs} and {InP}
  nanowires studied by terahertz spectroscopy}.
\newblock \emph{\bibinfo{journal}{Nanotechnology}}
  \textbf{\bibinfo{volume}{24}}, \bibinfo{pages}{214006}
  (\bibinfo{year}{2013}).
\newblock
  \urlprefix\url{http://stacks.iop.org/0957-4484/24/i=21/a=214006?key=crossref.3ddd730e84f137f4d4de09739ab00ec8}.

\bibitem{fox2001optical}
\bibinfo{author}{Fox, A.}
\newblock \emph{\bibinfo{title}{Optical Properties of Solids}}.
\newblock Oxford master series in condensed matter physics
  (\bibinfo{publisher}{Oxford University Press}, \bibinfo{year}{2001}).
\newblock \urlprefix\url{https://books.google.ca/books?id=-5bVBbAoaGoC}.

\bibitem{pathirane_hybrid_2015}
\bibinfo{author}{Pathirane, M.} \emph{et~al.}
\newblock \bibinfo{title}{Hybrid {ZnO} nanowire/a-{Si}:{H} thin-film radial
  junction solar cells using nanoparticle front contacts}.
\newblock \emph{\bibinfo{journal}{Applied Physics Letters}}
  \textbf{\bibinfo{volume}{107}}, \bibinfo{pages}{143903}
  (\bibinfo{year}{2015}).
\newblock \urlprefix\url{http://aip.scitation.org/doi/10.1063/1.4932649}.

\bibitem{ziwritsch_direct_2016}
\bibinfo{author}{Ziwritsch, M.} \emph{et~al.}
\newblock \bibinfo{title}{Direct {Time}-{Resolved} {Observation} of {Carrier}
  {Trapping} and {Polaron} {Conductivity} in {BiVO}$_4$}.
\newblock \emph{\bibinfo{journal}{ACS Energy Letters}}
  \textbf{\bibinfo{volume}{1}}, \bibinfo{pages}{888--894}
  (\bibinfo{year}{2016}).
\newblock
  \urlprefix\url{https://pubs.acs.org/doi/10.1021/acsenergylett.6b00423}.

\bibitem{yang_time-resolved_2018}
\bibinfo{author}{Yang, W.} \emph{et~al.}
\newblock \bibinfo{title}{Time-{Resolved} {Observations} of {Photo}-{Generated}
  {Charge}-{Carrier} {Dynamics} in {Sb}$_2${Se}$_3$ {Photocathodes} for
  {Photoelectrochemical} {Water} {Splitting}}.
\newblock \emph{\bibinfo{journal}{ACS Nano}} \textbf{\bibinfo{volume}{12}},
  \bibinfo{pages}{11088--11097} (\bibinfo{year}{2018}).
\newblock \urlprefix\url{http://pubs.acs.org/doi/10.1021/acsnano.8b05446}.

\bibitem{cinquanta_ultrafast_2019}
\bibinfo{author}{Cinquanta, E.} \emph{et~al.}
\newblock \bibinfo{title}{Ultrafast {THz} {Probe} of {Photoinduced} {Polarons}
  in {Lead}-{Halide} {Perovskites}}.
\newblock \emph{\bibinfo{journal}{Physical Review Letters}}
  \textbf{\bibinfo{volume}{122}}, \bibinfo{pages}{166601}
  (\bibinfo{year}{2019}).
\newblock
  \urlprefix\url{https://link.aps.org/doi/10.1103/PhysRevLett.122.166601}.

\bibitem{yu2010fundamentals}
\bibinfo{author}{Yu, P.} \& \bibinfo{author}{Cardona, M.}
\newblock \emph{\bibinfo{title}{Fundamentals of Semiconductors: Physics and
  Materials Properties}}.
\newblock Graduate Texts in Physics (\bibinfo{publisher}{Springer Berlin
  Heidelberg}, \bibinfo{year}{2010}).
\newblock \urlprefix\url{https://books.google.ca/books?id=5aBuKYBT\_hsC}.

\bibitem{sim_ultrafast_2014}
\bibinfo{author}{Sim, S.} \emph{et~al.}
\newblock \bibinfo{title}{Ultrafast terahertz dynamics of hot {Dirac}-electron
  surface scattering in the topological insulator {Bi}$_2${Se}$_3$}.
\newblock \emph{\bibinfo{journal}{Physical Review B}}
  \textbf{\bibinfo{volume}{89}}, \bibinfo{pages}{165137}
  (\bibinfo{year}{2014}).
\newblock \urlprefix\url{https://link.aps.org/doi/10.1103/PhysRevB.89.165137}.

\bibitem{joyce_review_2016}
\bibinfo{author}{Joyce, H.~J.}, \bibinfo{author}{Boland, J.~L.},
  \bibinfo{author}{Davies, C.~L.}, \bibinfo{author}{Baig, S.~A.} \&
  \bibinfo{author}{Johnston, M.~B.}
\newblock \bibinfo{title}{A review of the electrical properties of
  semiconductor nanowires: insights gained from terahertz conductivity
  spectroscopy}.
\newblock \emph{\bibinfo{journal}{Semiconductor Science and Technology}}
  \textbf{\bibinfo{volume}{31}}, \bibinfo{pages}{103003}
  (\bibinfo{year}{2016}).
\newblock
  \urlprefix\url{http://stacks.iop.org/0268-1242/31/i=10/a=103003?key=crossref.0ff1aa7713a8c6078d6a89906e0c9d3d}.

\bibitem{kuzel_terahertz_2020}
\bibinfo{author}{Kužel, P.} \& \bibinfo{author}{Němec, H.}
\newblock \bibinfo{title}{Terahertz {Spectroscopy} of {Nanomaterials}: a
  {Close} {Look} at {Charge}‐{Carrier} {Transport}}.
\newblock \emph{\bibinfo{journal}{Advanced Optical Materials}}
  \textbf{\bibinfo{volume}{8}}, \bibinfo{pages}{1900623}
  (\bibinfo{year}{2020}).
\newblock
  \urlprefix\url{https://onlinelibrary.wiley.com/doi/abs/10.1002/adom.201900623}.

\bibitem{baxter_conductivity_2006}
\bibinfo{author}{Baxter, J.~B.} \& \bibinfo{author}{Schmuttenmaer, C.~A.}
\newblock \bibinfo{title}{Conductivity of {ZnO} {Nanowires}, {Nanoparticles},
  and {Thin} {Films} {Using} {Time}-{Resolved} {Terahertz} {Spectroscopy}
  $^{\textrm{†}}$}.
\newblock \emph{\bibinfo{journal}{The Journal of Physical Chemistry B}}
  \textbf{\bibinfo{volume}{110}}, \bibinfo{pages}{25229--25239}
  (\bibinfo{year}{2006}).
\newblock \urlprefix\url{https://pubs.acs.org/doi/10.1021/jp064399a}.

\bibitem{walther_terahertz_2007}
\bibinfo{author}{Walther, M.} \emph{et~al.}
\newblock \bibinfo{title}{Terahertz conductivity of thin gold films at the
  metal-insulator percolation transition}.
\newblock \emph{\bibinfo{journal}{Physical Review B}}
  \textbf{\bibinfo{volume}{76}}, \bibinfo{pages}{125408}
  (\bibinfo{year}{2007}).
\newblock \urlprefix\url{https://link.aps.org/doi/10.1103/PhysRevB.76.125408}.

\bibitem{boland_high_2018}
\bibinfo{author}{Boland, J.~L.} \emph{et~al.}
\newblock \bibinfo{title}{High {Electron} {Mobility} and {Insights} into
  {Temperature}-{Dependent} {Scattering} {Mechanisms} in {InAsSb} {Nanowires}}.
\newblock \emph{\bibinfo{journal}{Nano Letters}} \textbf{\bibinfo{volume}{18}},
  \bibinfo{pages}{3703--3710} (\bibinfo{year}{2018}).
\newblock
  \urlprefix\url{https://pubs.acs.org/doi/10.1021/acs.nanolett.8b00842}.

\bibitem{poellmann_resonant_2015}
\bibinfo{author}{Poellmann, C.} \emph{et~al.}
\newblock \bibinfo{title}{Resonant internal quantum transitions and femtosecond
  radiative decay of excitons in monolayer {WSe2}}.
\newblock \emph{\bibinfo{journal}{Nature Materials}}
  \textbf{\bibinfo{volume}{14}}, \bibinfo{pages}{889--893}
  (\bibinfo{year}{2015}).
\newblock \urlprefix\url{http://www.nature.com/articles/nmat4356}.

\bibitem{luo_ultrafast_2017}
\bibinfo{author}{Luo, L.} \emph{et~al.}
\newblock \bibinfo{title}{Ultrafast terahertz snapshots of excitonic {Rydberg}
  states and electronic coherence in an organometal halide perovskite}.
\newblock \emph{\bibinfo{journal}{Nature Communications}}
  \textbf{\bibinfo{volume}{8}}, \bibinfo{pages}{15565} (\bibinfo{year}{2017}).
\newblock \urlprefix\url{http://www.nature.com/articles/ncomms15565}.

\bibitem{smith_classical_2001}
\bibinfo{author}{Smith, N.}
\newblock \bibinfo{title}{Classical generalization of the {Drude} formula for
  the optical conductivity}.
\newblock \emph{\bibinfo{journal}{Physical Review B}}
  \textbf{\bibinfo{volume}{64}}, \bibinfo{pages}{155106}
  (\bibinfo{year}{2001}).
\newblock \urlprefix\url{https://link.aps.org/doi/10.1103/PhysRevB.64.155106}.

\bibitem{nemec_far-infrared_2009}
\bibinfo{author}{Němec, H.}, \bibinfo{author}{Kužel, P.} \&
  \bibinfo{author}{Sundström, V.}
\newblock \bibinfo{title}{Far-infrared response of free charge carriers
  localized in semiconductor nanoparticles}.
\newblock \emph{\bibinfo{journal}{Physical Review B}}
  \textbf{\bibinfo{volume}{79}}, \bibinfo{pages}{115309}
  (\bibinfo{year}{2009}).
\newblock \urlprefix\url{https://link.aps.org/doi/10.1103/PhysRevB.79.115309}.

\bibitem{cocker_microscopic_2017}
\bibinfo{author}{Cocker, T.~L.} \emph{et~al.}
\newblock \bibinfo{title}{Microscopic origin of the {Drude}-{Smith} model}.
\newblock \emph{\bibinfo{journal}{Physical Review B}}
  \textbf{\bibinfo{volume}{96}}, \bibinfo{pages}{205439}
  (\bibinfo{year}{2017}).
\newblock \urlprefix\url{https://link.aps.org/doi/10.1103/PhysRevB.96.205439}.

\bibitem{cooke_ultrabroadband_2012}
\bibinfo{author}{Cooke, D.~G.}, \bibinfo{author}{Meldrum, A.} \&
  \bibinfo{author}{Uhd~Jepsen, P.}
\newblock \bibinfo{title}{Ultrabroadband terahertz conductivity of {Si}
  nanocrystal films}.
\newblock \emph{\bibinfo{journal}{Applied Physics Letters}}
  \textbf{\bibinfo{volume}{101}}, \bibinfo{pages}{211107}
  (\bibinfo{year}{2012}).
\newblock \urlprefix\url{http://aip.scitation.org/doi/10.1063/1.4767145}.

\bibitem{jensen_ultrafast_2013}
\bibinfo{author}{Jensen, S.~A.} \emph{et~al.}
\newblock \bibinfo{title}{Ultrafast {Photoconductivity} of {Graphene}
  {Nanoribbons} and {Carbon} {Nanotubes}}.
\newblock \emph{\bibinfo{journal}{Nano Letters}} \textbf{\bibinfo{volume}{13}},
  \bibinfo{pages}{5925--5930} (\bibinfo{year}{2013}).
\newblock \urlprefix\url{https://pubs.acs.org/doi/10.1021/nl402978s}.

\bibitem{laforge_conductivity_2014}
\bibinfo{author}{LaForge, J.~M.} \emph{et~al.}
\newblock \bibinfo{title}{Conductivity control of as-grown branched indium tin
  oxide nanowire networks}.
\newblock \emph{\bibinfo{journal}{Nanotechnology}}
  \textbf{\bibinfo{volume}{25}}, \bibinfo{pages}{035701}
  (\bibinfo{year}{2014}).
\newblock
  \urlprefix\url{http://stacks.iop.org/0957-4484/25/i=3/a=035701?key=crossref.4cfef47e75d4a3590da4f0b773250fb9}.

\bibitem{evers_high_2015}
\bibinfo{author}{Evers, W.~H.} \emph{et~al.}
\newblock \bibinfo{title}{High charge mobility in two-dimensional percolative
  networks of {PbSe} quantum dots connected by atomic bonds}.
\newblock \emph{\bibinfo{journal}{Nature Communications}}
  \textbf{\bibinfo{volume}{6}}, \bibinfo{pages}{8195} (\bibinfo{year}{2015}).
\newblock \urlprefix\url{http://www.nature.com/articles/ncomms9195}.

\bibitem{liu_ultrahigh_2016}
\bibinfo{author}{Liu, H.} \emph{et~al.}
\newblock \bibinfo{title}{Ultrahigh photoconductivity of bandgap-graded
  {CdS$_x$Se$_{1-x}$} nanowires probed by terahertz spectroscopy}.
\newblock \emph{\bibinfo{journal}{Scientific Reports}}
  \textbf{\bibinfo{volume}{6}}, \bibinfo{pages}{27387} (\bibinfo{year}{2016}).
\newblock \urlprefix\url{http://www.nature.com/articles/srep27387}.

\bibitem{titova_ultrafast_2016}
\bibinfo{author}{Titova, L.~V.} \emph{et~al.}
\newblock \bibinfo{title}{Ultrafast carrier dynamics and the role of grain
  boundaries in polycrystalline silicon thin films grown by molecular beam
  epitaxy}.
\newblock \emph{\bibinfo{journal}{Semiconductor Science and Technology}}
  \textbf{\bibinfo{volume}{31}}, \bibinfo{pages}{105017}
  (\bibinfo{year}{2016}).
\newblock
  \urlprefix\url{http://stacks.iop.org/0268-1242/31/i=10/a=105017?key=crossref.3e6510dd599dafba86e337f107925ba7}.

\bibitem{li_dynamical_2020}
\bibinfo{author}{Li, G.} \emph{et~al.}
\newblock \bibinfo{title}{Dynamical {Control} over {Terahertz}
  {Electromagnetic} {Interference} {Shielding} with {2D}
  {Ti}$_{\textrm{3}}${C}$_{\textrm{2}}${T}$_{\textrm{\textit{y} }}$ {MXene} by
  {Ultrafast} {Optical} {Pulses}}.
\newblock \emph{\bibinfo{journal}{Nano Letters}} \textbf{\bibinfo{volume}{20}},
  \bibinfo{pages}{636--643} (\bibinfo{year}{2020}).
\newblock
  \urlprefix\url{https://pubs.acs.org/doi/10.1021/acs.nanolett.9b04404}.

\bibitem{zhou_giant_2016}
\bibinfo{author}{Zhou, Y.} \emph{et~al.}
\newblock \bibinfo{title}{Giant photostriction in organic–inorganic lead
  halide perovskites}.
\newblock \emph{\bibinfo{journal}{Nature Communications}}
  \textbf{\bibinfo{volume}{7}}, \bibinfo{pages}{11193} (\bibinfo{year}{2016}).
\newblock \urlprefix\url{http://www.nature.com/articles/ncomms11193}.

\bibitem{zheng_ultrafast_2020}
\bibinfo{author}{Zheng, D.} \emph{et~al.}
\newblock \bibinfo{title}{Ultrafast lattice and electronic dynamics in
  single-walled carbon nanotubes}.
\newblock \emph{\bibinfo{journal}{Nanoscale Advances}}
  \textbf{\bibinfo{volume}{2}}, \bibinfo{pages}{2808--2813}
  (\bibinfo{year}{2020}).
\newblock \urlprefix\url{http://xlink.rsc.org/?DOI=D0NA00269K}.

\bibitem{neumann_raman_2015}
\bibinfo{author}{Neumann, C.} \emph{et~al.}
\newblock \bibinfo{title}{Raman spectroscopy as probe of nanometre-scale strain
  variations in graphene}.
\newblock \emph{\bibinfo{journal}{Nature Communications}}
  \textbf{\bibinfo{volume}{6}}, \bibinfo{pages}{8429} (\bibinfo{year}{2015}).
\newblock \urlprefix\url{http://www.nature.com/articles/ncomms9429}.

\bibitem{marcondes_importance_2018}
\bibinfo{author}{Marcondes, M.~L.}, \bibinfo{author}{Wentzcovitch, R.~M.} \&
  \bibinfo{author}{Assali, L.~V.}
\newblock \bibinfo{title}{Importance of van der {Waals} interaction on
  structural, vibrational, and thermodynamic properties of {NaCl}}.
\newblock \emph{\bibinfo{journal}{Solid State Communications}}
  \textbf{\bibinfo{volume}{273}}, \bibinfo{pages}{11--16}
  (\bibinfo{year}{2018}).
\newblock
  \urlprefix\url{https://linkinghub.elsevier.com/retrieve/pii/S0038109818300140}.

\bibitem{li_faraday_2018}
\bibinfo{author}{Li, M.}, \bibinfo{author}{Reimers, J.~R.},
  \bibinfo{author}{Dobson, J.~F.} \& \bibinfo{author}{Gould, T.}
\newblock \bibinfo{title}{Faraday cage screening reveals intrinsic aspects of
  the van der {Waals} attraction}.
\newblock \emph{\bibinfo{journal}{Proceedings of the National Academy of
  Sciences}} \textbf{\bibinfo{volume}{115}}, \bibinfo{pages}{E10295--E10302}
  (\bibinfo{year}{2018}).
\newblock
  \urlprefix\url{http://www.pnas.org/lookup/doi/10.1073/pnas.1811569115}.

\bibitem{beard_transient_2000}
\bibinfo{author}{Beard, M.~C.}, \bibinfo{author}{Turner, G.~M.} \&
  \bibinfo{author}{Schmuttenmaer, C.~A.}
\newblock \bibinfo{title}{Transient photoconductivity in {GaAs} as measured by
  time-resolved terahertz spectroscopy}.
\newblock \emph{\bibinfo{journal}{Physical Review B}}
  \textbf{\bibinfo{volume}{62}}, \bibinfo{pages}{15764--15777}
  (\bibinfo{year}{2000}).
\newblock \urlprefix\url{https://link.aps.org/doi/10.1103/PhysRevB.62.15764}.

\bibitem{larsen_finite-difference_2011}
\bibinfo{author}{Larsen, C.}, \bibinfo{author}{Cooke, D.~G.} \&
  \bibinfo{author}{Jepsen, P.~U.}
\newblock \bibinfo{title}{Finite-difference time-domain analysis of
  time-resolved terahertz spectroscopy experiments}.
\newblock \emph{\bibinfo{journal}{Journal of the Optical Society of America B}}
  \textbf{\bibinfo{volume}{28}}, \bibinfo{pages}{1308} (\bibinfo{year}{2011}).
\newblock
  \urlprefix\url{https://www.osapublishing.org/abstract.cfm?URI=josab-28-5-1308}.

\bibitem{uhd_jepsen_ultrafast_2001}
\bibinfo{author}{Uhd~Jepsen, P.} \emph{et~al.}
\newblock \bibinfo{title}{Ultrafast carrier trapping in microcrystalline
  silicon observed in optical pump–terahertz probe measurements}.
\newblock \emph{\bibinfo{journal}{Applied Physics Letters}}
  \textbf{\bibinfo{volume}{79}}, \bibinfo{pages}{1291--1293}
  (\bibinfo{year}{2001}).
\newblock \urlprefix\url{http://aip.scitation.org/doi/10.1063/1.1394953}.

\bibitem{sahota_many-body_2019}
\bibinfo{author}{Sahota, D.~G.} \emph{et~al.}
\newblock \bibinfo{title}{Many-body recombination in photoexcited insulating
  cuprates}.
\newblock \emph{\bibinfo{journal}{Physical Review Research}}
  \textbf{\bibinfo{volume}{1}}, \bibinfo{pages}{033214} (\bibinfo{year}{2019}).
\newblock
  \urlprefix\url{https://link.aps.org/doi/10.1103/PhysRevResearch.1.033214}.

\bibitem{yu_analyzing_2017}
\bibinfo{author}{Yu, Z.} \emph{et~al.}
\newblock \bibinfo{title}{Analyzing the {Carrier} {Mobility} in
  {Transition}-{Metal} {Dichalcogenide} {MoS} $_{\textrm{2}}$ {Field}-{Effect}
  {Transistors}}.
\newblock \emph{\bibinfo{journal}{Advanced Functional Materials}}
  \textbf{\bibinfo{volume}{27}}, \bibinfo{pages}{1604093}
  (\bibinfo{year}{2017}).
\newblock \urlprefix\url{http://doi.wiley.com/10.1002/adfm.201604093}.

\bibitem{gao_high_2015}
\bibinfo{author}{Gao, X.} \& \bibinfo{author}{Zhao, Z.}
\newblock \bibinfo{title}{High mobility organic semiconductors for field-effect
  transistors}.
\newblock \emph{\bibinfo{journal}{Science China Chemistry}}
  \textbf{\bibinfo{volume}{58}}, \bibinfo{pages}{947--968}
  (\bibinfo{year}{2015}).
\newblock \urlprefix\url{http://link.springer.com/10.1007/s11426-015-5399-5}.

\bibitem{huber_femtosecond_2005}
\bibinfo{author}{Huber, R.} \emph{et~al.}
\newblock \bibinfo{title}{Femtosecond {Formation} of {Coupled}
  {Phonon}-{Plasmon} {Modes} in {InP}: {Ultrabroadband} {THz} {Experiment} and
  {Quantum} {Kinetic} {Theory}}.
\newblock \emph{\bibinfo{journal}{Physical Review Letters}}
  \textbf{\bibinfo{volume}{94}}, \bibinfo{pages}{027401}
  (\bibinfo{year}{2005}).
\newblock
  \urlprefix\url{https://link.aps.org/doi/10.1103/PhysRevLett.94.027401}.

\bibitem{kubler_coherent_2007-1}
\bibinfo{author}{Kübler, C.} \emph{et~al.}
\newblock \bibinfo{title}{Coherent {Structural} {Dynamics} and {Electronic}
  {Correlations} during an {Ultrafast} {Insulator}-to-{Metal} {Phase}
  {Transition} in {VO} 2}.
\newblock \emph{\bibinfo{journal}{Physical Review Letters}}
  \textbf{\bibinfo{volume}{99}}, \bibinfo{pages}{116401}
  (\bibinfo{year}{2007}).
\newblock
  \urlprefix\url{https://link.aps.org/doi/10.1103/PhysRevLett.99.116401}.

\bibitem{gaal_internal_2007}
\bibinfo{author}{Gaal, P.} \emph{et~al.}
\newblock \bibinfo{title}{Internal motions of a quasiparticle governing its
  ultrafast nonlinear response}.
\newblock \emph{\bibinfo{journal}{Nature}} \textbf{\bibinfo{volume}{450}},
  \bibinfo{pages}{1210--1213} (\bibinfo{year}{2007}).
\newblock \urlprefix\url{http://www.nature.com/articles/nature06399}.

\bibitem{lan_ultrafast_2019}
\bibinfo{author}{Lan, Y.} \emph{et~al.}
\newblock \bibinfo{title}{Ultrafast correlated charge and lattice motion in a
  hybrid metal halide perovskite}.
\newblock \emph{\bibinfo{journal}{Science Advances}}
  \textbf{\bibinfo{volume}{5}}, \bibinfo{pages}{eaaw5558}
  (\bibinfo{year}{2019}).
\newblock
  \urlprefix\url{http://advances.sciencemag.org/lookup/doi/10.1126/sciadv.aaw5558}.

\bibitem{varga_coupling_1965}
\bibinfo{author}{Varga, B.~B.}
\newblock \bibinfo{title}{Coupling of {Plasmons} to {Polar} {Phonons} in
  {Degenerate} {Semiconductors}}.
\newblock \emph{\bibinfo{journal}{Physical Review}}
  \textbf{\bibinfo{volume}{137}}, \bibinfo{pages}{A1896--A1902}
  (\bibinfo{year}{1965}).
\newblock \urlprefix\url{https://link.aps.org/doi/10.1103/PhysRev.137.A1896}.

\bibitem{turchinovich_femtosecond_2017}
\bibinfo{author}{Turchinovich, D.}, \bibinfo{author}{D'Angelo, F.} \&
  \bibinfo{author}{Bonn, M.}
\newblock \bibinfo{title}{Femtosecond-timescale buildup of electron mobility in
  {GaAs} observed via ultrabroadband transient terahertz spectroscopy}.
\newblock \emph{\bibinfo{journal}{Applied Physics Letters}}
  \textbf{\bibinfo{volume}{110}}, \bibinfo{pages}{121102}
  (\bibinfo{year}{2017}).
\newblock \urlprefix\url{http://aip.scitation.org/doi/10.1063/1.4978648}.

\bibitem{lan_coherent_2019}
\bibinfo{author}{Lan, Y.} \emph{et~al.}
\newblock \bibinfo{title}{Coherent charge-phonon correlations and exciton
  dynamics in orthorhombic
  {CH}$_{\textrm{3}}${NH}$_{\textrm{3}}${PbI}$_{\textrm{3}}$ measured by
  ultrafast multi-{THz} spectroscopy}.
\newblock \emph{\bibinfo{journal}{The Journal of Chemical Physics}}
  \textbf{\bibinfo{volume}{151}}, \bibinfo{pages}{214201}
  (\bibinfo{year}{2019}).
\newblock \urlprefix\url{http://aip.scitation.org/doi/10.1063/1.5127992}.

\bibitem{chang_electrical_2012}
\bibinfo{author}{Chang, C.-C.} \emph{et~al.}
\newblock \bibinfo{title}{Electrical and {Optical} {Characterization} of
  {Surface} {Passivation} in {GaAs} {Nanowires}}.
\newblock \emph{\bibinfo{journal}{Nano Letters}} \textbf{\bibinfo{volume}{12}},
  \bibinfo{pages}{4484--4489} (\bibinfo{year}{2012}).
\newblock \urlprefix\url{https://pubs.acs.org/doi/10.1021/nl301391h}.

\bibitem{shatruk_first_1999}
\bibinfo{author}{Shatruk, M.~M.}, \bibinfo{author}{Kovnir, K.~A.},
  \bibinfo{author}{Shevelkov, A.~V.}, \bibinfo{author}{Presniakov, I.~A.} \&
  \bibinfo{author}{Popovkin, B.~A.}
\newblock \bibinfo{title}{First {Tin} {Pnictide} {Halides}
  {Sn}$_{\textrm{24}}${P}$_{\textrm{19.3}}${I}$_{\textrm{8}}$ and
  {Sn}$_{\textrm{24}}${As}$_{\textrm{19.3}}${I} $_{\textrm{8}}$: {Synthesis}
  and the {Clathrate}-{I} {Type} of the {Crystal} {Structure}}.
\newblock \emph{\bibinfo{journal}{Inorganic Chemistry}}
  \textbf{\bibinfo{volume}{38}}, \bibinfo{pages}{3455--3457}
  (\bibinfo{year}{1999}).
\newblock \urlprefix\url{https://pubs.acs.org/doi/10.1021/ic990153r}.

\bibitem{iwaszczuk_simultaneous_2009}
\bibinfo{author}{Iwaszczuk, K.}, \bibinfo{author}{Cooke, D.~G.},
  \bibinfo{author}{Fujiwara, M.}, \bibinfo{author}{Hashimoto, H.} \&
  \bibinfo{author}{Uhd~Jepsen, P.}
\newblock \bibinfo{title}{Simultaneous reference and differential waveform
  acquisition in time-resolved terahertz spectroscopy}.
\newblock \emph{\bibinfo{journal}{Optics Express}}
  \textbf{\bibinfo{volume}{17}}, \bibinfo{pages}{21969} (\bibinfo{year}{2009}).
\newblock
  \urlprefix\url{https://www.osapublishing.org/oe/abstract.cfm?uri=oe-17-24-21969}.

\end{thebibliography}


\begin{thebibliography}{10}
\expandafter\ifx\csname url\endcsname\relax
  \def\url#1{\texttt{#1}}\fi
\expandafter\ifx\csname urlprefix\endcsname\relax\def\urlprefix{URL }\fi
\providecommand{\bibinfo}[2]{#2}
\providecommand{\eprint}[2][]{\url{#2}}

\bibitem{Zhang_Experimental_2006}
\bibinfo{author}{Zhang, W.}
\newblock \emph{\bibinfo{title}{Experimental and Computational Analysis of
  Random Cylinder Packings with Applications}}.
\newblock Ph.D. thesis, \bibinfo{school}{Louisiana State University}
  (\bibinfo{year}{2006}).
\newblock \bibinfo{note}{AAI3245014}.

\bibitem{zicovich-wilson_calculation_2004}
\bibinfo{author}{Zicovich-Wilson, C.~M.} \emph{et~al.}
\newblock \bibinfo{title}{Calculation of the vibration frequencies of ?-quartz:
  {The} effect of {Hamiltonian} and basis set}.
\newblock \emph{\bibinfo{journal}{Journal of Computational Chemistry}}
  \textbf{\bibinfo{volume}{25}}, \bibinfo{pages}{1873--1881}
  (\bibinfo{year}{2004}).
\newblock \urlprefix\url{http://doi.wiley.com/10.1002/jcc.20120}.

\bibitem{ferrero_coupled_2008}
\bibinfo{author}{Ferrero, M.}, \bibinfo{author}{Rérat, M.},
  \bibinfo{author}{Orlando, R.} \& \bibinfo{author}{Dovesi, R.}
\newblock \bibinfo{title}{Coupled perturbed {Hartree}-{Fock} for periodic
  systems: {The} role of symmetry and related computational aspects}.
\newblock \emph{\bibinfo{journal}{The Journal of Chemical Physics}}
  \textbf{\bibinfo{volume}{128}}, \bibinfo{pages}{014110}
  (\bibinfo{year}{2008}).
\newblock \urlprefix\url{http://aip.scitation.org/doi/10.1063/1.2817596}.

\bibitem{erba_accurate_2013}
\bibinfo{author}{Erba, A.}, \bibinfo{author}{Ferrabone, M.},
  \bibinfo{author}{Orlando, R.} \& \bibinfo{author}{Dovesi, R.}
\newblock \bibinfo{title}{Accurate dynamical structure factors from \textit{ab
  initio} lattice dynamics: {The} case of crystalline silicon}.
\newblock \emph{\bibinfo{journal}{Journal of Computational Chemistry}}
  \textbf{\bibinfo{volume}{34}}, \bibinfo{pages}{346--354}
  (\bibinfo{year}{2013}).
\newblock \urlprefix\url{http://doi.wiley.com/10.1002/jcc.23138}.

\bibitem{bergren_ultrafast_2014}
\bibinfo{author}{Bergren, M.~R.} \emph{et~al.}
\newblock \bibinfo{title}{Ultrafast {Electrical} {Measurements} of {Isolated}
  {Silicon} {Nanowires} and {Nanocrystals}}.
\newblock \emph{\bibinfo{journal}{The Journal of Physical Chemistry Letters}}
  \textbf{\bibinfo{volume}{5}}, \bibinfo{pages}{2050--2057}
  (\bibinfo{year}{2014}).
\newblock \urlprefix\url{https://pubs.acs.org/doi/10.1021/jz500863a}.

\bibitem{cocker_microscopic_2017}
\bibinfo{author}{Cocker, T.~L.} \emph{et~al.}
\newblock \bibinfo{title}{Microscopic origin of the {Drude}-{Smith} model}.
\newblock \emph{\bibinfo{journal}{Physical Review B}}
  \textbf{\bibinfo{volume}{96}}, \bibinfo{pages}{205439}
  (\bibinfo{year}{2017}).
\newblock \urlprefix\url{https://link.aps.org/doi/10.1103/PhysRevB.96.205439}.

\bibitem{laforge_conductivity_2014}
\bibinfo{author}{LaForge, J.~M.} \emph{et~al.}
\newblock \bibinfo{title}{Conductivity control of as-grown branched indium tin
  oxide nanowire networks}.
\newblock \emph{\bibinfo{journal}{Nanotechnology}}
  \textbf{\bibinfo{volume}{25}}, \bibinfo{pages}{035701}
  (\bibinfo{year}{2014}).
\newblock
  \urlprefix\url{http://stacks.iop.org/0957-4484/25/i=3/a=035701?key=crossref.4cfef47e75d4a3590da4f0b773250fb9}.

\bibitem{titova_ultrafast_2016}
\bibinfo{author}{Titova, L.~V.} \emph{et~al.}
\newblock \bibinfo{title}{Ultrafast carrier dynamics and the role of grain
  boundaries in polycrystalline silicon thin films grown by molecular beam
  epitaxy}.
\newblock \emph{\bibinfo{journal}{Semiconductor Science and Technology}}
  \textbf{\bibinfo{volume}{31}}, \bibinfo{pages}{105017}
  (\bibinfo{year}{2016}).
\newblock
  \urlprefix\url{http://stacks.iop.org/0268-1242/31/i=10/a=105017?key=crossref.3e6510dd599dafba86e337f107925ba7}.

\bibitem{ostroverkhova_ultrafast_2006}
\bibinfo{author}{Ostroverkhova, O.} \emph{et~al.}
\newblock \bibinfo{title}{Ultrafast carrier dynamics in pentacene,
  functionalized pentacene, tetracene, and rubrene single crystals}.
\newblock \emph{\bibinfo{journal}{Applied Physics Letters}}
  \textbf{\bibinfo{volume}{88}}, \bibinfo{pages}{162101}
  (\bibinfo{year}{2006}).
\newblock \urlprefix\url{http://aip.scitation.org/doi/10.1063/1.2193801}.

\bibitem{de_filippis_crossover_2015}
\bibinfo{author}{De~Filippis, G.} \emph{et~al.}
\newblock \bibinfo{title}{Crossover from {Super}- to {Subdiffusive} {Motion}
  and {Memory} {Effects} in {Crystalline} {Organic} {Semiconductors}}.
\newblock \emph{\bibinfo{journal}{Physical Review Letters}}
  \textbf{\bibinfo{volume}{114}}, \bibinfo{pages}{086601}
  (\bibinfo{year}{2015}).
\newblock
  \urlprefix\url{https://link.aps.org/doi/10.1103/PhysRevLett.114.086601}.

\bibitem{fratini_transient_2016}
\bibinfo{author}{Fratini, S.}, \bibinfo{author}{Mayou, D.} \&
  \bibinfo{author}{Ciuchi, S.}
\newblock \bibinfo{title}{The {Transient} {Localization} {Scenario} for
  {Charge} {Transport} in {Crystalline} {Organic} {Materials}}.
\newblock \emph{\bibinfo{journal}{Advanced Functional Materials}}
  \textbf{\bibinfo{volume}{26}}, \bibinfo{pages}{2292--2315}
  (\bibinfo{year}{2016}).
\newblock \urlprefix\url{http://doi.wiley.com/10.1002/adfm.201502386}.

\bibitem{joyce_electronic_2013}
\bibinfo{author}{Joyce, H.~J.} \emph{et~al.}
\newblock \bibinfo{title}{Electronic properties of {GaAs}, {InAs} and {InP}
  nanowires studied by terahertz spectroscopy}.
\newblock \emph{\bibinfo{journal}{Nanotechnology}}
  \textbf{\bibinfo{volume}{24}}, \bibinfo{pages}{214006}
  (\bibinfo{year}{2013}).
\newblock
  \urlprefix\url{http://stacks.iop.org/0957-4484/24/i=21/a=214006?key=crossref.3ddd730e84f137f4d4de09739ab00ec8}.

\bibitem{parkinson_transient_2007}
\bibinfo{author}{Parkinson, P.} \emph{et~al.}
\newblock \bibinfo{title}{Transient {Terahertz} {Conductivity} of {GaAs}
  {Nanowires}}.
\newblock \emph{\bibinfo{journal}{Nano Letters}} \textbf{\bibinfo{volume}{7}},
  \bibinfo{pages}{2162--2165} (\bibinfo{year}{2007}).
\newblock \urlprefix\url{https://pubs.acs.org/doi/10.1021/nl071162x}.

\bibitem{kuzel_terahertz_2014}
\bibinfo{author}{Kužel, P.} \& \bibinfo{author}{Němec, H.}
\newblock \bibinfo{title}{Terahertz conductivity in nanoscaled systems:
  effective medium theory aspects}.
\newblock \emph{\bibinfo{journal}{Journal of Physics D: Applied Physics}}
  \textbf{\bibinfo{volume}{47}}, \bibinfo{pages}{374005}
  (\bibinfo{year}{2014}).
\newblock
  \urlprefix\url{https://iopscience.iop.org/article/10.1088/0022-3727/47/37/374005}.

\bibitem{myroshnychenko_modelling_2008}
\bibinfo{author}{Myroshnychenko, V.} \emph{et~al.}
\newblock \bibinfo{title}{Modelling the optical response of gold
  nanoparticles}.
\newblock \emph{\bibinfo{journal}{Chemical Society Reviews}}
  \textbf{\bibinfo{volume}{37}}, \bibinfo{pages}{1792} (\bibinfo{year}{2008}).
\newblock \urlprefix\url{http://xlink.rsc.org/?DOI=b711486a}.

\bibitem{pfister_inorganic_2016}
\bibinfo{author}{Pfister, D.} \emph{et~al.}
\newblock \bibinfo{title}{Inorganic {Double} {Helices} in {Semiconducting}
  {SnIP}}.
\newblock \emph{\bibinfo{journal}{Advanced Materials}}
  \textbf{\bibinfo{volume}{28}}, \bibinfo{pages}{9783--9791}
  (\bibinfo{year}{2016}).
\newblock \urlprefix\url{http://doi.wiley.com/10.1002/adma.201603135}.

\bibitem{mics_density-dependent_2013}
\bibinfo{author}{Mics, Z.}, \bibinfo{author}{D'Angio, A.},
  \bibinfo{author}{Jensen, S.~A.}, \bibinfo{author}{Bonn, M.} \&
  \bibinfo{author}{Turchinovich, D.}
\newblock \bibinfo{title}{Density-dependent electron scattering in photoexcited
  {GaAs} in strongly diffusive regime}.
\newblock \emph{\bibinfo{journal}{Applied Physics Letters}}
  \textbf{\bibinfo{volume}{102}}, \bibinfo{pages}{231120}
  (\bibinfo{year}{2013}).
\newblock \urlprefix\url{http://aip.scitation.org/doi/10.1063/1.4810756}.

\bibitem{markel_introduction_2016}
\bibinfo{author}{Markel, V.~A.}
\newblock \bibinfo{title}{Introduction to the {Maxwell} {Garnett}
  approximation: tutorial}.
\newblock \emph{\bibinfo{journal}{Journal of the Optical Society of America A}}
  \textbf{\bibinfo{volume}{33}}, \bibinfo{pages}{1244} (\bibinfo{year}{2016}).
\newblock
  \urlprefix\url{https://www.osapublishing.org/abstract.cfm?URI=josaa-33-7-1244}.

\bibitem{zhao_monitoring_2019}
\bibinfo{author}{Zhao, D.} \emph{et~al.}
\newblock \bibinfo{title}{Monitoring {Electron}–{Phonon} {Interactions} in
  {Lead} {Halide} {Perovskites} {Using} {Time}-{Resolved} {THz}
  {Spectroscopy}}.
\newblock \emph{\bibinfo{journal}{ACS Nano}} \textbf{\bibinfo{volume}{13}},
  \bibinfo{pages}{8826--8835} (\bibinfo{year}{2019}).
\newblock \urlprefix\url{https://pubs.acs.org/doi/10.1021/acsnano.9b02049}.

\bibitem{uhd_jepsen_ultrafast_2001}
\bibinfo{author}{Uhd~Jepsen, P.} \emph{et~al.}
\newblock \bibinfo{title}{Ultrafast carrier trapping in microcrystalline
  silicon observed in optical pump–terahertz probe measurements}.
\newblock \emph{\bibinfo{journal}{Applied Physics Letters}}
  \textbf{\bibinfo{volume}{79}}, \bibinfo{pages}{1291--1293}
  (\bibinfo{year}{2001}).
\newblock \urlprefix\url{http://aip.scitation.org/doi/10.1063/1.1394953}.

\bibitem{ott_flexible_2019}
\bibinfo{author}{Ott, C.} \emph{et~al.}
\newblock \bibinfo{title}{Flexible and {Ultrasoft} {Inorganic} 1d
  {Semiconductor} and {Heterostructure} {Systems} {Based} on {SnIP}}.
\newblock \emph{\bibinfo{journal}{Advanced Functional Materials}}
  \textbf{\bibinfo{volume}{29}}, \bibinfo{pages}{1900233}
  (\bibinfo{year}{2019}).
\newblock
  \urlprefix\url{https://onlinelibrary.wiley.com/doi/abs/10.1002/adfm.201900233}.

\bibitem{shi_room-temperature_2018}
\bibinfo{author}{Shi, X.} \emph{et~al.}
\newblock \bibinfo{title}{Room-temperature ductile inorganic semiconductor}.
\newblock \emph{\bibinfo{journal}{Nature Materials}}
  \textbf{\bibinfo{volume}{17}}, \bibinfo{pages}{421--426}
  (\bibinfo{year}{2018}).
\newblock \urlprefix\url{http://www.nature.com/articles/s41563-018-0047-z}.

\bibitem{qin_raman_2020}
\bibinfo{author}{Qin, J.-K.} \emph{et~al.}
\newblock \bibinfo{title}{Raman response and transport properties of tellurium
  atomic chains encapsulated in nanotubes}.
\newblock \emph{\bibinfo{journal}{Nature Electronics}}
  \textbf{\bibinfo{volume}{3}}, \bibinfo{pages}{141--147}
  (\bibinfo{year}{2020}).
\newblock \urlprefix\url{https://doi.org/10.1038/s41928-020-0365-4}.

\bibitem{kumar_elastic_2015}
\bibinfo{author}{Kumar, V.}, \bibinfo{author}{Singh, J.~K.} \&
  \bibinfo{author}{Prasad, G.~M.}
\newblock \bibinfo{title}{Elastic properties of elemental, binary and ternary
  semiconductor materials}.
\newblock \emph{\bibinfo{journal}{Indian Journal of Pure \& Applied Physics}}
  \textbf{\bibinfo{volume}{53}}, \bibinfo{pages}{7} (\bibinfo{year}{2015}).

\bibitem{feng_mechanical_2014}
\bibinfo{author}{Feng, J.}
\newblock \bibinfo{title}{Mechanical properties of hybrid organic-inorganic
  {CH}$_{\textrm{3}}${NH}$_{\textrm{3}}${BX}$_{\textrm{3}}$ ({B} = {Sn}, {Pb};
  {X} = {Br}, {I}) perovskites for solar cell absorbers}.
\newblock \emph{\bibinfo{journal}{APL Materials}} \textbf{\bibinfo{volume}{2}},
  \bibinfo{pages}{081801} (\bibinfo{year}{2014}).
\newblock \urlprefix\url{http://aip.scitation.org/doi/10.1063/1.4885256}.

\bibitem{ferreira_elastic_2018}
\bibinfo{author}{Ferreira, A.} \emph{et~al.}
\newblock \bibinfo{title}{Elastic {Softness} of {Hybrid} {Lead} {Halide}
  {Perovskites}}.
\newblock \emph{\bibinfo{journal}{Physical Review Letters}}
  \textbf{\bibinfo{volume}{121}}, \bibinfo{pages}{085502}
  (\bibinfo{year}{2018}).
\newblock
  \urlprefix\url{https://link.aps.org/doi/10.1103/PhysRevLett.121.085502}.

\bibitem{keller_effect_1977}
\bibinfo{author}{Keller, R.}, \bibinfo{author}{Holzapfel, W.~B.} \&
  \bibinfo{author}{Schulz, H.}
\newblock \bibinfo{title}{Effect of pressure on the atom positions in {Se} and
  {Te}}.
\newblock \emph{\bibinfo{journal}{Physical Review B}}
  \textbf{\bibinfo{volume}{16}}, \bibinfo{pages}{4404--4412}
  (\bibinfo{year}{1977}).
\newblock \urlprefix\url{https://link.aps.org/doi/10.1103/PhysRevB.16.4404}.

\bibitem{bandyopadhyay_pressure_1999}
\bibinfo{author}{Bandyopadhyay, A.~K.} \& \bibinfo{author}{Singh, D.~B.}
\newblock \bibinfo{title}{Pressure induced phase transformations and band
  structure of different high pressure phases in tellurium}.
\newblock \emph{\bibinfo{journal}{Pramana}} \textbf{\bibinfo{volume}{52}},
  \bibinfo{pages}{303--319} (\bibinfo{year}{1999}).
\newblock \urlprefix\url{http://link.springer.com/10.1007/BF02828893}.

\bibitem{oehzelt_crystal_2006}
\bibinfo{author}{Oehzelt, M.} \emph{et~al.}
\newblock \bibinfo{title}{Crystal structure of oligoacenes under high
  pressure}.
\newblock \emph{\bibinfo{journal}{Physical Review B}}
  \textbf{\bibinfo{volume}{74}}, \bibinfo{pages}{104103}
  (\bibinfo{year}{2006}).
\newblock \urlprefix\url{https://link.aps.org/doi/10.1103/PhysRevB.74.104103}.

\bibitem{wu_strain_2016}
\bibinfo{author}{Wu, Y.} \emph{et~al.}
\newblock \bibinfo{title}{Strain effects on the work function of an organic
  semiconductor}.
\newblock \emph{\bibinfo{journal}{Nature Communications}}
  \textbf{\bibinfo{volume}{7}}, \bibinfo{pages}{10270} (\bibinfo{year}{2016}).
\newblock \urlprefix\url{http://www.nature.com/articles/ncomms10270}.

\bibitem{wang_effect_2017}
\bibinfo{author}{Wang, P.}, \bibinfo{author}{Zhao, R.}, \bibinfo{author}{Wu,
  L.} \& \bibinfo{author}{Zhang, M.}
\newblock \bibinfo{title}{Effect of {Y} doping on high-pressure behavior of
  {Ag}$_{\textrm{2}}${S} nanocrystals}.
\newblock \emph{\bibinfo{journal}{RSC Advances}} \textbf{\bibinfo{volume}{7}},
  \bibinfo{pages}{35105--35110} (\bibinfo{year}{2017}).
\newblock \urlprefix\url{http://xlink.rsc.org/?DOI=C7RA05327D}.

\bibitem{cartz_effect_1979}
\bibinfo{author}{Cartz, L.}, \bibinfo{author}{Srinivasa, S.~R.},
  \bibinfo{author}{Riedner, R.~J.}, \bibinfo{author}{Jorgensen, J.~D.} \&
  \bibinfo{author}{Worlton, T.~G.}
\newblock \bibinfo{title}{Effect of pressure on bonding in black phosphorus}.
\newblock \emph{\bibinfo{journal}{The Journal of Chemical Physics}}
  \textbf{\bibinfo{volume}{71}}, \bibinfo{pages}{1718--1721}
  (\bibinfo{year}{1979}).
\newblock \urlprefix\url{http://aip.scitation.org/doi/10.1063/1.438523}.

\bibitem{liu_mechanically_2019}
\bibinfo{author}{Liu, B.} \emph{et~al.}
\newblock \bibinfo{title}{Mechanically {Strengthened} {Amorphous} {Silicon} by
  {Phosphorus}: {A} {Molecular} {Dynamics} {Simulation} and {Experimental}
  {Study}}.
\newblock \emph{\bibinfo{journal}{DEStech Transactions on Engineering and
  Technology Research}}  (\bibinfo{year}{2019}).
\newblock
  \urlprefix\url{http://dpi-proceedings.com/index.php/dtetr/article/view/26884}.

\bibitem{lang_mechanical_2009}
\bibinfo{author}{Lang, U.}, \bibinfo{author}{Naujoks, N.} \&
  \bibinfo{author}{Dual, J.}
\newblock \bibinfo{title}{Mechanical characterization of {PEDOT}:{PSS} thin
  films}.
\newblock \emph{\bibinfo{journal}{Synthetic Metals}}
  \textbf{\bibinfo{volume}{159}}, \bibinfo{pages}{473--479}
  (\bibinfo{year}{2009}).
\newblock
  \urlprefix\url{https://linkinghub.elsevier.com/retrieve/pii/S0379677908003731}.

\bibitem{uddin_enhanced_2017}
\bibinfo{author}{Uddin, M.~S.} \& \bibinfo{author}{Ju, J.}
\newblock \bibinfo{title}{Enhanced {Coarse}-{Graining} of {Thermoplastic}
  {Polyurethane} {Elastomer} for {Multiscale} {Modeling}}.
\newblock \emph{\bibinfo{journal}{Journal of Engineering Materials and
  Technology}} \textbf{\bibinfo{volume}{139}}, \bibinfo{pages}{011001}
  (\bibinfo{year}{2017}).
\newblock
  \urlprefix\url{https://asmedigitalcollection.asme.org/materialstechnology/article/doi/10.1115/1.4034328/472959/Enhanced-CoarseGraining-of-Thermoplastic}.

\bibitem{madelung_semiconductors_1996}
\bibinfo{editor}{Madelung, O.} (ed.) \emph{\bibinfo{title}{Semiconductors —
  {Basic} {Data}}} (\bibinfo{publisher}{Springer Berlin Heidelberg},
  \bibinfo{address}{Berlin, Heidelberg}, \bibinfo{year}{1996}).
\newblock \urlprefix\url{http://link.springer.com/10.1007/978-3-642-97675-9}.

\bibitem{roessler_cdte_2009}
\bibinfo{author}{Gutowski, J.}, \bibinfo{author}{Sebald, K.} \&
  \bibinfo{author}{Voss, T.}
\newblock \bibinfo{title}{{CdTe}: mobility}.
\newblock In \bibinfo{editor}{Roessler, U.} (ed.) \emph{\bibinfo{booktitle}{New
  {Data} and {Updates} for {II}-{VI} {Compounds}}}, vol. \bibinfo{volume}{44B},
  \bibinfo{pages}{142--147} (\bibinfo{publisher}{Springer Berlin Heidelberg},
  \bibinfo{address}{Berlin, Heidelberg}, \bibinfo{year}{2009}).
\newblock
  \urlprefix\url{http://materials.springer.com/lb/docs/sm_lbs_978-3-540-74392-7_86}.
\newblock \bibinfo{note}{ISSN: 1616-9549 Series Title: Landolt-Börnstein -
  Group III Condensed Matter}.

\bibitem{herz_charge-carrier_2017}
\bibinfo{author}{Herz, L.~M.}
\newblock \bibinfo{title}{Charge-{Carrier} {Mobilities} in {Metal} {Halide}
  {Perovskites}: {Fundamental} {Mechanisms} and {Limits}}.
\newblock \emph{\bibinfo{journal}{ACS Energy Letters}}
  \textbf{\bibinfo{volume}{2}}, \bibinfo{pages}{1539--1548}
  (\bibinfo{year}{2017}).
\newblock
  \urlprefix\url{https://pubs.acs.org/doi/10.1021/acsenergylett.7b00276}.

\bibitem{bhaskar_mobility_2017}
\bibinfo{author}{Bhaskar, P.}, \bibinfo{author}{Achtstein, A.~W.},
  \bibinfo{author}{Diedenhofen, S.~L.} \& \bibinfo{author}{Siebbeles, L. D.~A.}
\newblock \bibinfo{title}{Mobility and {Decay} {Dynamics} of {Charge}
  {Carriers} in {One}-{Dimensional} {Selenium} van der {Waals} {Solid}}.
\newblock \emph{\bibinfo{journal}{The Journal of Physical Chemistry C}}
  \textbf{\bibinfo{volume}{121}}, \bibinfo{pages}{18917--18921}
  (\bibinfo{year}{2017}).
\newblock \urlprefix\url{https://pubs.acs.org/doi/10.1021/acs.jpcc.7b05183}.

\bibitem{jurchescu_interface_controlled_2007}
\bibinfo{author}{Jurchescu, O.}, \bibinfo{author}{Popinciuc, M.},
  \bibinfo{author}{van Wees, B.} \& \bibinfo{author}{Palstra, T.}
\newblock \bibinfo{title}{Interface-controlled, high-mobility organic
  transistors}.
\newblock \emph{\bibinfo{journal}{Advanced Materials}}
  \textbf{\bibinfo{volume}{19}}, \bibinfo{pages}{688--692}
  (\bibinfo{year}{2007}).
\newblock
  \urlprefix\url{https://onlinelibrary.wiley.com/doi/abs/10.1002/adma.200600929}.
\newblock
  \eprint{https://onlinelibrary.wiley.com/doi/pdf/10.1002/adma.200600929}.

\bibitem{yamagishi_high-mobility_2007}
\bibinfo{author}{Yamagishi, M.} \emph{et~al.}
\newblock \bibinfo{title}{High-mobility double-gate organic single-crystal
  transistors with organic crystal gate insulators}.
\newblock \emph{\bibinfo{journal}{Applied Physics Letters}}
  \textbf{\bibinfo{volume}{90}}, \bibinfo{pages}{182117}
  (\bibinfo{year}{2007}).
\newblock \urlprefix\url{http://aip.scitation.org/doi/10.1063/1.2736208}.

\bibitem{reese_high-performance_2006}
\bibinfo{author}{Reese, C.}, \bibinfo{author}{Chung, W.-J.},
  \bibinfo{author}{Ling, M.-m.}, \bibinfo{author}{Roberts, M.} \&
  \bibinfo{author}{Bao, Z.}
\newblock \bibinfo{title}{High-performance microscale single-crystal
  transistors by lithography on an elastomer dielectric}.
\newblock \emph{\bibinfo{journal}{Applied Physics Letters}}
  \textbf{\bibinfo{volume}{89}}, \bibinfo{pages}{202108}
  (\bibinfo{year}{2006}).
\newblock \urlprefix\url{http://aip.scitation.org/doi/10.1063/1.2388151}.

\bibitem{rudenko_intrinsic_2016}
\bibinfo{author}{Rudenko, A.}, \bibinfo{author}{Brener, S.} \&
  \bibinfo{author}{Katsnelson, M.}
\newblock \bibinfo{title}{Intrinsic {Charge} {Carrier} {Mobility} in
  {Single}-{Layer} {Black} {Phosphorus}}.
\newblock \emph{\bibinfo{journal}{Physical Review Letters}}
  \textbf{\bibinfo{volume}{116}}, \bibinfo{pages}{246401}
  (\bibinfo{year}{2016}).
\newblock
  \urlprefix\url{https://link.aps.org/doi/10.1103/PhysRevLett.116.246401}.

\bibitem{gao_high_2015}
\bibinfo{author}{Gao, X.} \& \bibinfo{author}{Zhao, Z.}
\newblock \bibinfo{title}{High mobility organic semiconductors for field-effect
  transistors}.
\newblock \emph{\bibinfo{journal}{Science China Chemistry}}
  \textbf{\bibinfo{volume}{58}}, \bibinfo{pages}{947--968}
  (\bibinfo{year}{2015}).
\newblock \urlprefix\url{http://link.springer.com/10.1007/s11426-015-5399-5}.

\bibitem{wang_high_2018}
\bibinfo{author}{Wang, X.} \emph{et~al.}
\newblock \bibinfo{title}{High electrical conductivity and carrier mobility in
  {oCVD} {PEDOT} thin films by engineered crystallization and acid treatment}.
\newblock \emph{\bibinfo{journal}{Science Advances}}
  \textbf{\bibinfo{volume}{4}}, \bibinfo{pages}{eaat5780}
  (\bibinfo{year}{2018}).
\newblock
  \urlprefix\url{https://advances.sciencemag.org/lookup/doi/10.1126/sciadv.aat5780}.

\bibitem{park_highly_2019}
\bibinfo{author}{Park, D.~H.} \emph{et~al.}
\newblock \bibinfo{title}{Highly {Stretchable}, {High}‐{Mobility},
  {Free}‐{Standing} {All}‐{Organic} {Transistors} {Modulated} by
  {Solid}‐{State} {Elastomer} {Electrolytes}}.
\newblock \emph{\bibinfo{journal}{Advanced Functional Materials}}
  \textbf{\bibinfo{volume}{29}}, \bibinfo{pages}{1808909}
  (\bibinfo{year}{2019}).
\newblock
  \urlprefix\url{https://onlinelibrary.wiley.com/doi/abs/10.1002/adfm.201808909}.

\bibitem{shatruk_first_1999}
\bibinfo{author}{Shatruk, M.~M.}, \bibinfo{author}{Kovnir, K.~A.},
  \bibinfo{author}{Shevelkov, A.~V.}, \bibinfo{author}{Presniakov, I.~A.} \&
  \bibinfo{author}{Popovkin, B.~A.}
\newblock \bibinfo{title}{First {Tin} {Pnictide} {Halides}
  {Sn}$_{\textrm{24}}${P}$_{\textrm{19.3}}${I}$_{\textrm{8}}$ and
  {Sn}$_{\textrm{24}}${As}$_{\textrm{19.3}}${I}$_{\textrm{8}}$: {Synthesis} and
  the {Clathrate}-{I} {Type} of the {Crystal} {Structure}}.
\newblock \emph{\bibinfo{journal}{Inorganic Chemistry}}
  \textbf{\bibinfo{volume}{38}}, \bibinfo{pages}{3455--3457}
  (\bibinfo{year}{1999}).
\newblock \urlprefix\url{https://pubs.acs.org/doi/10.1021/ic990153r}.

\bibitem{glover_conductivity_1957}
\bibinfo{author}{Glover, R.~E.} \& \bibinfo{author}{Tinkham, M.}
\newblock \bibinfo{title}{Conductivity of {Superconducting} {Films} for
  {Photon} {Energies} between 0.3 and 40 k{T}$_c$}.
\newblock \emph{\bibinfo{journal}{Physical Review}}
  \textbf{\bibinfo{volume}{108}}, \bibinfo{pages}{243--256}
  (\bibinfo{year}{1957}).
\newblock \urlprefix\url{https://link.aps.org/doi/10.1103/PhysRev.108.243}.

\bibitem{lyddane_polar_1941}
\bibinfo{author}{Lyddane, R.~H.}, \bibinfo{author}{Sachs, R.~G.} \&
  \bibinfo{author}{Teller, E.}
\newblock \bibinfo{title}{On the {Polar} {Vibrations} of {Alkali} {Halides}}.
\newblock \emph{\bibinfo{journal}{Physical Review}}
  \textbf{\bibinfo{volume}{59}}, \bibinfo{pages}{673--676}
  (\bibinfo{year}{1941}).
\newblock \urlprefix\url{https://link.aps.org/doi/10.1103/PhysRev.59.673}.

\bibitem{born1988dynamical}
\bibinfo{author}{Born, M.} \& \bibinfo{author}{Huang, K.}
\newblock \emph{\bibinfo{title}{Dynamical Theory of Crystal Lattices}}.
\newblock International series of monographs on physics
  (\bibinfo{publisher}{Clarendon Press}, \bibinfo{year}{1988}).
\newblock \urlprefix\url{https://books.google.ca/books?id=5q9iRttaaDAC}.

\bibitem{huber_femtosecond_2005}
\bibinfo{author}{Huber, R.} \emph{et~al.}
\newblock \bibinfo{title}{Femtosecond {Formation} of {Coupled}
  {Phonon}-{Plasmon} {Modes} in {InP}: {Ultrabroadband} {THz} {Experiment} and
  {Quantum} {Kinetic} {Theory}}.
\newblock \emph{\bibinfo{journal}{Physical Review Letters}}
  \textbf{\bibinfo{volume}{94}}, \bibinfo{pages}{027401}
  (\bibinfo{year}{2005}).
\newblock
  \urlprefix\url{https://link.aps.org/doi/10.1103/PhysRevLett.94.027401}.

\end{thebibliography}

\begin{acknowledgments}
	We wish to acknowledge funding from the Natural Sciences and Engineering Research Council of Canada (NSERC), the Alberta / Technical University of Munich International Graduate School for Hybrid Functional Materials (ATUMS), the Canada Foundation for Innovation (CFI), the Alberta Innovates Technology Futures (AITF) Strategic Chairs Program, and the German Science Foundation DFG Grant No. Ni 1095/8-1. We also wish to acknowledge helpful discussions with D. G. Cooke and T. L. Cocker, technical support from B. Shi and G. Popowich, and S. Xu and the University of Alberta NanoFAB for the helium ion microscope measurements. 
\end{acknowledgments}

\section{Competing Interests}
The authors declare no competing interests.
\vfill

\end{document}

\newcommand{\uofa}{Department of Physics, University of Alberta,Edmonton, Alberta T6G 2E1, Canada}

\newcommand{\tum}{Department of Chemistry, Technical University of Munich, Garching bei M\"unchen 85748, Germany}

\author{David N. Purschke}
\email[Corresponding Author: ]{purschke@ualberta.ca}

\affiliation{\uofa}

\author{Markus R. P. Pielmeier}
\affiliation{\tum}
\author{Ebru \"Uzer}
\affiliation{\tum}
\author{Claudia Ott}
\affiliation{\tum}
\author{Charles Jensen}
\affiliation{\uofa}
\author{Annabelle Degg}
\affiliation{\tum}
\author{Anna Vogel}
\affiliation{\tum}
\author{Naaman Amer}
\affiliation{\uofa}
\author{Tom Nilges}
\affiliation{\tum}
\author{Frank A. Hegmann}
\affiliation{\uofa}%

\title{
	Supplementary information for\\
	Ultrafast photoconductivity and terahertz vibrational dynamics \linebreak in double-helix SnIP nanowires}

\maketitle

\section{\supl\noteMorph}
\begin{figure}[htb]
	\begin{center}
		\includegraphics[scale=0.9]{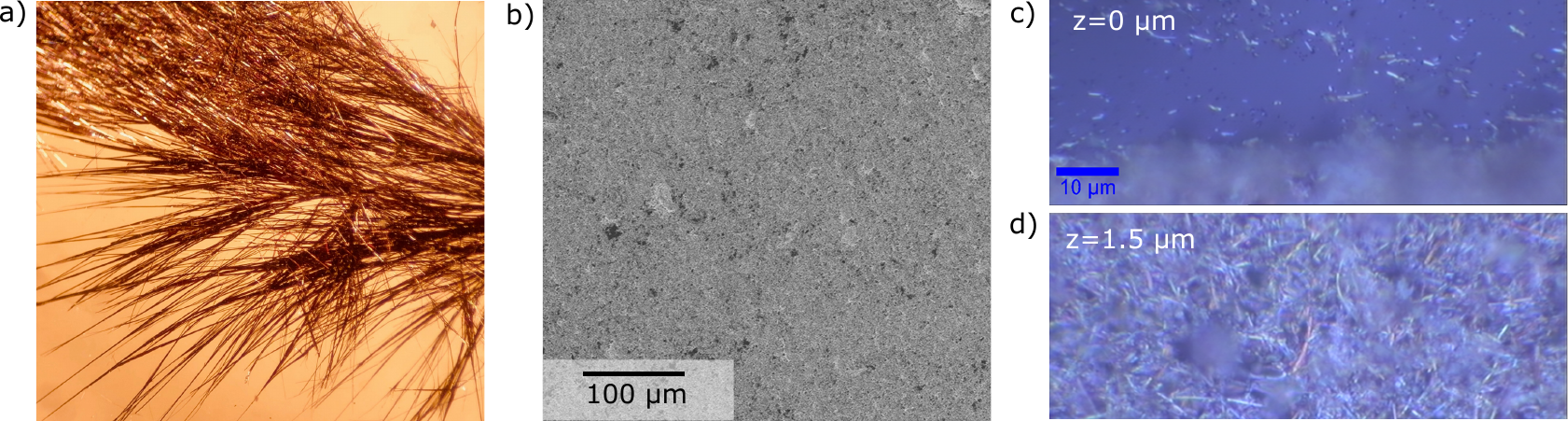}
		\caption{\textbf{SnIP bulk needle and nanowire thin film morphologies. a} Bulk SnIP needles with large aspect ratio. \textbf{b} Helium-ion microscope image showing the large-scale structure and uniformity of the SnIP film. 100x magnification optical microscope image of the SnIP nanowire film with the focus height, z, fixed to set to \textbf{c} the surface of the quartz substrate on a cleaned area and \textbf{d} the SnIP nanowire in a region translated a short distance away from the image in \textbf{c}.}
		\label{morphology}
	\end{center}
\end{figure}

Supplementary Fig. \ref{morphology}a shows an example of SnIP needles annealed over a long period of time (months) during the bake cycle. Supplementary Fig. \ref{morphology}b shows the large scale structure of the ultrasonicated SnIP nanowire thin film. A number of small gaps are present, however, in general there is a high degree of coverage. 
\par

Supplementary Fig. \ref{morphology}c, d show optical microscope images of the substrate and thin film, respectively, with a 100x magnification optical microscope. For a given objective height above the substrate focus, only a fraction of the nanowires are in focus. Choosing the height with the most nanowires in focus in comparison to the height with the substrate in focus allows us to estimate the average film thickness, which we find to be $1.5\,\pm\,0.5\mu m$.
\clearpage

\section{\supl\noteDist}
\begin{figure}[htb]
	\begin{center}
		\includegraphics[scale=0.9]{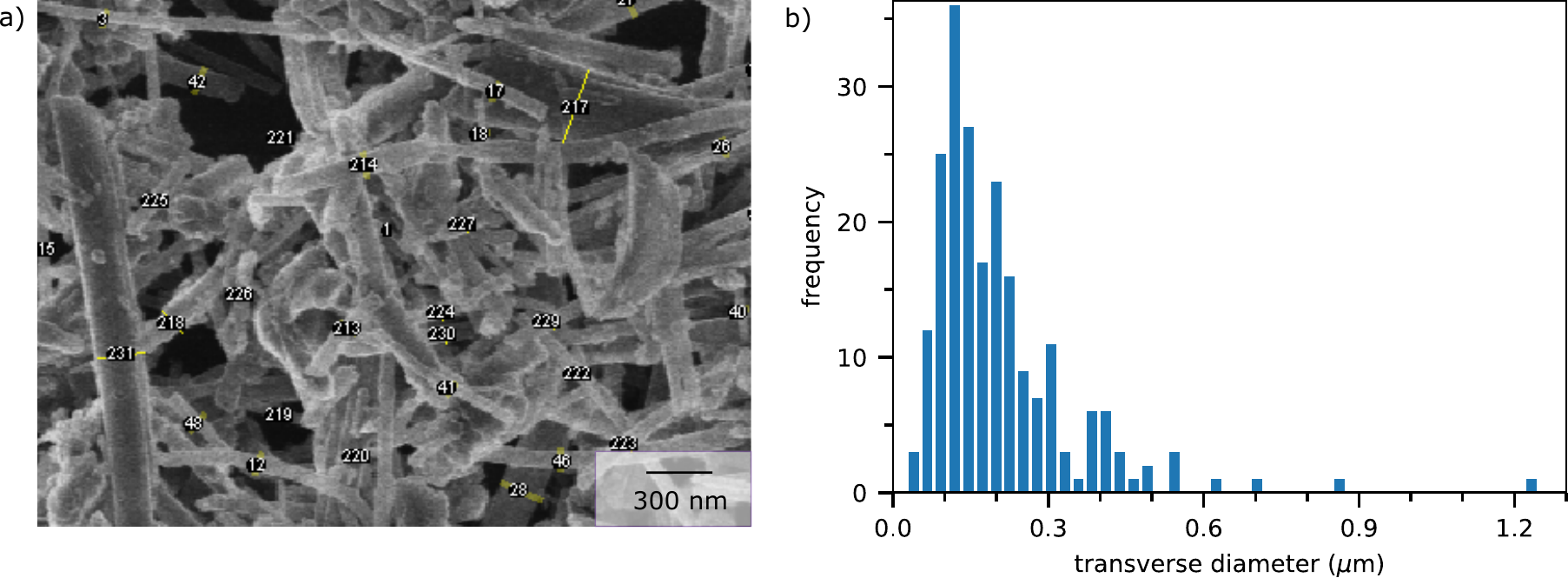}
		\caption{\textbf{Estimation of average nanowire diameter. a} An enlarged section of the helium ion microscope image in Fig. 1d from the main text. The numbered lines indicate nanowires that were measured using ImageJ. \textbf{b} A histogram of the diameter of 230 nanowires sampled randomly from the image in Fig. 1d of the main text.}
		\label{needleDist}
	\end{center}
\end{figure}
In this Supplementary Note, we study the distribution of nanowire diameters. Supplementary Fig. \ref{needleDist}a shows a magnified section of the helium ion microscope image from Fig. 1d of the main text. Approximately 230 nanowires were chosen visually (effort was made to obtain a random selection representative of the full distribution) and their diameter was measured using ImageJ. Almost all of nanowires have diameter less than 500 nm and, ignoring outliers (which we define as diameter greater than 500 nm), we find the average, median, and standard deviation of the distribution are 190 nm, 160 nm, and 100 nm, respectively. The most common nanowire size in the histogram was in the 100 to 110 nm bin. The smallest nanowire measured was 40 nm and the largest 1200 nm. We expect the nanowires to be approximately cylindrical.
\par

It is also useful to estimate the filling fraction of the thin film. While we can see that the area filling fraction is high, it is more difficult to estimate the volume filling fraction from images alone. For a densely packed system of cylinders, the highest volume filling fraction attainable is 0.9, however, this requires highly ordered packing. Examination of Supplementary Fig. \ref{needleDist}a shows that our film consists of randomly oriented nanowires with longitudinal axis in the plane. Numerical calculations of random packing for cylinders with large aspect ratio, similar to the nanowires in our films, yield a volume filling fraction of 0.35 to 0.45 \cite{Zhang_Experimental_2006}. In the following sections, we will use 0.35 as an estimate of the volume filling fraction in our films.

\clearpage

\section{\supl\noteEpsDFT} 
\begin{figure}[htb]
	\begin{center}
		\includegraphics[scale=0.8]{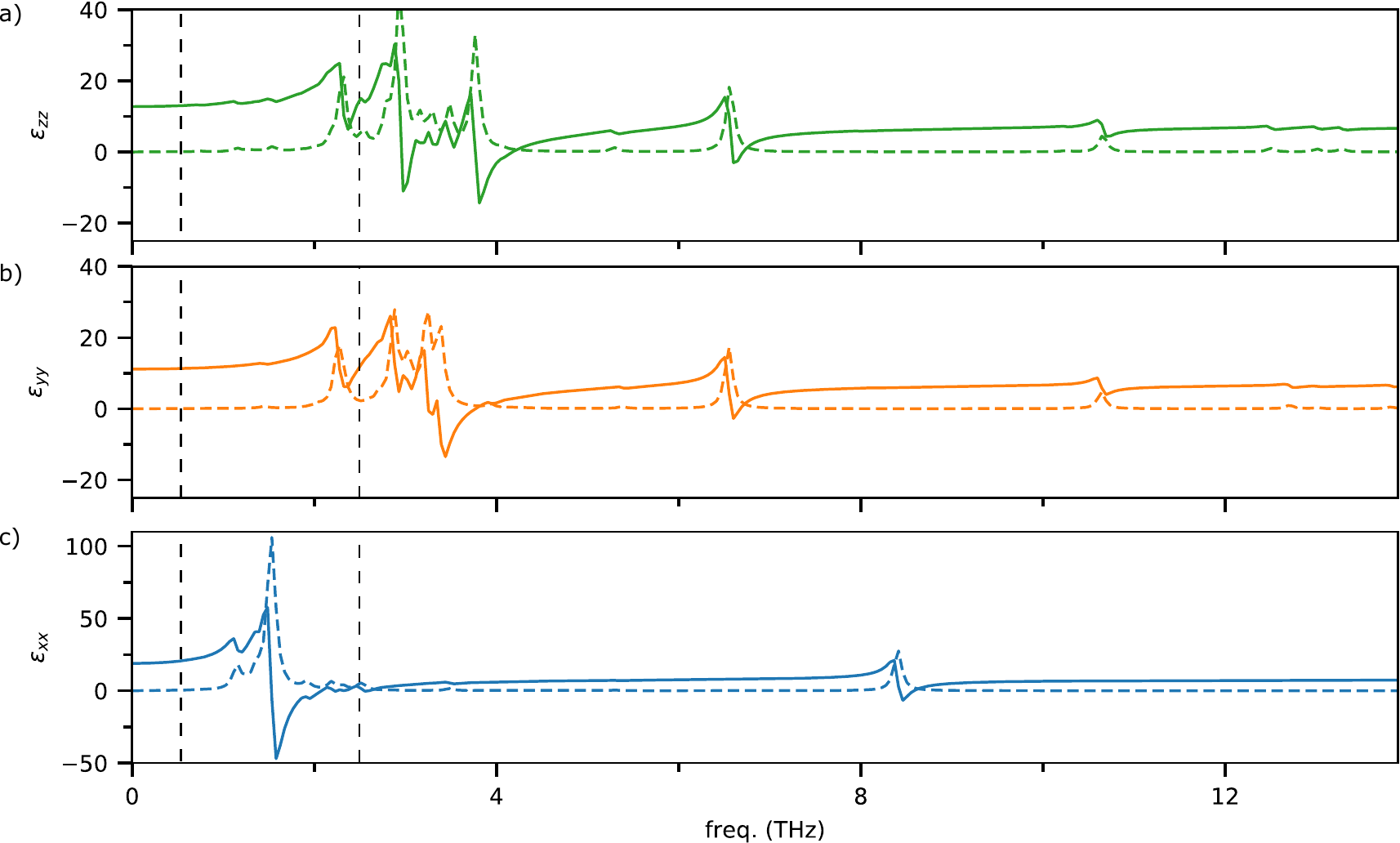}
		\caption{\textbf{On diagonal components of the dielectric function from DFT. a},\textbf{b}, and \textbf{c}, elements of the dielectric function, $\epsilon_{zz}$, $\epsilon_{yy}$, $\epsilon_{x}$, respectively. The dielectric function was calculated using the oscillator strength, calculated from the Born effective charges extracted from DFT, and assuming a Lorentzian lineshape with a broadening of 0.1 THz. In each plot the dashed line is the imaginary part of the susceptibility while the solid line is the real part.}
		\label{polEps}
	\end{center}
\end{figure}
In the framework of DFT, optically active modes are characterized by their Born effective charge vector, $\vec{Z}_{p,i}$, where $p$ is the mode index and $i$ is the Cartesian direction \cite{zicovich-wilson_calculation_2004,ferrero_coupled_2008,erba_accurate_2013}. The intensity of the $p^{th}$ mode, which is plotted in Fig. 1e of the main text and Supplementary Fig. \ref{intColored}, is defined as,
\begin{equation}
I_p=\frac{\pi N_A}{3c^2}\cdot d_p\cdot|\vec{Z}_p|^2,
\end{equation}
where $N_A$ is Avogadro's number and $d_p$ is the degeneracy of the mode. Instead of the intensity, we can use the mass-weighted Born effective-charge vector to calculate the oscillator strength tensor,
\begin{equation}
f_{p,ij}=\frac{4\pi}{\Omega}\frac{\vec{Z}_{p,i}\vec{Z}_{p,j}}{\omega_p^2},
\end{equation}
where $\Omega$ is the unit cell volume and $\omega_p$ is the mode angular frequency. Using the oscillator strength, we can then calculate the frequency dependent dielectric tensor,
\begin{equation}
\epsilon_{ij}(\omega)=\epsilon_{\infty,ij}+\sum_{p=1}^{p_{max}}\frac{f_{p,ij}\omega_p^2}{\omega_p^2-\omega^2-i\gamma\omega},
\label{dielectricTensor}
\end{equation}
where $\epsilon_{\infty,ij}$ is the high-frequency permittivity due to electronic transitions, $\omega$ is the angular frequency, and $\gamma$ is a phenomenological damping constant, which we assume is the same for every mode although it can in principle vary. The high-frequency permittivity, also calculated from DFT using the coupled-perturbed Kohn-Sham method, is,
\begin{equation}
\epsilon_{\infty,ij} = 
\begin{pmatrix}
7.7283 & 0      & 0.2167 \\
0      & 6.8028 & 0      \\
0.2167 &   0    & 7.2405 \\
\end{pmatrix},
\end{equation}
which indicates similar electronic polarizability in each direction and a small off-diagonal coupling between $x$ and $z$. Using the Born charges for each mode (see Supplemental Note 4 and attachment Vibration\_table.xlsx), along with the static dielectric tensor we can now use eq. \ref{dielectricTensor} to calculate the polarization-dependent dielectric function, as shown in Supplementary Fig. \ref{polEps}a-c for the $zz$, $yy$, and $xx$ components of the dielectric function. The $\hat{x}$, $\hat{y}$, and $\hat{z}$ axes are defined with respect to the crystal structure in Supplementary Fig. \ref{bands}a or Fig. 1a of the main text. We can see that the $yy$ and $zz$ components are qualitatively similar while the $x$ direction is significantly different, which reflects the obvious structural asymmetry parallel versus perpendicular to the double-helix axis. As in $\epsilon_{\infty}$, there is a small off-diagonal component, $\epsilon_{xz}$ (not shown), however, it is significantly smaller than the on-diagonal components.
\par

We additionally plot the average dielectric function, 
\begin{equation}
\epsilon_{avg}=0.5\,\epsilon_{xx}+0.25\,\epsilon_{yy}+0.25\,\epsilon_{zz},
\label{dielAvg}
\end{equation}
which is a simple approximation that assumes linear weighting of the various contributions. As for the average effective mass calculation, we have assumed an equal proportion of nanowires are oriented parallel and perpendicular to the THz field and, of the perpendicularly oriented nanowires an equal proportion are oriented along the $\hat{y}$ and $\hat{z}$ axes. This is more directly comparable to the experimentally measured dielectric function that probes an ensemble of all directions. We see that the dominant contribution to the averaged dielectric constant is from the $\epsilon_{xx}$ component.
\par

From the averaged dielectric constant, we see that the static dielectric constant, $\epsilon_{st}$, is approximately 15. In contrast, the effective dielectric constant extracted from THz-TDS was found to be approximately 4, which is significantly smaller. The largest contribution to this discrepancy is likely the low filling fraction of the film, estimated to be 0.35 in \supl\noteDist.
\begin{figure}[htb]
	\begin{center}
		\includegraphics[scale=0.8]{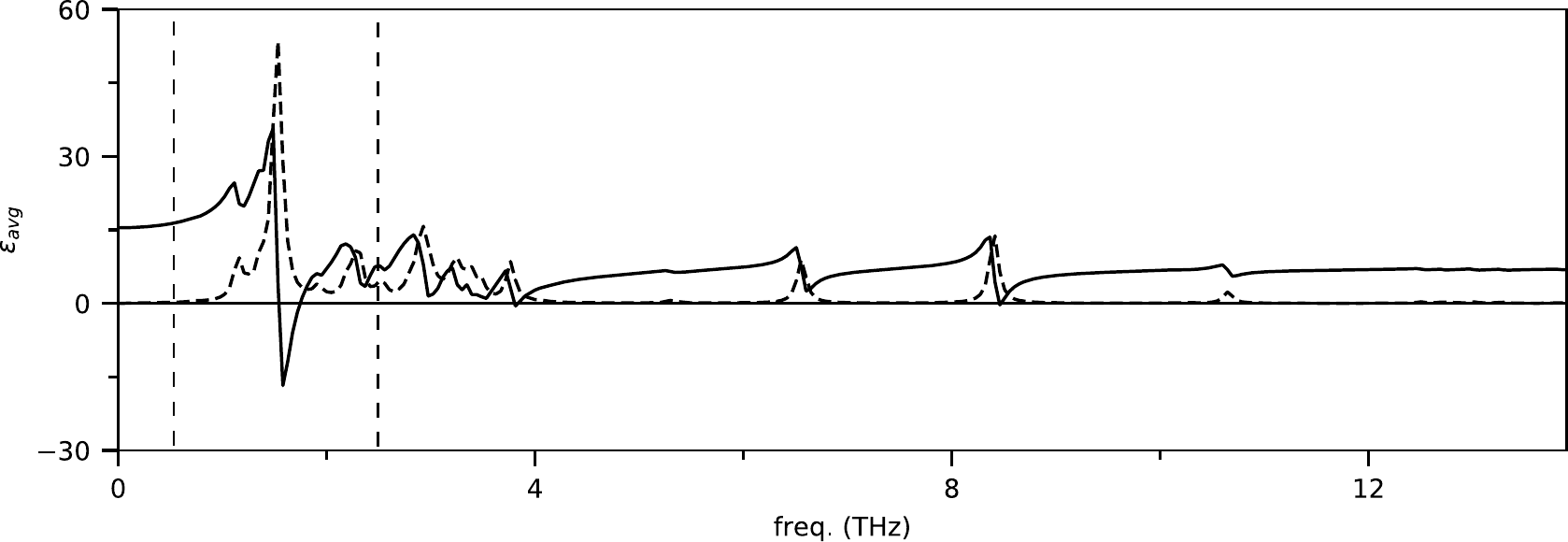}
		\caption{\textbf{Average dielectric function.} Dielectric function calculated from an average of the 3 polarizations as defined by Supplementary eq. \ref{dielAvg}}
		\label{polEps}
	\end{center}
\end{figure}

\clearpage

\section{\supl\noteVibVecs}
The attachment Vibration\_table.xlsx contains a spreadsheet with the mode number, wavenumber, frequency, irreducible representation, IR and Raman activity, mode intensity, and effective Born charge for each of the normal modes at the gamma point calculated using the coupled-perturbed Kohn-Sham method. 
\par

Additionally attached are animations of the following mode 9, 23, 24, 43, 63, 92, and 99, which are highlighted in Supplementary Fig. \ref{intColored} with projections of their normal mode vectors to the a and b axes shown in Supplementary Figs. \ref{mode9}-\ref{mode99}. Modes 9, 23, and 24 were chosen as they are the most likely to correspond to the modes observed in our THz spectra, while modes 43, 63, 92, and 99 were chosen due to their high intensities and representative of the types of motion in each frequency range. The mode numbers are indexed according to the number in the attachment Vibration\_table.xlsx.

\begin{figure}[htb]
	\begin{center}
		\includegraphics[scale=1]{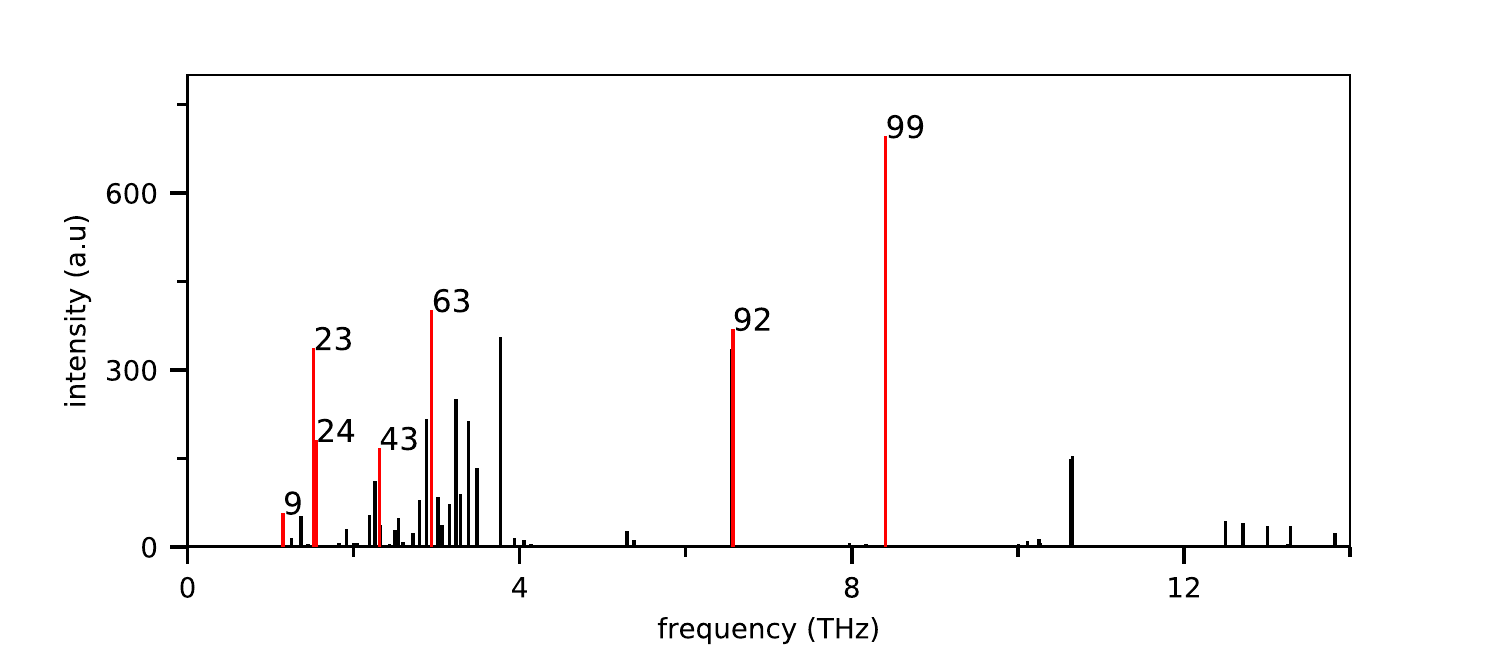}
		\caption{\textbf{Labeling normal modes for the provided animations.} Intensity of each normal mode as a function of its frequency, calculated using Crystal17. The lines colored red and numbered indicate the modes that we have provided animations and displacement vectors for.}
		\label{intColored}
	\end{center}
\end{figure}

\begin{figure}[htb]
	\begin{center}
		\includegraphics[scale=0.28]{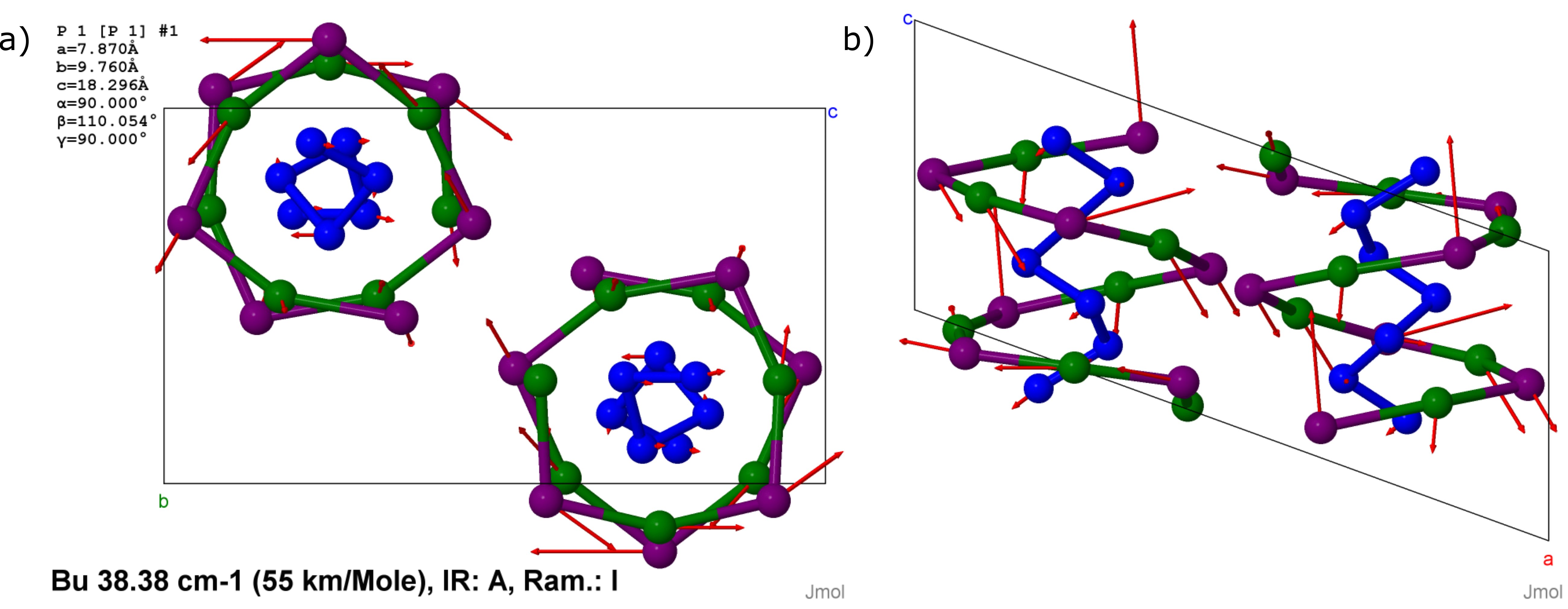}
		\caption{\textbf{Normal mode displacement vectors for mode 9. a} Projected along the a-axis and \textbf{b} projected along the b-axis. The corresponding animation is contained in the Supplementary file mode\_9.avi.}
		\label{mode9}
	\end{center}
\end{figure}

\begin{figure}[htb]
	\begin{center}
		\includegraphics[scale=0.28]{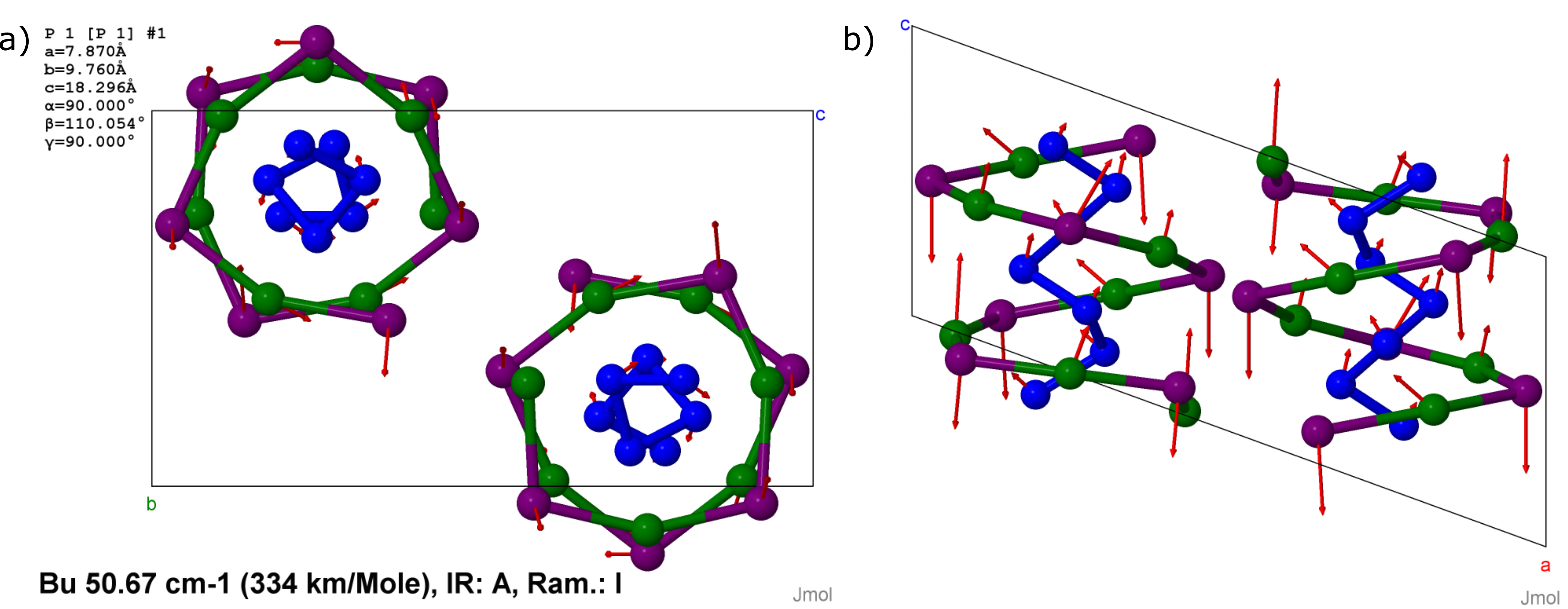}
		\caption{\textbf{Normal mode displacement vectors for mode 23. a} Projected along the a-axis and \textbf{b} projected along the b-axis. The corresponding animation is contained in the Supplementary file mode\_23.avi.}
		\label{mode23}
	\end{center}
\end{figure}

\begin{figure}[htb]
	\begin{center}
		\includegraphics[scale=0.28]{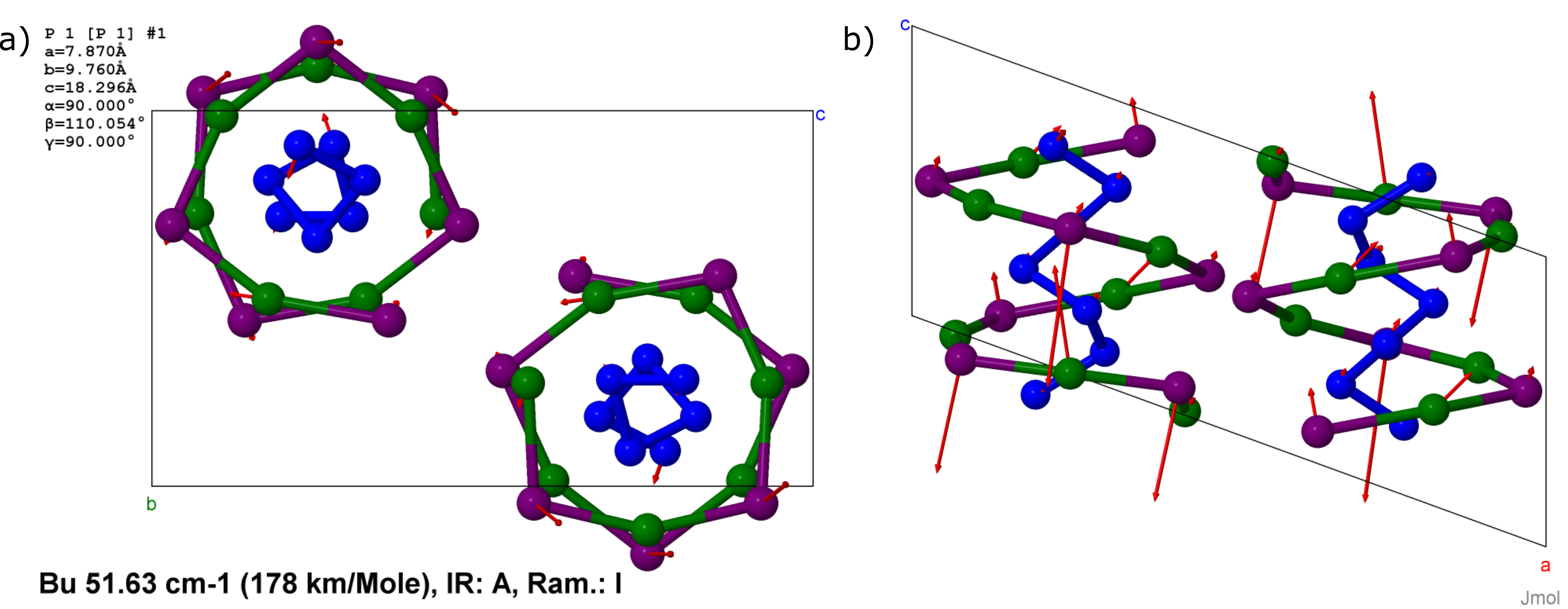}
		\caption{\textbf{Normal mode displacement vectors for mode 24. a} Projected along the a-axis and \textbf{b} projected along the b-axis. The corresponding animation is contained in the Supplementary file mode\_24.avi.}
		\label{mode23}
	\end{center}
\end{figure}

\begin{figure}[htb]
	\begin{center}
		\includegraphics[scale=0.28]{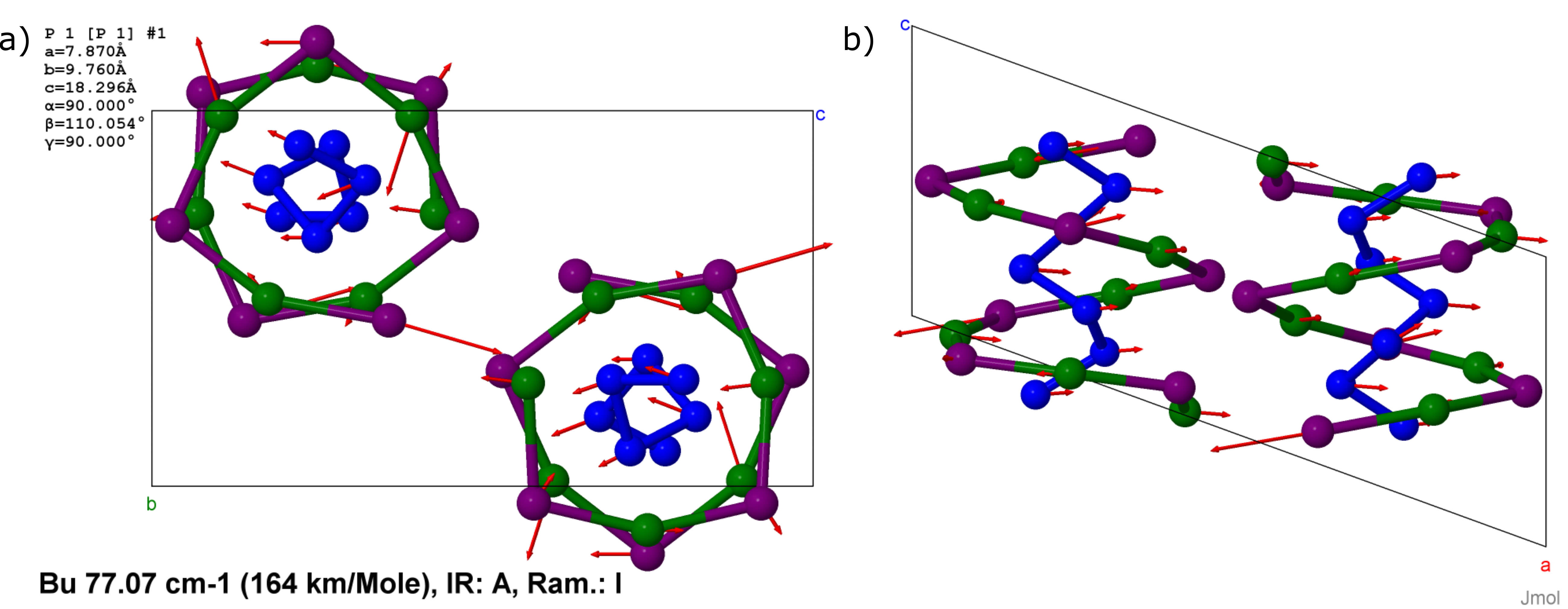}
		\caption{\textbf{Normal mode displacement vectors for mode 43. a} Projected along the a-axis and \textbf{b} projected along the b-axis. The corresponding animation is contained in the Supplementary file mode\_43.avi.}
		\label{mode40}
	\end{center}
\end{figure}

\begin{figure}[htb]
	\begin{center}
		\includegraphics[scale=0.28]{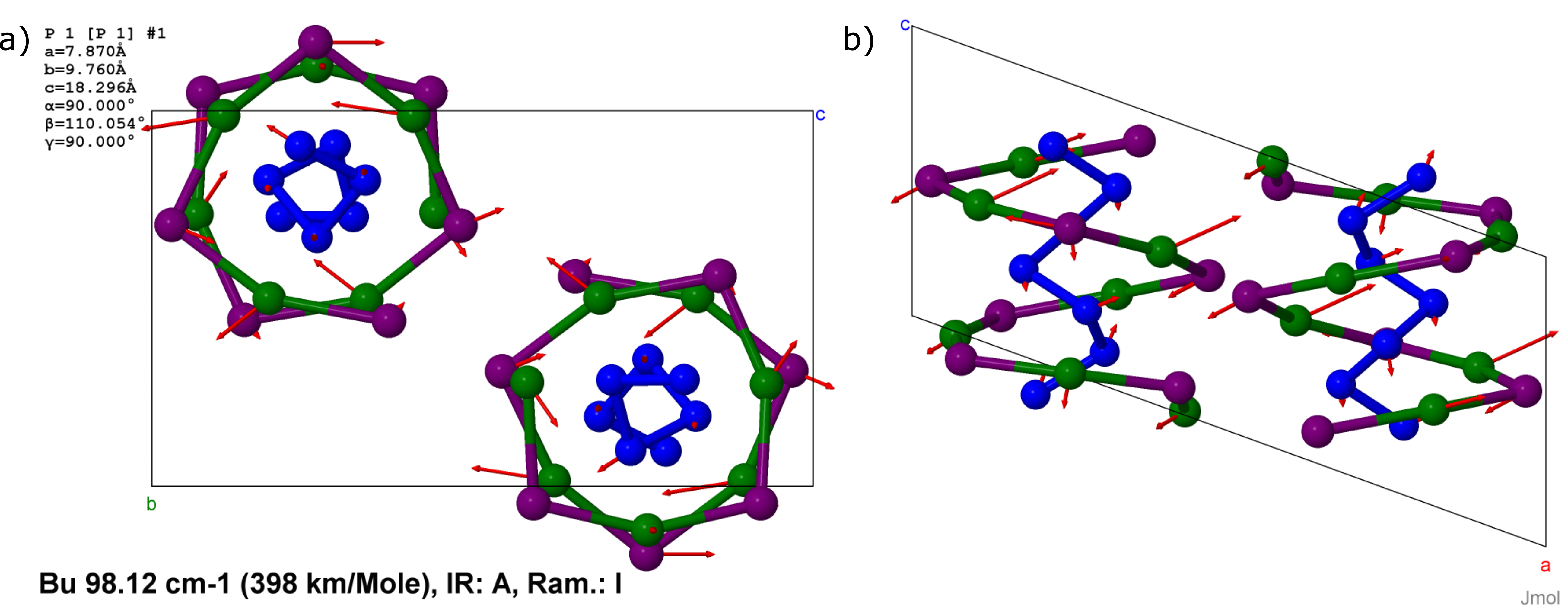}
		\caption{\textbf{Normal mode displacement vectors for mode 63. a} Projected along the a-axis and \textbf{b} projected along the b-axis. The corresponding animation is contained in the Supplementary file mode\_63.avi.}
		\label{mode40}
	\end{center}
\end{figure}

\begin{figure}
	\begin{center}
		\includegraphics[scale=0.28]{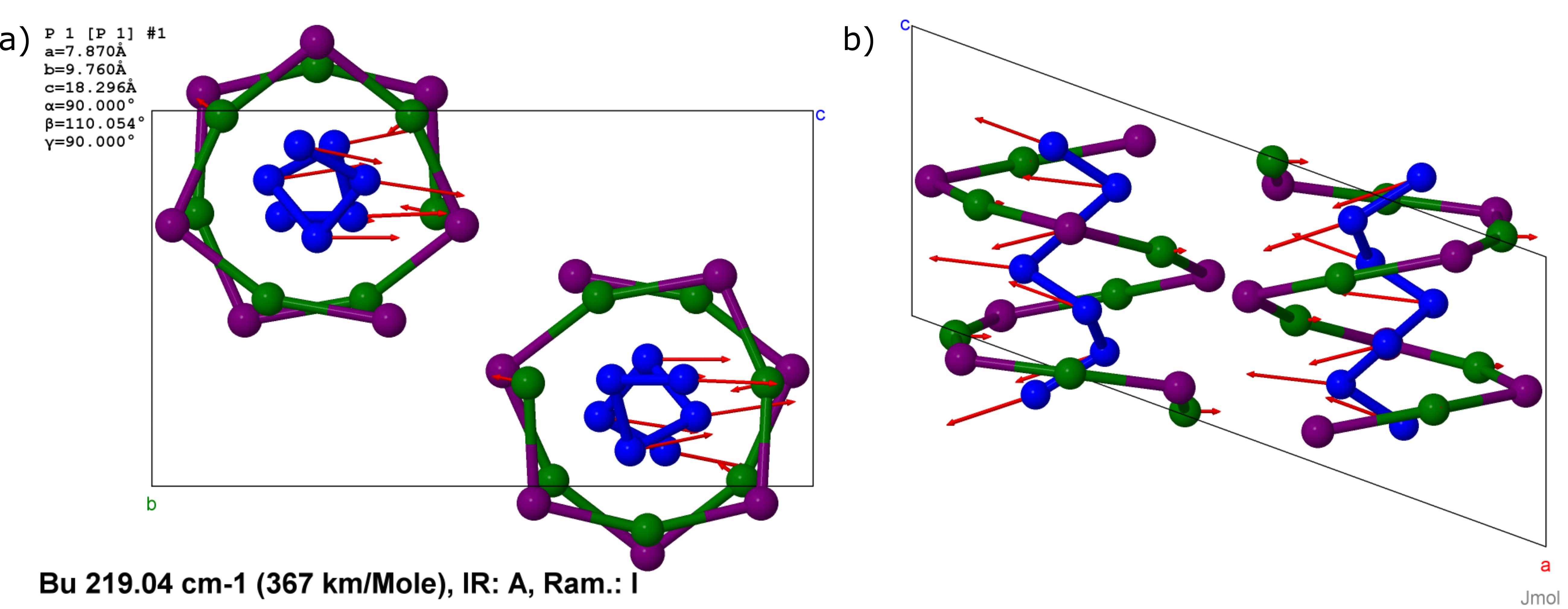}
		\caption{\textbf{Normal mode displacement vectors for mode 92. a} Projected along the a-axis and \textbf{b} projected along the b-axis. The corresponding animation is contained in the Supplementary file mode\_92.avi.}
		\label{mode92}
	\end{center}
\end{figure}

\begin{figure}[htb]
	\begin{center}
		\includegraphics[scale=0.28]{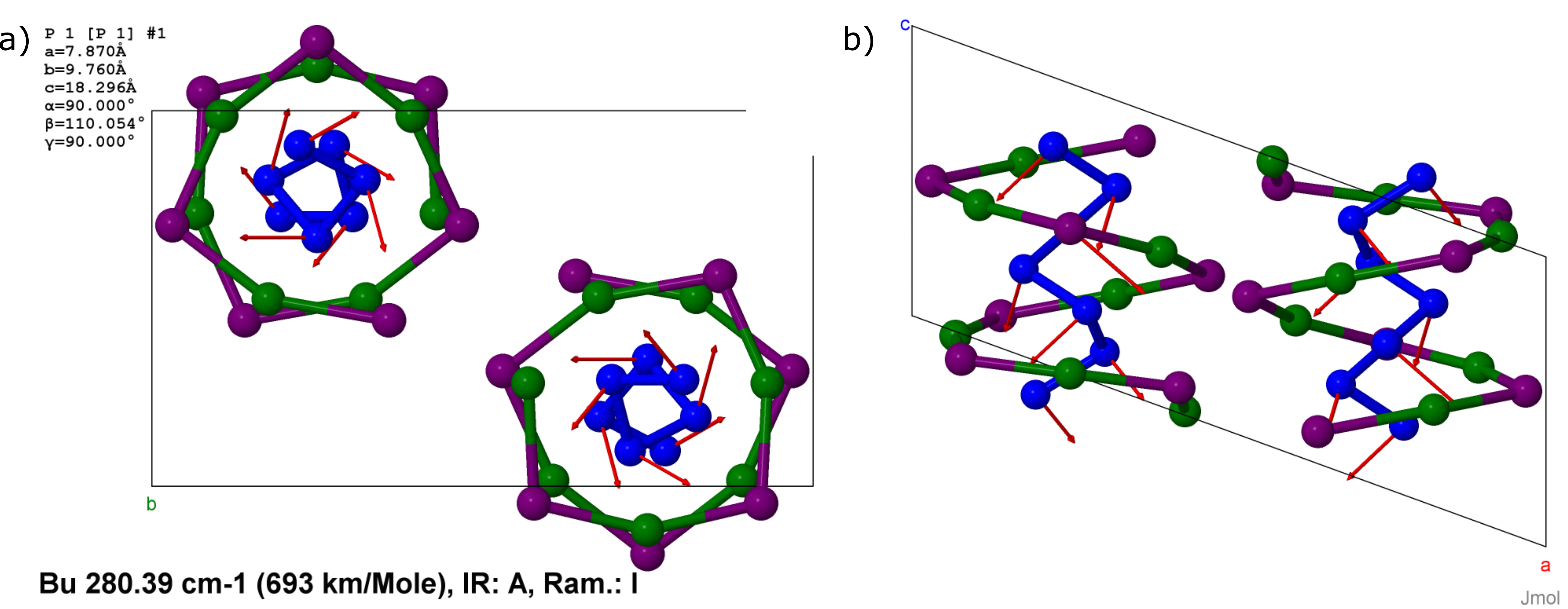}
		\caption{\textbf{Normal mode displacement vectors for mode 99. a} Projected along the a-axis and \textbf{b} projected along the b-axis. The corresponding animation is contained in the Supplementary file mode\_99.avi.}
		\label{mode99}
	\end{center}
\end{figure}
\vfill
\clearpage

\section{\supl\noteDifTime}
\begin{figure}[htb]
	\begin{center}
		\includegraphics[scale=0.8]{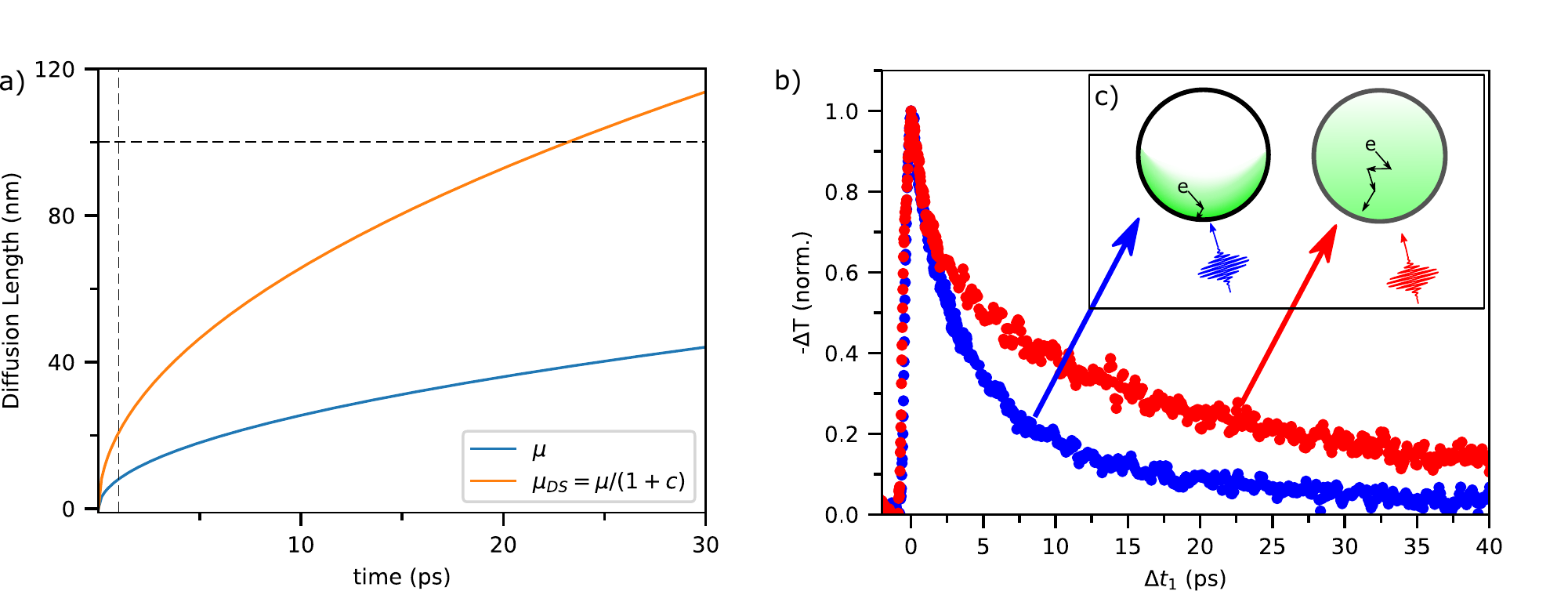}
		\caption{\textbf{Estimation of carrier diffusion times. a} Calculation of the average diffusion length using the long-range (blue) and short-range (orange) mobilities from the Drude-Smith model, $\mu$ and $\mu_{DS}$, respectively, as defined in the main text. The horizontal dashed line indicates the average nanowire radius while the vertical dashed line indicates 1 ps for the timescale of half an oscillation at 0.5 THz (the lower bound of the experimental bandwidth). \textbf{b} Comparison of the carrier lifetime for 400 (blue) \textit{vs} 800 nm (red) excitation to show the increased lifetime with lower wavelength. \textbf{c} Schematic illustration of the difference in excitation profile, where carriers are generated closer to the surface with 400 nm excitation, which leads to a shorter lifetime.}
		\label{diffusionTime}
	\end{center}
\end{figure}
In this note, we consider the timescales associated with carrier diffusion to help understand origin of carrier localization observed in THz spectroscopy and the carrier lifetime in the high fluence condition where bulk traps are saturated. The timescale for carriers to diffuse a distance $x$, for 1D diffusion, is given by $t=x^2/D$. The diffusion constant is given by, 
\begin{equation}
D=\mu_{diff} k_BT/e,
\end{equation}
where $k_B$ is the Boltzmann constant. We find $D=0.64\,cm^2\,s^{-1}$ if we consider $\mu_{diff}=\mu_{long-range}=e\tau_{DS}(1+c)/m^*_{z'}$ for transport in the $\hat{z}'$-direction (approximately transverse to the double-helix/nanowire axis) and a temperature of 300 K. For nanowires on the small-diameter (80 nm) and large-diameter (300 nm) sides of the distribution, the timescale to drift one nanowore radius is 25 ps and 350 ps, respectively. 
\par

This implies, for the distribution of nanowires on average, a timescale of decay significantly longer than the photoconductive decays from the main text (re-plotted in Supplementary Fig. \ref{diffusionTime}b, for convenience). If we instead consider $\mu_{diff}=\mu=e\tau_{DS}/m^*_{z'}$, we find $D=4.3\,cm^2\,s^{-1}$ and the diffusive timescale is reduced to 3.7 ps and 52 ps for 80 nm and 300 nm diameter nanowires, respectively, which is closer to the experimentally observed timescales. We note that this is a simplified description of 2D diffusion in a cylindrical geometry, which would be a more precise description of the experiment (see supporting information in ref. \cite{bergren_ultrafast_2014}). Nevertheless, the discussion using the 1D approximation provides semi-quantitative validation of the relevant timescales and supports diffusion to the surface as the likely mechanism for recombination.
\par

\begin{figure}[htb]
	\begin{center}
		\includegraphics[scale=0.8]{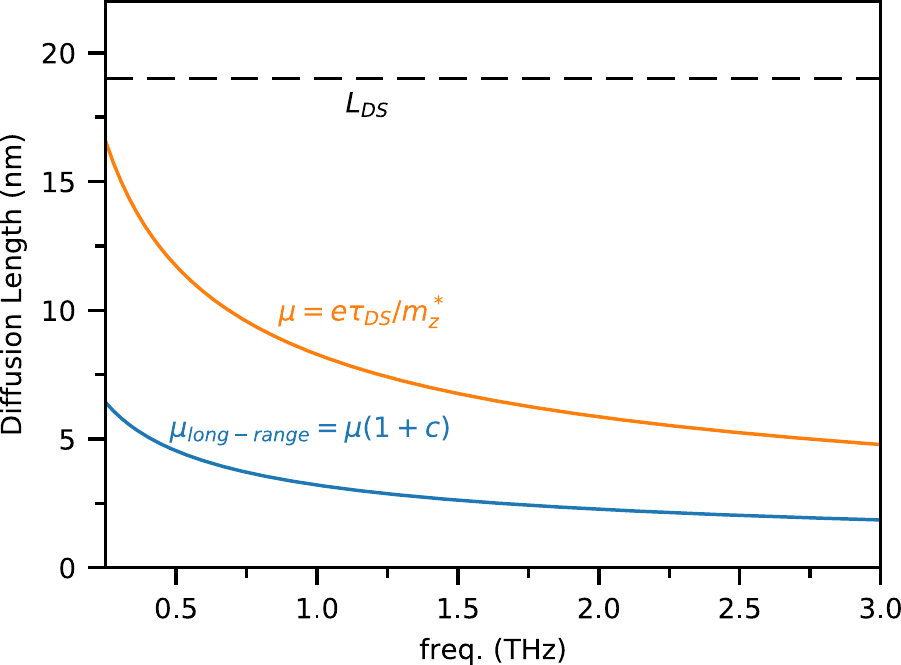}
		\caption{\textbf{Length scale probed by THz spectroscopy.} Diffusion length on a timescale $t_{diff}=1/(2\pi f)$ as a function function of frequency for the long-range mobility, $\mu_{long-range}$ (blue), and short range mobility, $\mu$ (orange). The horizontal dashed black line indicates the localization length estimated from the Drude-Smith model.}
		\label{difLenFreq}
	\end{center}
\end{figure}

We can use similar length scale arguments to speculate on the origin of the large $c$ parameter from the Drude-Smith fits. A commonly used estimate for the length scale probed by an AC field of frequency $f$ is given by $x=\sqrt{D/(2\pi f)}$, which can be thought of as the diffusion length on a timescale of approximately 1/6$^{th}$ of a cycle. Shown in Supplementary Fig. \ref{difLenFreq} is the length scale probed as a function of frequency using this estimate. Alternatively, Cocker \textit{et al.}, found an analytic relationship between the length-scale of localization and the Drude-Smith scattering time using the so-called modified Drude-Smith model, where the conductivity is given by \cite{cocker_microscopic_2017},
\begin{equation}
\sigma_{DS-mod}=\frac{Ne^2\tau_{DS}/m}{1-i\omega\tau_{DS}}\left(1+\frac{c}{1-i\omega\tau_{diff}}\right),
\end{equation}
where $\tau_{diff}$ is the characteristic timescale of a diffusive restoring force that gives rise to the suppression of low frequency conductivity and the other parameters are defined similarly to the usual Drude-Smith model. Fitting our data to the modified Drude-Smith model, we find $\tau_{diff}\approx\tau_{DS}$, \textit{i.e.}, the modified Drude-Smith model reduces to the usual Drude-Smith model as defined in the main text. For this special case, the length scale of localization is given by (see section V in ref. \cite{cocker_microscopic_2017}),
\begin{equation}
L_{DS}=\frac{9\tau_{DS}}{2}\sqrt{\frac{k_BT}{m^*}}.
\end{equation}
We find a value of $L_{DS}=19\,nm$. For reference, a horizontal dashed line is plotted at $L_{DS}$ in Supplementary Fig. \ref{difLenFreq}, where we see that the length scale estimate from the relation $x=\sqrt{D/(2\pi f)}$ approaches the value of $L_{DS}$ at low frequencies. 
\par

The approximately $20\,nm$ length scale suggested by the modified Drude-Smith model is smaller than the most common nanowire size of 110 nm. It is possible that this length scale estimate, which assumes a rectangular box, does not hold for the cylindrical geometry. However, it seems unlikely that this explanation could account for such a large difference between estimated and observed length scale. Another explanation is the presence of morphological features on length scales smaller than those observable by helium-ion microscopy such as grain boundaries \cite{laforge_conductivity_2014, titova_ultrafast_2016}.
\par

Another possibility is that spectral signatures of carrier localization are an intrinsic feature of transport in SnIP, as in potentially the case in highly-polar solids such as rubrene \cite{ostroverkhova_ultrafast_2006,Zhang_Experimental_2006}. Generally, theories that predict these spectral signatures require large electron-phonon coupling\cite{de_filippis_crossover_2015,fratini_transient_2016}, while in SnIP we estimate a Fr\"ohlich constant of $\alpha=0.7$ (see \enquote{Methods} in the main text), which is likely too small for such theories to apply. We therefore consider the presence of microscopic morphology to be the likely origin of the Drude-Smith conductivity. 
\clearpage

\section{\supl\noteOptLin}
\begin{figure}[htb]
	\begin{center}
		\includegraphics{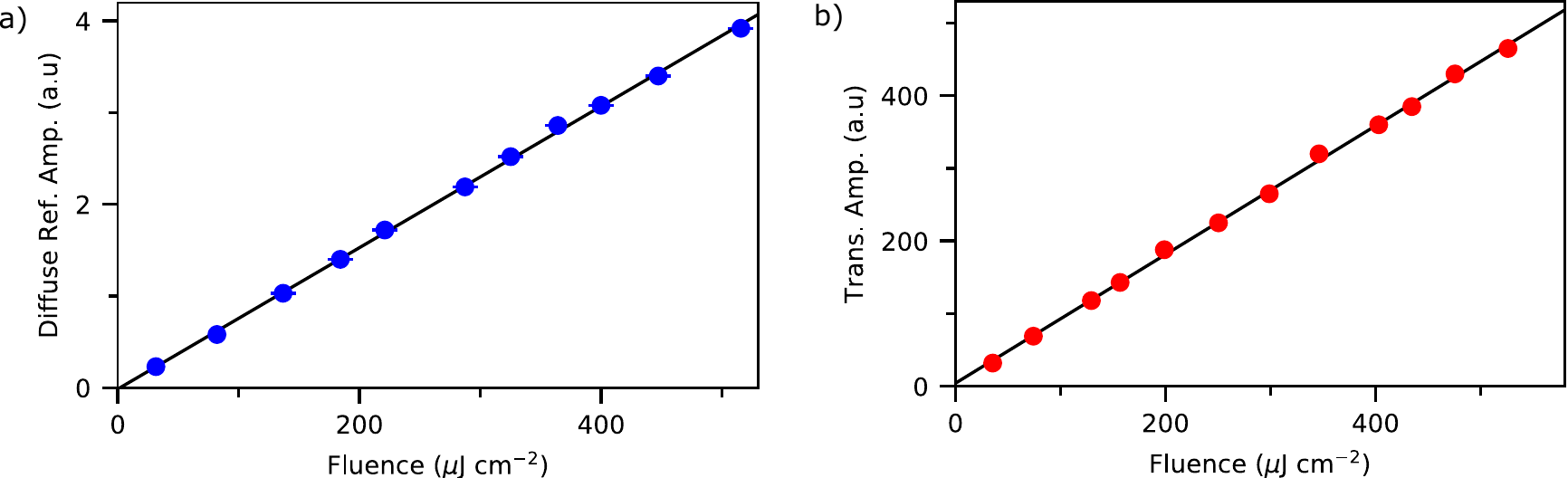}
		\caption{\textbf{Linearity of optical excitation. a} Diffuse reflectance with 400 nm excitation. \textbf{b} Transmitted power with 800 nm excitation.}
		\label{refl_trans}
	\end{center}
\end{figure}
Shown in Supplementary Fig. \ref{refl_trans}a is  the diffuse reflectance for 400 nm excitation. Reflection was chosen rather than transmission because the only transmitted power was through small gaps in the film. Care was taken to avoid the specularly reflected portion of the beam as it was composed partly of the front-surface reflection from the quartz substrate due to micron-scale gaps in the film (see \supl\noteMorph). The signal is linear within the sensitivity of the measurement, which suggests that absorption saturation is not the dominant mechanism of the saturation behavior in the photoconductivity with 400 nm excitation. 
\par

Shown in Supplementary Fig. \ref{refl_trans}b is the transmitted power with 800 nm excitation. Here, the transmitted power is significantly larger than can be accounted for by gaps in the film so transmission mode was chosen. Again, the signal is linear with incident power through the whole range of fluence. The decrease in transmission expected for 2-photon absorption is not seen, which supports our conclusion of defect-tail-to-band absorption as the dominant mechanism of photoconductivity with 800 nm excitation.
\clearpage

\section{\supl\noteModDep}
\begin{figure}[htb]
	\begin{center}
		\includegraphics{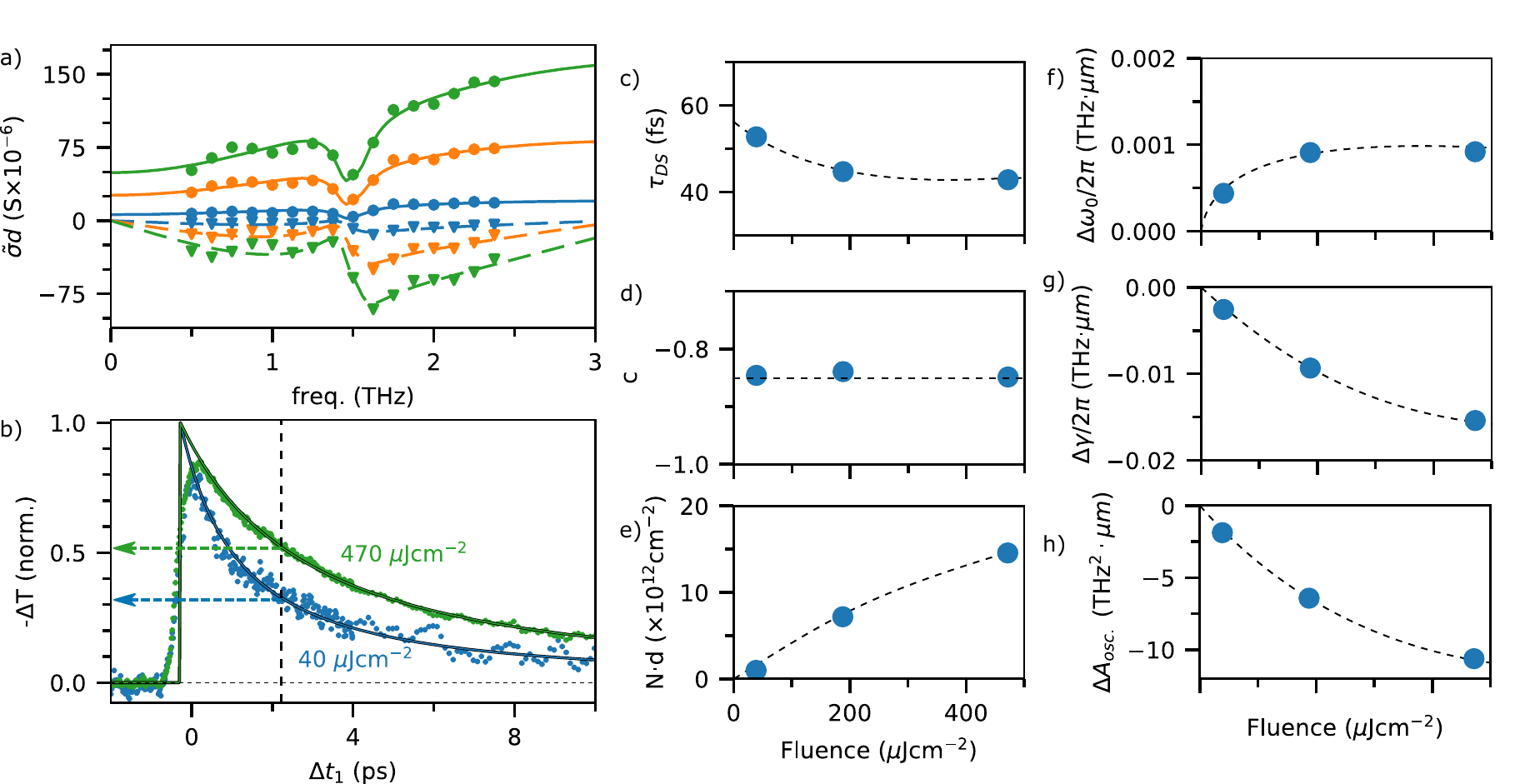}
		\caption{\textbf{Fluence dependent THz photoconductivity. a} THz photoconductivity spectrum with 40 $\mu J\,cm^{-2}$ (blue), 190 $\mu J\,cm^{-2}$ (orange), and 470 $\mu J\,cm^{-2}$ (green) photoexcitation fluence. The circles and triangles represent the measured $\sigma_1$ and $\sigma_2$, respectively, while the solid lines are fits to the model described in the text. The \textbf{b} Time-decay of photoconductivity for 2 of the excitation fluences in \textbf{a}. The vertical dashed line indicates the time delay of the measurements in \textbf{a}. \textbf{c} to \textbf{e} Excitation fluence dependent Drude-Smith scattering time, localization parameter, and sheet density, and \textbf{f} to \textbf{h} Lorentz-oscillator center frequency, damping, and amplitude. The dashed lines are drawn as a guide to the eye.}
		\label{powDep}
	\end{center}
\end{figure}
Supplementary Fig. \ref{powDep} contains a study of the fluence dependence of the THz photoconductivity. The conductivity was acquired 2.2 ps after the peak of the differential transmission for 40 $\mu J\,cm^{-2}$, 190 $\mu J\,cm^{-2}$, and 470 $\mu J\,cm^{-2}$, as shown in Supplementary Fig. \ref{powDep}a. By deconvoluting the bi-exponential decay using a Gaussian response function, we estimate that up to 50\% of the carriers have decayed by this time at at 470 $\mu J\,cm^{-2}$ (similar for 190 $\mu J\,cm^{-2}$) and up to 70\% for 40 $\mu J\,cm^{-2}$, as seen in Supplementary Fig. \ref{powDep}b. 
\par

The fit parameters extracted from the model are detailed in Supplementary Fig. \ref{powDep}c-h. The dashed lines are lines drawn as a guide to the eye. From low to high fluence, we see a slight reduction in scattering time, implying a reduction in mobility along the double-helix axis axis from 53$\, cm^2V^{-1}s^{-1}$ to 42$\, cm^2V^{-1}s^{-1}$. The localization parameter, $c$, remains essentially unchanged, while the charge density increases sub-linearly with fluence similarly to the peak -$\Delta$T, as seen in Fig. 5a of the main text.
\par

Using the density extracted from the Drude-Smith fits and accounting for the fraction of carriers that have decayed as indicated previously, we find that an areal density of electron-hole pairs is 1.4$\times 10^{12}cm^{-2}$, 7.2$\times 10^{12}cm^{-2}$, and 14$\times 10^{12}cm^{-2}$, for incident photon fluxes of 8$\times 10^{13}cm^{-2}$, 38$\times 10^{13}cm^{-2}$, and 95$\times 10^{13}cm^{-2}$, respectively. This suggests a photocarrier generation efficiency of 1.8\%, 3.8\%, and 3.1\% compared to the incident photon flux for the low, medium, and high fluence measurements, respectively. There are a number of factors, such as reflection losses and transmission (which is small for 400 nm excitation), that contribute to this low efficiency. Importantly, trapping on timescales faster than the 0.4 ps response time of the system cannot be accounted for by our deconvolution procedure. Moreover, we are injecting carriers high into the conduction bands where the effective mass approximation breaks down. Quantitatively accounting for these effects is challenging and beyond the scope of this study, however, we can say that the 3.8\% (in the medium fluence case) is a lower bound on the photocarrier generation efficiency.
\par

As an alternative to the Drude-Smith conductivity, the plasmon model has been used to describe the THz conductivity of nanowires \cite{joyce_electronic_2013}. The plasmon model is characterized by a Drude-Lorentz conductivity,
\begin{equation}
\Delta\sigma=\frac{A_D}{1-i\omega\tau_{Plas}}+\frac{i\omega A_L}{\omega_0^2-\omega^2-i\omega/\tau_{Plas}},
\label{eqDrLor}
\end{equation}
where $A_D$, $A_L$, $\tau_{Plas}$, and $\omega_0$ are the Drude amplitude, Lorentz amplitude, carrier scattering time, and resonant frequency, respectively. The Drude and Lorentz weights are proportional to carrier density contributing to the free carrier and plasmonic responses, respectively. In early studies of nanowires using TRTS, the presence of a free carrier (Drude) response in nanowire systems was originally attributed to transport along the longitudinal axis, where the resonant frequency is predicted to be quite low, while the plasmonic (Lorentzian) response results from a transverse resonance \cite{parkinson_transient_2007}. More recently it has been suggested that the Drude component instead results from percolation pathways in the thin film \cite{kuzel_terahertz_2014}. Equation \ref{eqDrLor} is essentially equivalent to the generalized effective-medium theory of reference \cite{kuzel_terahertz_2014} with a Drude conductivity as the differential conductivity of the nanowire.
\par

\begin{figure}[h]
	\begin{center}
		\includegraphics{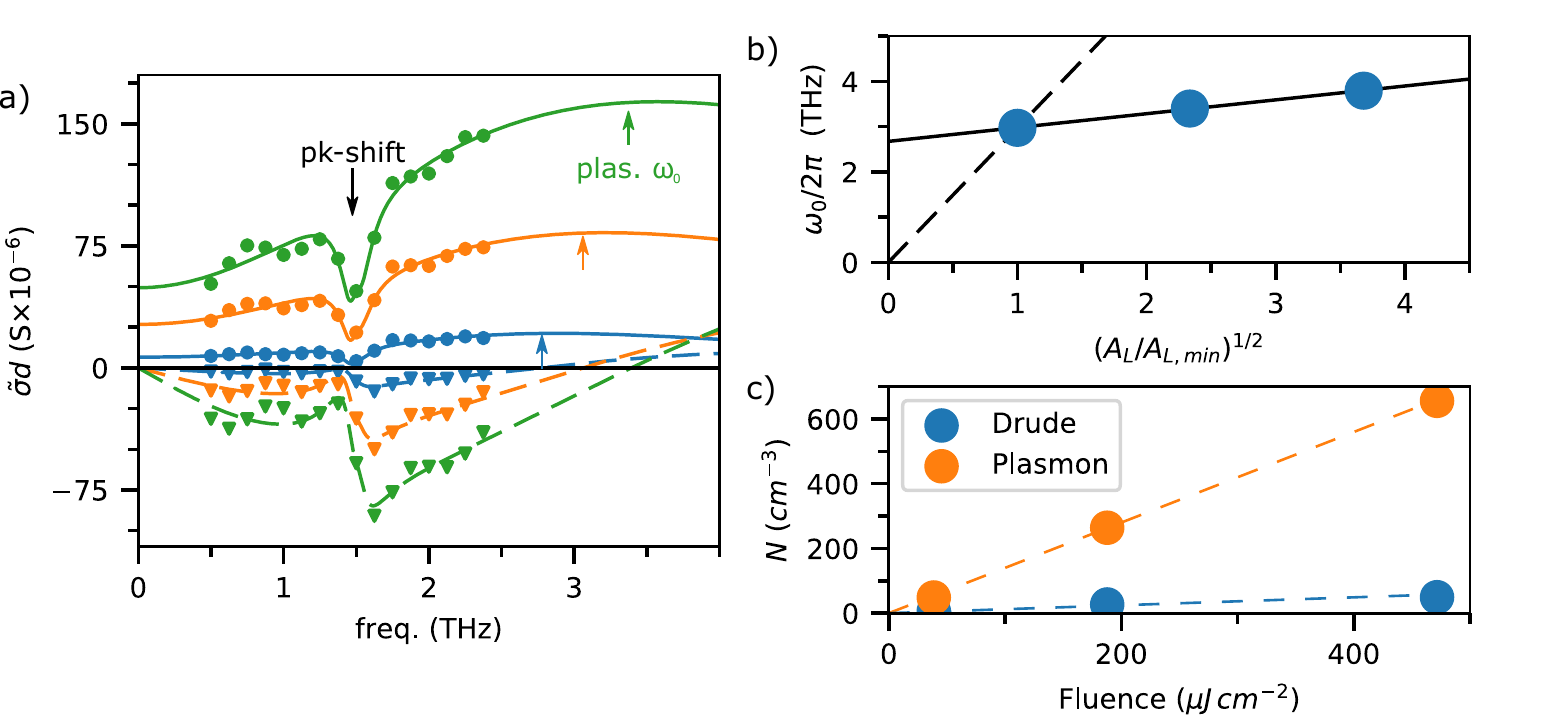}
		\caption{\textbf{Plasmon model fitting. a} THz conductivity spectra for the same data as Supplementary Fig. \ref{powDep} fit using a Drude-Lorentz model with a peak-shift term to account for differential lineshape of mode 2 after photoexcitation. The colored arrows indicate the resonant frequency extracted from the Drude-Lorentz model while the black arrow indicates the peak-shift feature. \textbf{b} Relationship between resonant frequency and square root of the oscillator amplitude normalized to the minimum value. The solid line is a linear fit while the dashed line shows the scaling predicted by the plasmon model. \textbf{c} Extracted density of the Drude and Lorentz parts of the fits as a function of excitation fluence.}
		\label{plasmonFits}
	\end{center}
\end{figure}

In Supplementary Fig. \ref{plasmonFits}a we show fits to the experimental spectra using the Drude-Lorentz model, which shows excellent agreement with data, similarly to the Drude-Smith model. Unfortunately, as a result of the excellent agreement we are not able to rule out either model due to fit quality alone, which is somewhat expected as the two fit functions produce virtually identical spectra when the Drude-Smith localization parameter is close to -1. However, we note that an important feature of the plasmon model is that the resonant frequency scales with the bound-charge density according to,
\begin{equation}
\omega_0^2=\frac{g\,N_fe^2}{\epsilon_r\epsilon_0m^*},
\label{plasmonScaling}
\end{equation}
where $\epsilon_r$ is the dielectric function of the host medium, generally taken to be 1 in the low density limit, $N_f$ is the free carrier density in the nanowire, $m^*$ is the effective mass, and $g$ is a geometric constant \cite{myroshnychenko_modelling_2008}. In our work, there is large uncertainty in the absolute density so we cannot directly verify eq. \ref{plasmonScaling}. Instead, we note that the ratio of resonant frequency to square root of density should be a constant, \textit{i.e.},
\begin{equation}
\frac{\omega_{0}(N_1)}{\omega_{0}(N_2)}=\sqrt{\frac{N_1}{N_2}}=\sqrt{\frac{A_{L1}}{A_{L2}}},
\label{plasmonRatio}
\end{equation}
where $\omega_{0}$ is that frequency at density $N_i$. $A_i$ is the amplitude, which is proportional to the density in the Lorentz model.
\par

\begin{figure}[htb]
	\begin{center}
		\includegraphics{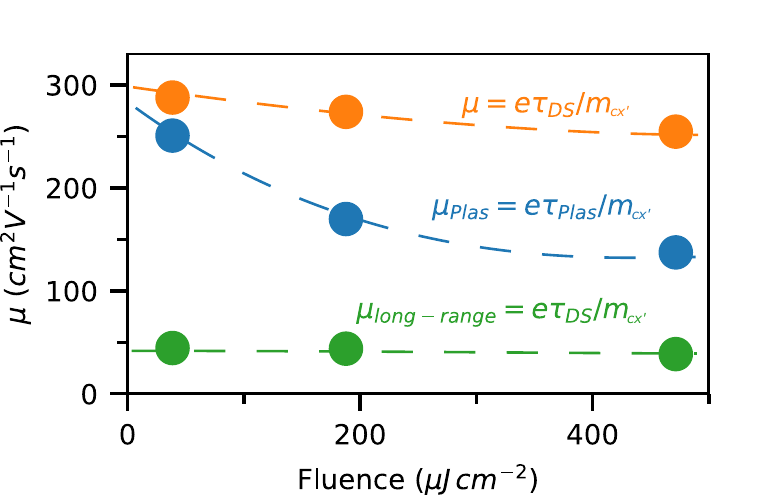}
		\caption{\textbf{Comparison of mobilities for Drude-Lorentz and Drude-Smith fits.} Orange curve: microscopic Drude-Smith mobility calculated using the scattering time without the 1+$c$ factor. Blue: Drude-Lorentz mobility calculated from the scattering rate. Green: Drude-Smith mobility calculated from the scattering time and including the 1+$c$ factor. }
		\label{mobilityComparison}
	\end{center}
\end{figure}

Plotted in Supplementary Fig. \ref{plasmonFits}b is the resonant frequency as a function of $(A_L/A_{L,1})^{1/2}$, where $A_{L,1}$ is the oscillator amplitude extracted from the 40 $\mu J\,cm^{-2}$ spectrum. Based on the analysis of eq. \ref{plasmonRatio}, if the plasmon model applies to this system we expect a linear relationship with close to zero intercept, due to the low intrinsic carrier concentration, and a slope given by $\omega_0(A_{L,min})$. Moreover, while it is possible that background doping of the nanowires could lead to a resonant-frequency scaling similar to what is observed here, this would require a large background conductivity that is inconsistent with conductance measurements\cite{pfister_inorganic_2016} and not observed in our TDS measurements. We therefore conclude that the plasmon model does not apply to our data. 
\par

Despite the fact that the resonant frequency does not show the correct scaling expected from the plasmon model, it is still interesting to analyze the carrier mobility extracted from this model,
\begin{equation}
\mu_{Plas}=\frac{e\tau_{Plas}}{m},
\end{equation}
to see how it compares to that of the Drude-Smith model. Here, $m$ is the effective in any direction as disussed in the main text. Shown in Supplementary Fig. \ref{mobilityComparison} is the mobility from the Drude-Lorentz and Drude-Smith fits using the electron effective mass along the double-helix axis. The low and high mobilities from the Drude-Smith model appear to lie above and below the plasmon model mobility, respectively, at all fluences. Both the high and low mobility from the Drude-Smith model decrease with fluence, with a larger reduction seen in the high mobility. The Drude-Lorentz mobility shows the most dramatic change with fluence and at low fluence tends towards the value of the high mobility from the Drude-Smith model. The general trend of reduced mobility at high fluence is often observed in semiconductors \cite{mics_density-dependent_2013}.
\par

\begin{figure}[htb]
	\begin{center}
		\includegraphics{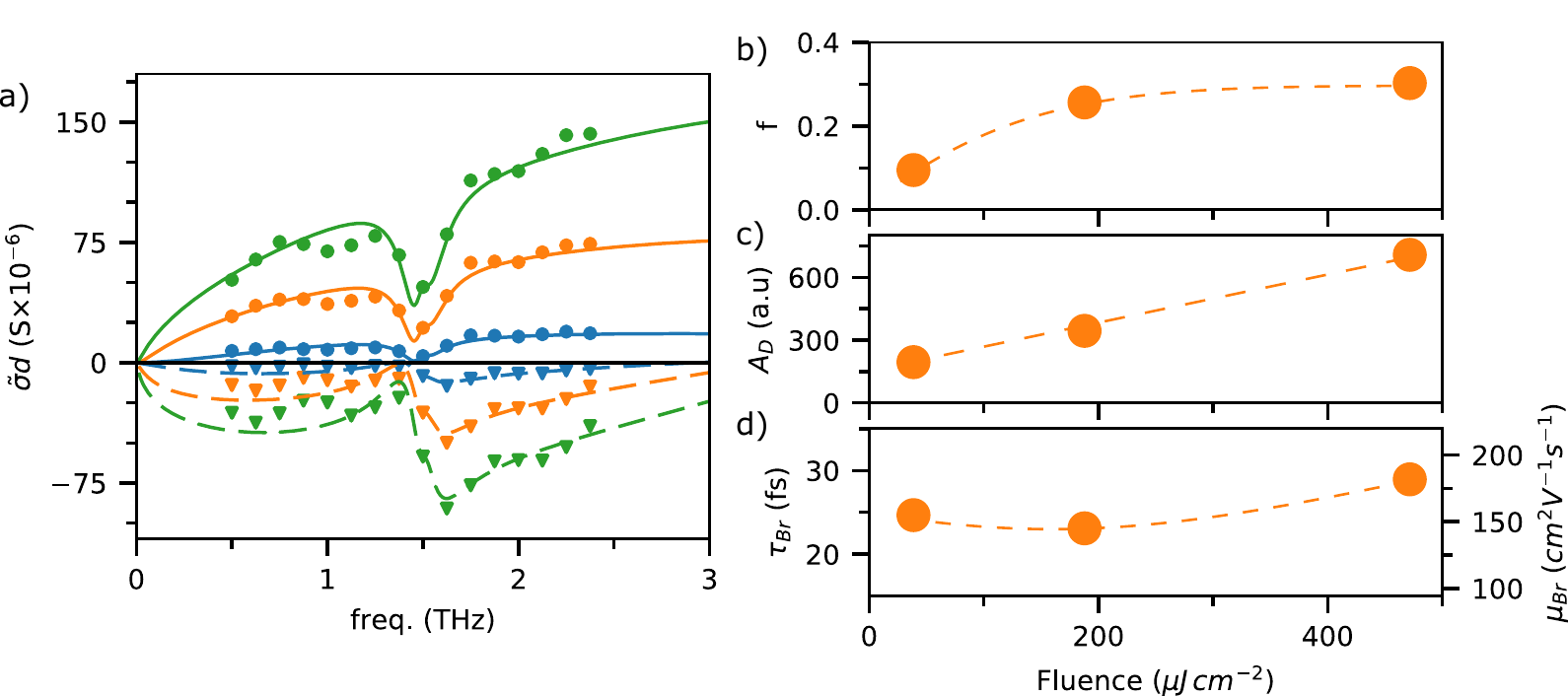}
		\caption{\textbf{Bruggeman model fitting. a} THz conductivity spectra for the same data as Supplementary Fig. \ref{powDep} fit using a Drude conductivity within the Bruggeman effective-medium approach along with a peak-shift term to account for differential lineshape of mode 2 after photoexcitation. \textbf{b}, \textbf{c}, and \textbf{d}, fluence dependence of the filling fraction, Drude amplitude, and scattering time/mobility extracted from the fitting in \textbf{a}. The dashed lines in \textbf{b}-\textbf{d} are drawn as guides to the eye}
		\label{BruggemanFits}
	\end{center}
\end{figure}

It is not surprising that the resonant frequency does not show the correct scaling for the plasmon model. The plasmon model applies strictly to the case of isolated nanowires and therefore only applies in the low-filling-fraction regime, as with Maxwell Garnett effective medium theory \cite{kuzel_terahertz_2014, markel_introduction_2016}. In the high-filling-fraction regime, Bruggeman effective-medium theory is often used to describe the conductivity of inhomogeneous systems. 
\par

In Bruggeman theory the effective dielectric function, $\epsilon_{eff}$, is related to the inclusion dielectric function, $\epsilon_{inc}$, by the equation,
\begin{equation}
f\frac{\epsilon_{inc}-\epsilon_{BG}}{\epsilon_{inc}+2\epsilon_{BG}}+(1-f)\frac{1-\epsilon_{BG}}{1+2\epsilon_{BG}}=0,
\end{equation}
where $f$ is the filling fraction of the inclusion \cite{markel_introduction_2016}. Shown in Supplementary Fig. \ref{BruggemanFits}a are fits to the fluence-dependent conductivity in SnIP using a Drude dielectric function input to the Bruggeman effective-medium theory along with a peak-shift term. Note that the Drude conductivity is given by the first term in the right-hand side of eq. \ref{eqDrLor} and we use the relationship between the optical conductivity and dielectric function (see Methods from the main text) to convert the conductivity to the dielectric function and vice-versa.
\par

While the Bruggeman fit quality is reasonable, both the Drude-Lorentz+peak-shift or Drude-Smith+peak-shift fits are of higher quality, especially in the low frequency region of the imaginary conductivity. In addition to the poor fit quality, there are several features of the fitting that do not make sense physically. First, we note that the filling fraction, shown in Supplementary Fig. \ref{BruggemanFits}b, must be adjusted independently for each fluence. Second, the Drude amplitude, shown in Supplementary Fig. \ref{BruggemanFits}c, does not extrapolate to zero in the low fluence regime. This is unusual as the Drude amplitude is proportional to the density of photoexcited electron-hole pairs, which should go to zero at low fluence. Finally, we see that the mobility, shown in Supplementary Fig. \ref{BruggemanFits}d, is highest at the highest fluence, which is opposite to what is expected in most materials. In Supplementary Fig. \ref{BruggemanFits}d, we also plot the Bruggeman scattering time, $\tau_{Br}$, which is the Drude scattering time extracted from the Bruggeman fits, and the Bruggeman mobility, $\mu_{Br}=e\,\tau_{Br}/m$.
\par

To summarize, both the Drude-Smith and Drude-Lorentz models yield excellent fits to the data, while the Bruggeman fits with a Drude conductivity fail in the low frequency regime. Additionally, the Bruggeman model yields a filling fraction, Drude amplitude, and mobility that show unusual scaling with fluence. Most interestingly, the mobility from the Drude-Lorentz and Bruggeman models lie between the higher and lower mobility from the Drude-Smith model. We see that the Drude-Lorentz mobility approaches the higher value of the Drude-Smith mobility at low fluence, which leads us to speculate that the high value from the Drude-Smith model provides an upper bound on the carrier mobility in SnIP.
\clearpage

\section{\supl\noteEffM}
\begin{figure}[htb]
	\begin{center}
		\includegraphics[scale=0.8]{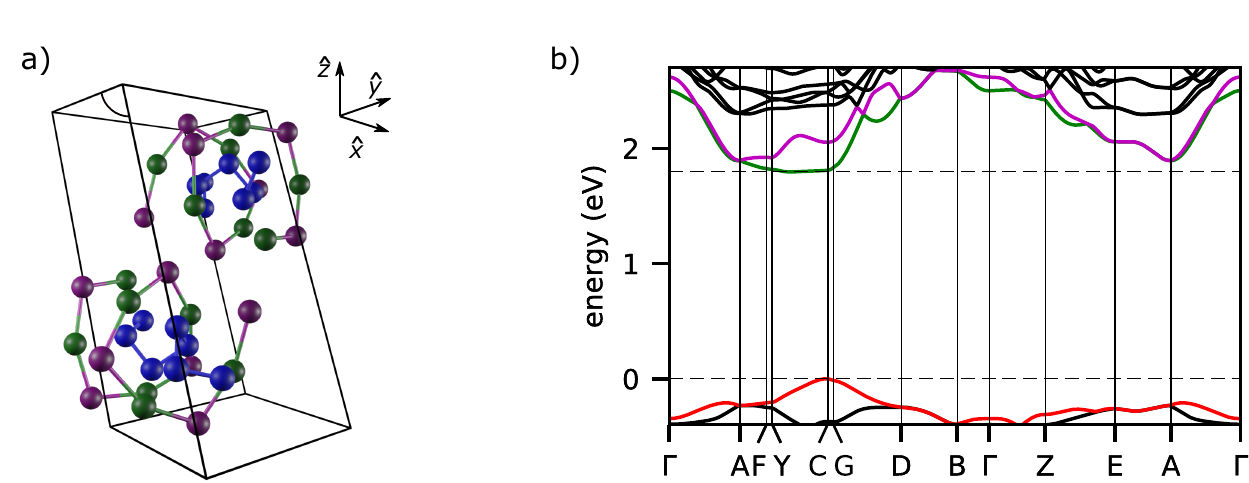}
		\caption{\textbf{SnIP unit cell and band structure. a} SnIP unit cell with the Cartesian axes indicated. \textbf{b} Band structure of several bands near the Fermi level. The constant energy surfaces of the red (highest valence), green (lowest conduction) and purple (second lowest conduction) bands are studied in this note.}
		\label{bands}
	\end{center}
\end{figure}
In this note we provide a more detailed study of the electronic structure of the bands near the Fermi level and a study of the range of validity of the effective mass approximation. In Supplementary Fig. \ref{bands}a, we show the unit cell of SnIP and in Supplementary Fig. \ref{bands}b is the band structure of several bands near the Fermi level. We focus on only the 2 lowest conduction bands (green and purple in Supplementary Fig. \ref{bands}b) and the highest valence band (red in Supplementary Fig. \ref{bands}b).
\par

In Supplementary Fig. \ref{condCES1} we show the constant energy surfaces (CES) for 3 energies above the first conduction band edge along with the equivalent surfaces in the effective-mass approximation (EMA). We see that for 5 and 15 meV above the band edge, the EMA provides good agreement in the $\hat{x}'$ and $\hat{z}'$ directions and provides reasonable agreement for 35 meV. On the other hand, in the $\hat{y}$ direction the EMA already deviates at 5 meV. In principle, this should be taken into account in calculating the average effective mass used in the Drude-Smith model. However, the $\hat{y}$ direction is the heaviest mass (2.0$m_e$ \textit{vs} 0.28$m_e$ and 0.51$m_e$, respectively) and therefore contributes the least to conductivity. A more detailed consideration of the dispersion should therefore only introduce small corrections to the density extracted from TRTS and would not affect our interpretations. We also see that in the $\hat{y}$ direction the surfaces become cylindrical over a large range of the Brillouin zone.
\par 

\begin{figure}[htb]
	\begin{center}
		\includegraphics[scale=0.8]{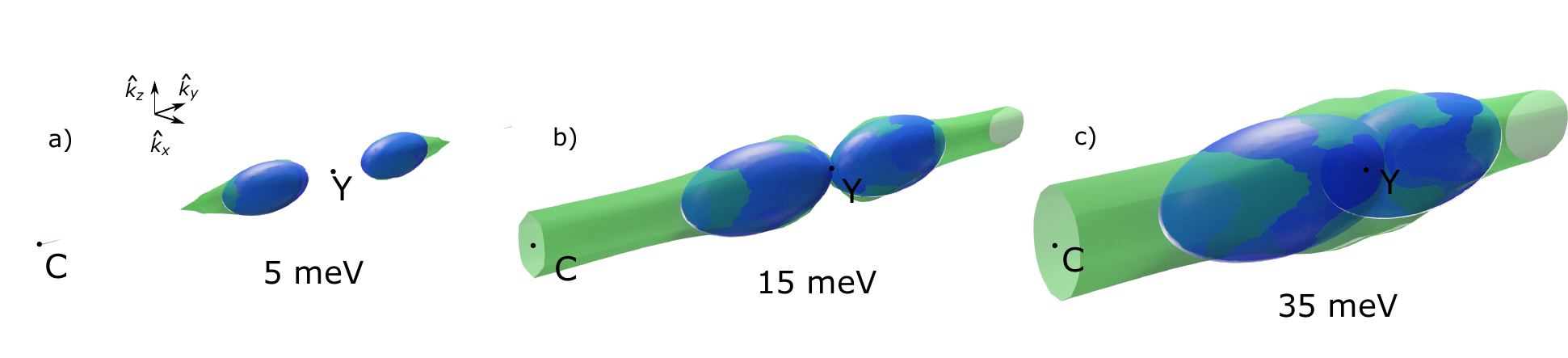}
		\caption{\textbf{Validity of effective-mass approximation in the conduction band. a-c} Conduction band CES for 5, 15, and 35 meV above the lowest-lying conduction band minimum. The blue surface indicates the EFA while the green surface indicates the calculation from DFT.}
		\label{condCES1}
	\end{center}
\end{figure}
We see that this trend continues at higher energies in Supplementary Fig. \ref{condCES2}a-c, where we plot the CES 60, 110, and 160 meV above the band edge. The flatness of these bands implies that transport in the $\hat{y}-$direction essentially freezes out for a large fraction of the states in the conduction band. At 110 meV above the band edge, a new minimum forms at the A point. At 160 meV, the geometry of this band also becomes cylindrical, however, in the $\hat{z}$-direction as opposed to the $\hat{y}$-direction. This implies a strong valley dependence to the electronic structure anisotropy, which could have significant consequences on early time dynamics for carriers that are injected to higher-lying bands, which they can occupy for several picoseconds as they relax to the bottom of the conduction band. 
\par

\begin{figure}[h]
	\begin{center}
		\includegraphics[scale=0.8]{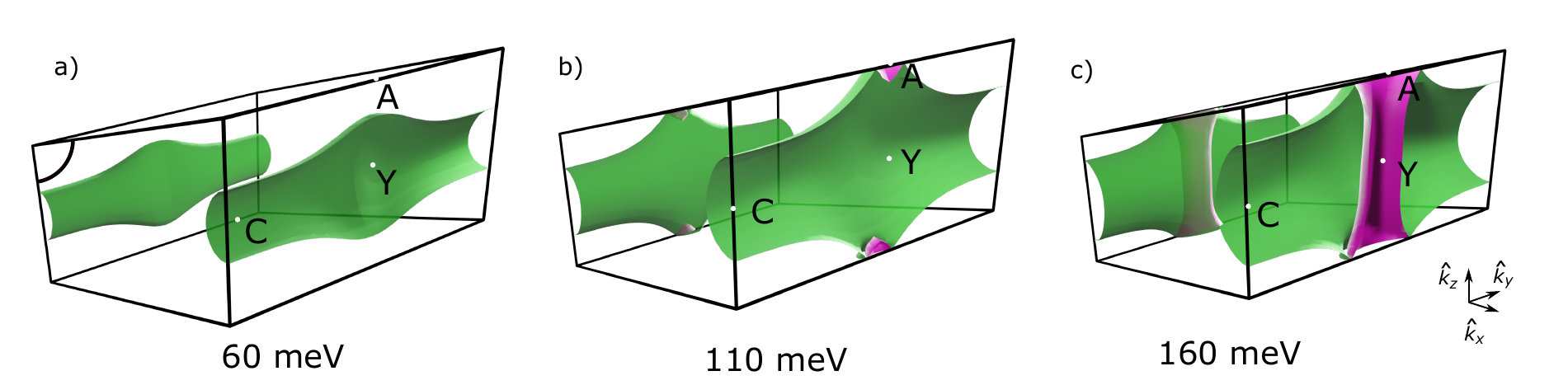}
		\caption{\textbf{Behavior of next-highest conduction band. a-c} CES of the two lowest lying conduction bands for 60, 110, and 160 meV above the lowest-lying conduction band edge. The surfaces are color coded with respect to the bands in Supplementary Fig. \ref{bands}b.}
		\label{condCES2}
	\end{center}
\end{figure}
As opposed to the conduction band, the effective-mass approximation works very well in the valence band. For completeness, in Supplementary Fig. \ref{valCES}a, b we show the CES for the valence band at 10 meV and 40 meV above the band edge, respectively. The valence band plotted is highlighted in red in Supplementary Fig. \ref{bands}b. We see that for 10 meV, the CES from DFT and the EFA approximately overlap. Even at 40 meV, which is significantly higher than the thermal energy, there are only small deviations from the EFA. 
\par

\begin{figure}[htb]
	\begin{center}
		\includegraphics[scale=0.8]{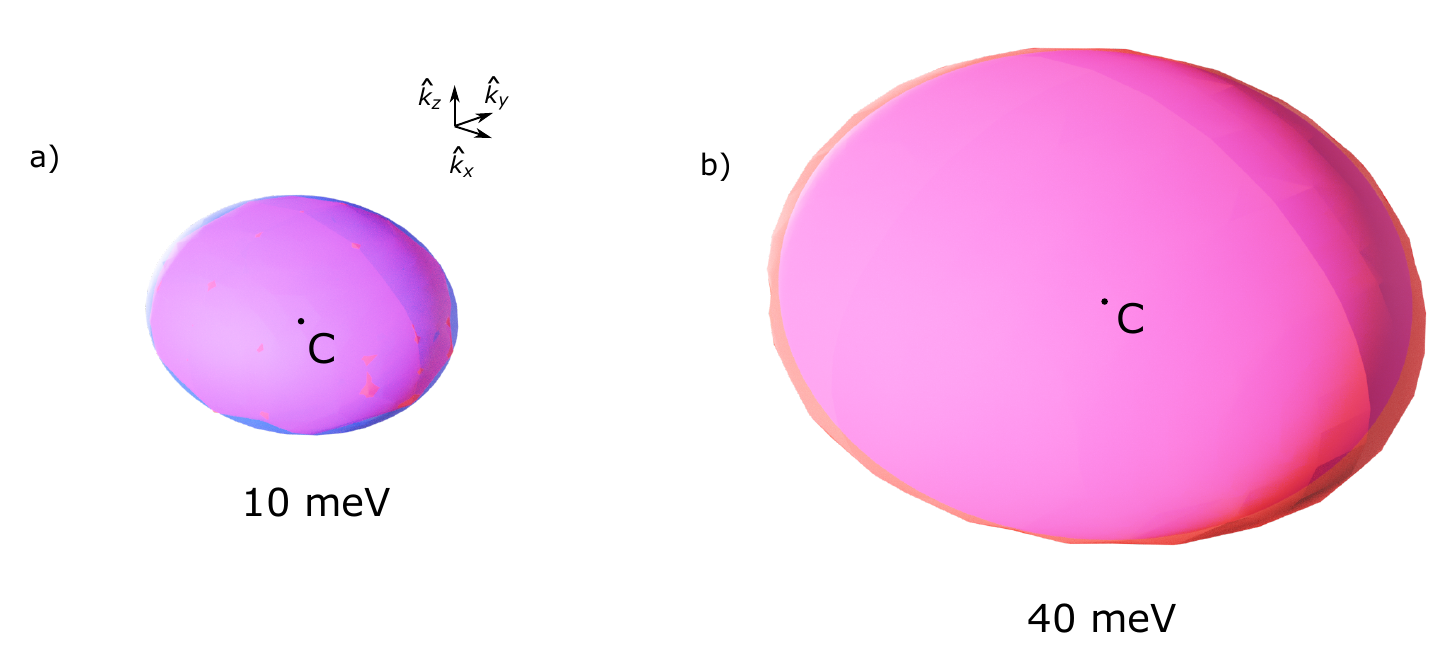}
		\caption{\textbf{Validity of the effective mass approximation in the valence band.} CES of the highest lying valence band (red) along with the CES in the effective mass approximation (blue) for \textbf{a} 10 and \textbf{b} 40 meV below the band maximum.}
		\label{valCES}
	\end{center}
\end{figure}
\clearpage

\section{\supl\noteHess}
As discussed in the main text, the Hessian matrix, which is given by,
\begin{equation*}
H^i_{m,n} = 
\begin{pmatrix}
\frac{\partial^2E^i}{\partial k_x^2} & \frac{\partial^2E^i}{\partial k_x\partial k_y} & \frac{\partial^2E^i}{\partial k_x\partial k_z}\\
\frac{\partial^2E^i}{\partial k_y \partial k_x} & \frac{\partial^2E^i}{\partial k_y^2} & \frac{\partial^2E^i}{\partial k_y\partial k_z}\\
\frac{\partial^2E^i}{\partial k_z\partial k_x} & \frac{\partial^2E^i}{\partial k_z\partial k_y} & \frac{\partial^2E^i}{\partial k_z^2}
\end{pmatrix}
,
\end{equation*}
evaluated at the extrema of band i is related to the inverse effective mass tensor of the corresponding valley,
\begin{equation}
M^i_{mn}=\left(\hbar^2H^i_{mn}|_{k_0}\right)^{-1}.
\end{equation}
The eigenvalues and eigenvectors of the effective mass matrix for one valley of the conduction band are given by,
\begin{equation}
m_1 = 0.28\,m_e,\,\,\hat{x}'_c=
\begin{pmatrix} 
0.982 \\
0 \\
-0.191
\end{pmatrix},
\end{equation}
\begin{equation}
m_2 = 0.51\,m_e,\,\hat{y}'_c=\hat{y}=
\begin{pmatrix} 
0 \\
1 \\
0 
\end{pmatrix}.
\end{equation}
\begin{equation}
m_3 = 2.0\,m_e,\,\,\hat{z}'_c=
\begin{pmatrix} 
0.191 \\
0 \\
-0.982 
\end{pmatrix},
\end{equation}
For the valence band, the eigenvalues and eigenvectors of the effective mass tensor are given by,
\begin{equation}
m_1 = -0.71\,m_e,\,\,\hat{x}'_v=
\begin{pmatrix} 
0.998 \\
0 \\
0.067
\end{pmatrix},
\end{equation}
\begin{equation}
m_2 = -0.33\,m_e,\,\,\hat{y}'_v=\hat{y}
\begin{pmatrix} 
0 \\
1 \\
0 
\end{pmatrix}.
\end{equation}
\begin{equation}
m_3 = -0.66\,m_e,\,\,\hat{z}'_v=
\begin{pmatrix} 
0.067 \\
0 \\
0.998 
\end{pmatrix},
\end{equation}
The effective masses are summarized in Supplementary Table \ref{tab:effMass}. Additionally included is a summary of the directionally-dependent mobility, $\mu$, and the long-range mobility, $\mu_{long-range}$, as defined in the main text, for each direction in the conduction and valence bands. We see that the smallest conduction-band effective mass is along the $\hat{x}'$ direction (approximately along the double-helix axis) while the minimum valence-band effective mass is along the $\hat{z}'$ direction. As discussed in the main text, transport in the $\hat{z}$ direction corresponds to hopping between planes of opposite helicity while transport in the $\hat{y}$ direction corresponds to hopping between planes composed of alternating-helicity strands. The $\hat{y}$-direction mass is larger than the $\hat{x}'$-direction mass, by factor of 2 for the valence band and a factor of 4 for the conduction band. 
\begin{table}[htb]
	
	\begin{center}
		\caption{Anisotropic effective masses and mobilities for the highest-lying valence (lowest-lying conduction) bands, defined as $m^*_v$ ($m^*_c$) and $\mu_v$ ($\mu_c$), respectively. $m_e$ is the free-space electron mass. Also shown is the directionally-dependent long-range mobility, $\mu_{long-range}=\mu(1+c)$. The directions are defined with respect to the axes in Fig. 3b,c of the main text. Mobilities are reported for the scattering time extracted from the data with 400 nm excitation in the main text.}
		\def\arraystretch{1.5}%
		\setlength{\tabcolsep}{10pt}
		\begin{tabular}{|c|c|c|c|}
			\hline\label{tab:effMass}
			&         $\hat{x}'$     &     $\hat{y}$      & $ \hat{z}'$ \\
			\hline
			$m^*_{c}/m_e$                        &        0.28           &    2.0             &    0.51 \\
			$m^*_{v}/m_e$                        &        0.71           &    0.66            &    0.33 \\
			$\mu_{long-range,c}\,\,(cm^2V^{-1}s^{-1})$    &        45			   &    6               &     25      \\
			$\mu_{long-range,v}\,\,(cm^2V^{-1}s^{-1})$    &        18			   &    19              &     38      \\
			$\mu_{c}\,\,(cm^2V^{-1}s^{-1})$ &        281			   &    39              &     154      \\
			$\mu_{v}\,\,(cm^2V^{-1}s^{-1})$ &        111			   &    119             &     238      \\
			
			\hline
		\end{tabular}
	\end{center}
\end{table}
\clearpage

\section{\supl\notePS}
In this note we discuss the peak-shift model of Zhao \textit{et al.} and its range of validity in the context of the vibrational modes in SnIP \cite{zhao_monitoring_2019}. Additionally, we estimate the absolute change in oscillator parameters extracted from fitting the experimental spectra to the peak-shift model. From eq. 14 in the main text, the differential lineshape of the oscillator is given by,
\begin{equation}
\Delta\tilde{\sigma}_{PS}(\omega)=\sum_{i=1}^{3}\frac{\partial\sigma_{LO,2}}{\partial x_i}\Delta x_i,
\label{PSmodel}
\end{equation}
where the parameters, $\Delta x_i$, are the change in center frequency, $\omega_{0,2}$, amplitude, $A_{osc,2}$, and damping, $\gamma_2$, of mode 2 and the oscillator conductivity, $\tilde{\sigma}_{LO,2}$ is given by,
\begin{equation}
\tilde{\sigma}_{LO,2}(\omega)=-i\omega\epsilon_0\cdot\left(\frac{A_{osc,2}}{{\omega}_{0,2}^2-\omega^2+i\omega\gamma_2}+\epsilon_\infty-1\right).
\end{equation}
The differential lineshape is then given explicitly by,
\begin{equation}
\Delta\tilde{\sigma}_{PS}=\frac{i\omega\epsilon_0}{\omega_{0,2}^2-\omega^2-i\gamma_2\omega}\cdot\left(-\Delta A_{osc,2}+
\frac{2\omega_{0,2}A_{osc,2}}{\omega_{0,2}^2-\omega^2- i\gamma_2\omega}\cdot\Delta\omega_{0,2}-
\frac{i\omega A_{osc,2}}{\omega_{0,2}^2-\omega^2-i\gamma_2\omega}\cdot\Delta\gamma_2 \right)
\label{eq:PSmodel}
\end{equation}
\par

Due to the linearity of the conductivity with respect to the oscillator amplitude, the differential form is exact with respect to $\Delta A_{osc,2}$. However, the situation is more complicated for the damping and center freqeuncy. To examine the range of validity we plot the differential lineshape, eq. \ref{eq:PSmodel}, and the exact difference function, eq. 13 from the main text. Shown in Supplementary Fig. \ref{dw0} is a study of the accuracy of the peak-shift model with respect to a change in center frequency. The line width was chosen to represent the relative linewidth in SnIP with respect to the center frequency.
\par
\begin{figure}[h]
	\begin{center}
		\includegraphics[scale=0.9]{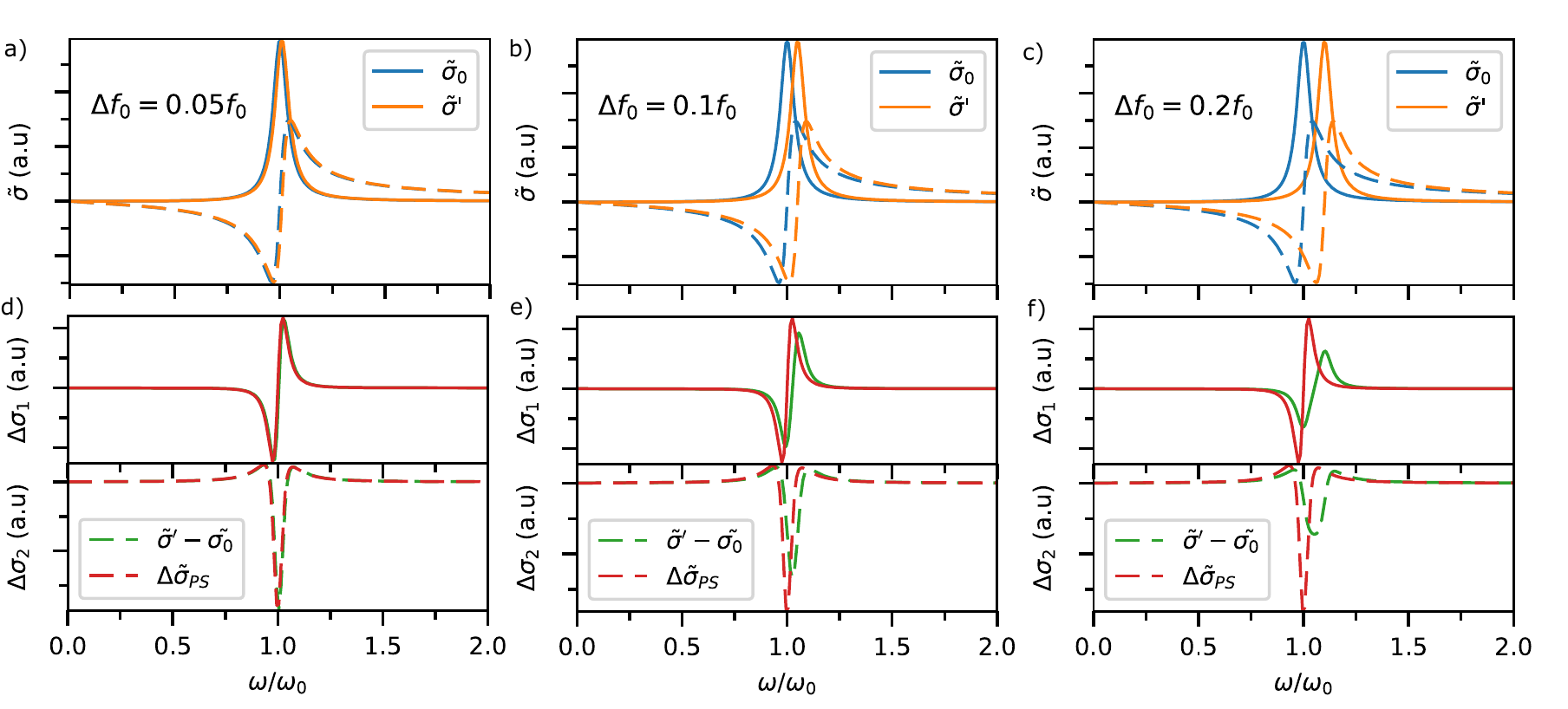}
		\caption{\textbf{Accuracy of peak-shift model with $\Delta\omega_0$} Complex conductivity of an oscillator with (orange) and without (blue) a perturbation to the resonant frequency, $\omega_0$, of \textbf{a} 0.01$\omega_0$, \textbf{b} 0.05$\omega_0$, and \textbf{c} $0.1\omega_0$. \textbf{d-f} the corresponding difference in oscillator conductivity (blue) and differential lineshape from the peak-shift model (orange). }
		\label{dw0}
	\end{center}
\end{figure}

For a 1\% change in $\omega_0$, the peak-shift model accurately reproduces the difference in conductivity, however, at 5\% modulation the peak-shift model already fails to reproduce the correct differential lineshape quantitatively. By 10\% difference, the peak-shift model does not work at all. Somewhat counter intuitively, it is actually the percent change in resonant frequency with respect to the linewidth that is the determining factor for the range of validity rather than with respect to the center frequency itself. For larger values of damping, the range of validity is larger. 
\par

Alternatively, we can examine the accuracy of the peak-shift model with respect to change in damping, as shown in Supplementary Fig. \ref{dgam}. In this case, the range of validity is significantly larger and the differential lineshape shows reasonable quantitative agreement all the way to 20\% change in $\gamma$. Even at 40\% change in $\gamma$, the peak-shift model correctly reproduces the lineshape qualitatively. 
\par

It is also interesting to note the similarity between the lineshape of the $\Delta\gamma$ term and the $\Delta A_{osc}$, which are qualitatively similar. However, we find that the lineshape is significantly narrower than that of the $\Delta A_{osc}$ term and it contains peaks to either side of the center frequency with opposite sign to the main peak, which makes it possible for fitting algorithms to distinguish between the two terms. Furthermore, we note the similarity between the real and imaginary parts of the $\Delta\gamma$ and $\Delta\omega_0$ terms as they are nearly complex conjugates. This coincidence is due to the functional form of the Lorentzian denominator, $\omega_0^2-\omega^2-i\omega\gamma$, where $\omega_0^2$ and $\gamma$ are nearly complex conjugates other than a factor of $\omega$, which does not vary significantly in a narrow spectral range.
\par

\begin{figure}
	\begin{center}
		\includegraphics[scale=0.9]{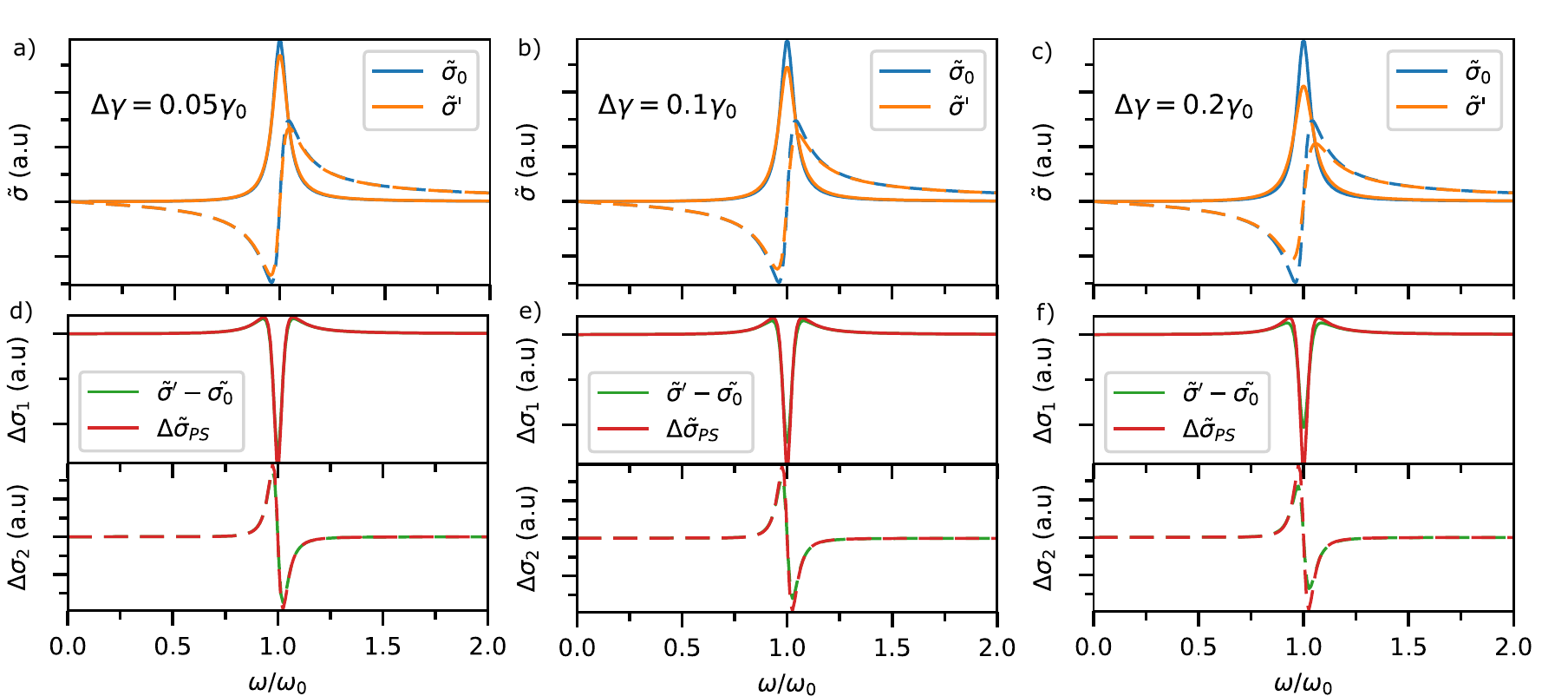}
		\caption{\textbf{Accuracy of peak-shift model with $\Delta\gamma$} Complex conductivity of an oscillator with (orange,$\tilde{\sigma}'$) and without (blue, $\tilde{\sigma_0}$) a perturbation to the damping constant, $\gamma$, of \textbf{a} 0.1$\gamma$, \textbf{b} 0.2$\gamma$, and \textbf{c} 0.4$\gamma$. \textbf{d-f} the corresponding difference in oscillator conductivity (blue) and differential lineshape from the peak-shift model (orange). }
		\label{dgam}
	\end{center}
\end{figure}

Finally, we estimate the absolute change in oscillator parameters. From the fits to experimental data in the main text we measured the products $\Delta\omega_{0,2}/2\pi\cdot d=9\times 10^{-4}\,THz\cdot\mu m$, $\Delta\gamma_2/2\pi\cdot d=-9.1\,\times 10^{-3}\,THz\cdot\mu m$, and $\Delta A_{osc,2}\cdot d=-6.7\,THz^2\cdot\mu m$, where $d$ is the photoexcited film thickness. In the main text we used the estimate of $d=100\, nm$ based on the direct gap and 400 nm excitation wavelength. Using this thickness, a direct calculation of $\Delta A_{osc,2}$ from $\Delta A_{osc,2}\cdot d$ yields a negative oscillator amplitude after photoexcitation. 
\par

This inconsistency could result from several possible origins: An overestimate of the film thickness for TDS, an underestimate of the 400 nm penetration depth, or complications due to the inhomogeneous film morphology. While it is easy to see how the film and penetration depth thickness could result in an underestimate of $A_{osc,2}$ and overestimate of $\Delta A_{osc,2}$, respectively, however, it is not as easy to see how the inhomogeneous morphology would affect the result. To understand this, we note that the the feature sizes are of comparable size to the excitation wavelength. As a result, the excitation pulse does not see an effective medium in the same way as the THz and can excite a film with a larger areal filling fraction for an infinitesimal slice of the film. To be more specific, the photoexcitation pulse can penetrate the air gaps present in the first several layers of the film to excite nanowires deeper in the film.
\par

To take this into account, we suggest that as a first approximation when calculating the absolute change in oscillator amplitude the effective film thickness should be taken as the penetration depth divided by the filling fraction. With 0.35 as an estimate of the volume filling fraction of our films (\supl\noteDist) and with a 100 nm penetration depth we find an absolute change in oscillator amplitude of $\Delta A_{osc,2}=-6.7 THz^2\mu m/(100 nm/0.35)=-23 THz^2$, or approximately 50\% of the absolute oscillator amplitude. Additionally, we find $\Delta\gamma_2/2\pi=3.2\times 10^{-2}$ and $\Delta\omega_{0,2}/2\pi=3\times 10^{-3}$ or a modulation of 20\% and 0.2\%, respectively. To visualize this change, in Supplementary Fig. \ref{absOsc}a, b we plot the oscillator conductivity before and after photoexcitation. In Supplementary Fig. \ref{absOsc}b we also include the Drude-Smith conductivity to see the relative scale. The absolute Drude-Smith conductivity was also calculated using a 100 nm film thickness and 0.35 filling fraction. 

\begin{figure}
	\begin{center}
		\includegraphics[scale=0.9]{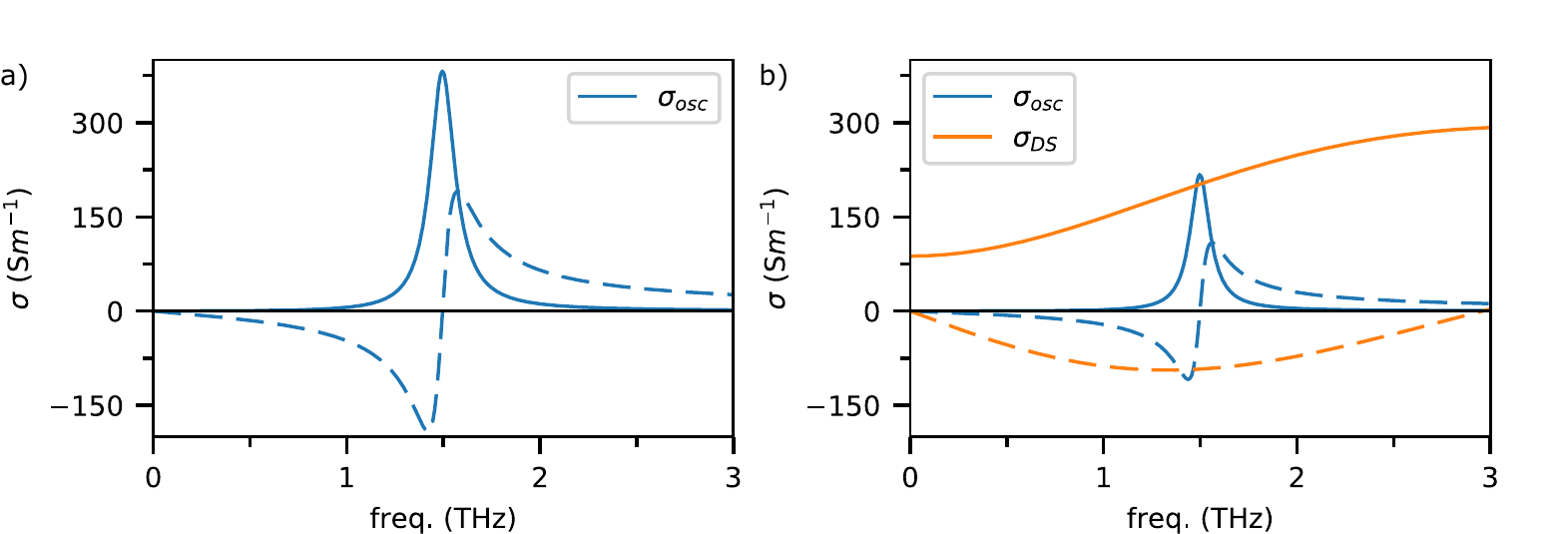}
		\caption{\textbf{Lineshape of oscillator after photoexcitation. a} Real (solid) and imaginary (dashed) oscillator conductivity of mode 2 using the parameters extracted by THz-TDS. \textbf{b} Real (solid) and imaginary (dashed) conductivity of the oscillator (blue) after photoexcitation, using the parameters estimated in this section, along with the Drude-Smith conductivity for reference.}
		\label{absOsc}
	\end{center}
\end{figure}

\clearpage

\section{\supl\noteBiExp}
\begin{figure}[htb]
	\begin{center}
		\includegraphics[scale=0.8]{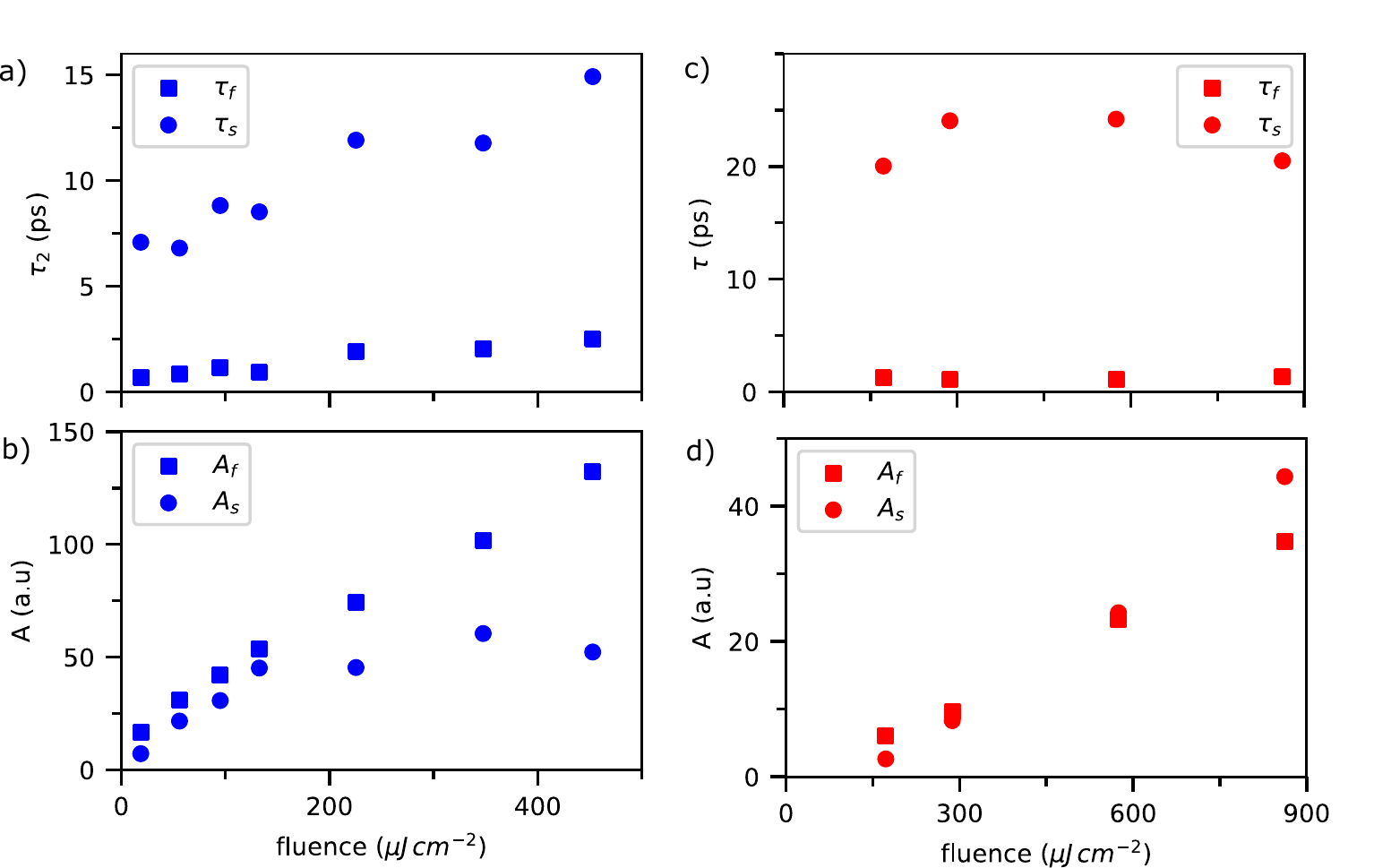}
		\caption{\textbf{Excitation-fluence dependent bi-exponential fit parameters. a} Fast and slow lifetime, $\tau_f$ (squares) and $\tau_s$ (circles), respectively, and \textbf{b} amplitude of the fast (squares) and slow (circles) lifetime for 400 nm excitation. \textbf{c} and \textbf{d} The corresponding lifetimes and amplitudes for 800 nm excitation.}
		\label{valCES}
	\end{center}
\end{figure}

\clearpage

\section{\supl\noteRateMod}
\begin{figure}[h]
	\begin{center}
		\includegraphics[scale=0.9]{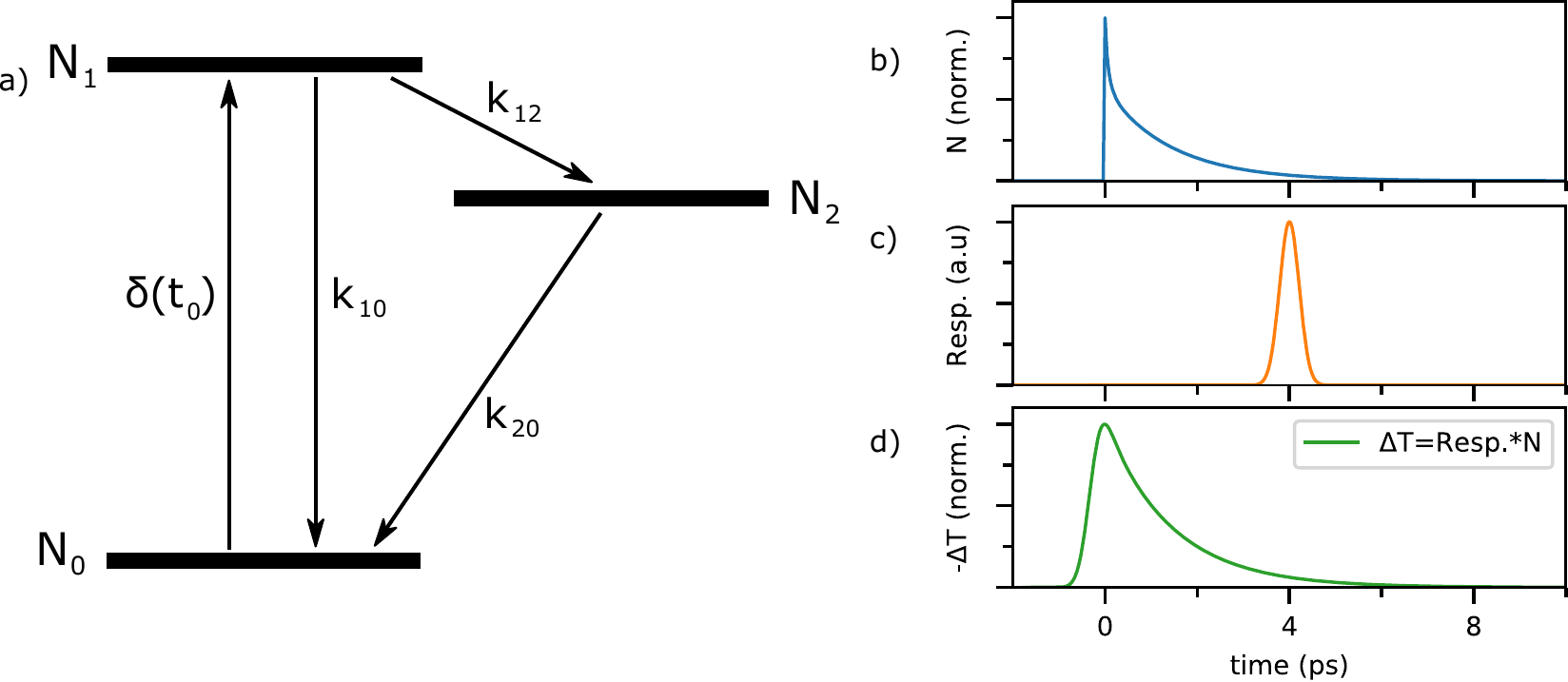}
		\caption{\textbf{Trapping dynamics from rate equations. a} Level diagram showing the conducting state (N$_1$) and non-conducting states (ground state N$_0$ and saturable trap state N$_2$). \textbf{b} Numerical calculation of the system in \textbf{a} with a delta function input. \textbf{c} Gaussian system response function. \textbf{d} Convolution of the system response and numerically calculated relaxation dynamics.}
		\label{rateLevels}
	\end{center}
\end{figure}
In this note, we discuss the use of a rate equation model to help understand the role of saturation of traps in the fluence dependent peak photoconductivity with 800 nm excitation (Fig. 5d in the main text). We use a minimal formulation with a three-level system consisting of a ground state, N$_0$, a conductive state, N$_1$, and a saturable trap state (non-conducting), N$_2$, as picture in Supplementary Fig. \ref{rateLevels}a. Excitation consists of a delta function source at $t=0$ while rates of coupling between each level are parameterized by $k_{ij}$. We note that this formulation is insufficient to reproduce the relaxation dynamics and fit the photoconductivity decay curves quantitatively. This is unsurprising, as we know that diffusion and surface recombination are important. Nevertheless, by parameterizing the observed fast lifetime with k$_{10}$ and the sub-response time trapping with k$_{12}$ we can gain insight into the early time behavior.
\par

The system of equations governing the relaxation is given by,
\begin{align}
\frac{dN_1}{dt}=N_{1,0}\delta(0)-(k_{eff,12}+k_{10})N_1, \\
\frac{dN_2}{dt}=k_{eff,12}N_1-k_{21}N_2, \\
\end{align}
where $N_{1,0}$ is the initial density and we have defined the effective coupling rate,
\begin{equation}
k_{eff,12}=\left(1-\frac{N_2}{N_{2,sat}}\right)k_{12},
\end{equation}
where $N_{2,sat}$ is the density at which all traps are filled \cite{uhd_jepsen_ultrafast_2001}.
\par

Supplementary Fig. \ref{rateLevels}b shows the density as a function of time with  $k_{21}=0$, $k_{12}=8\,ps^{-1}$, $k_{10}=0.7\,ps^{-1}$ and $N_{2,sat}=0.2$ in units where $N=1$ is the density normalized to the maximum fluence with 800 nm excitation (850 $\mu J\,cm^{-2}$). We note that the linearity of the system of equations allows us to work in normalized units in this way without redefining rate constants (lifetimes). The rate $k_{10}$ was taken to match the experimentally observed time-constant at early times, $k_{12}$ and $N_{2,sat}$ were found by fitting the fluence-dependent $\Delta$T signal (Fig. 5d in the main text), and $k_{20}$ was chosen for simplicity. 
\par

To go from the time-dependent density to the differential transmission, we must take into account the system response, $R$, approximated here as a Gaussian (Supplementary Fig. \ref{rateLevels}c), and include a weighting factor between simulation density and differential transmission,
\begin{equation}
\Delta T=\eta N_1 * R,
\end{equation}
where $*$ indicates a convolution and $\eta$ is a linear weighting factor. Deviations from linearity could occur due to a time- or fluence-dependent scattering rate or localization parameter, however, the assumption of linearity is reasonable as a first approximation because the lineshape does not change significantly with time or density (see \supl\noteModDep and Fig. 4 from the main text). The convoluted signal is shown in Supplementary Fig. \ref{rateLevels}d, where we see that the sub-response time trapping is smeared out by the system response. 
\par

\begin{figure}
	\begin{center}
		\includegraphics[scale=0.9]{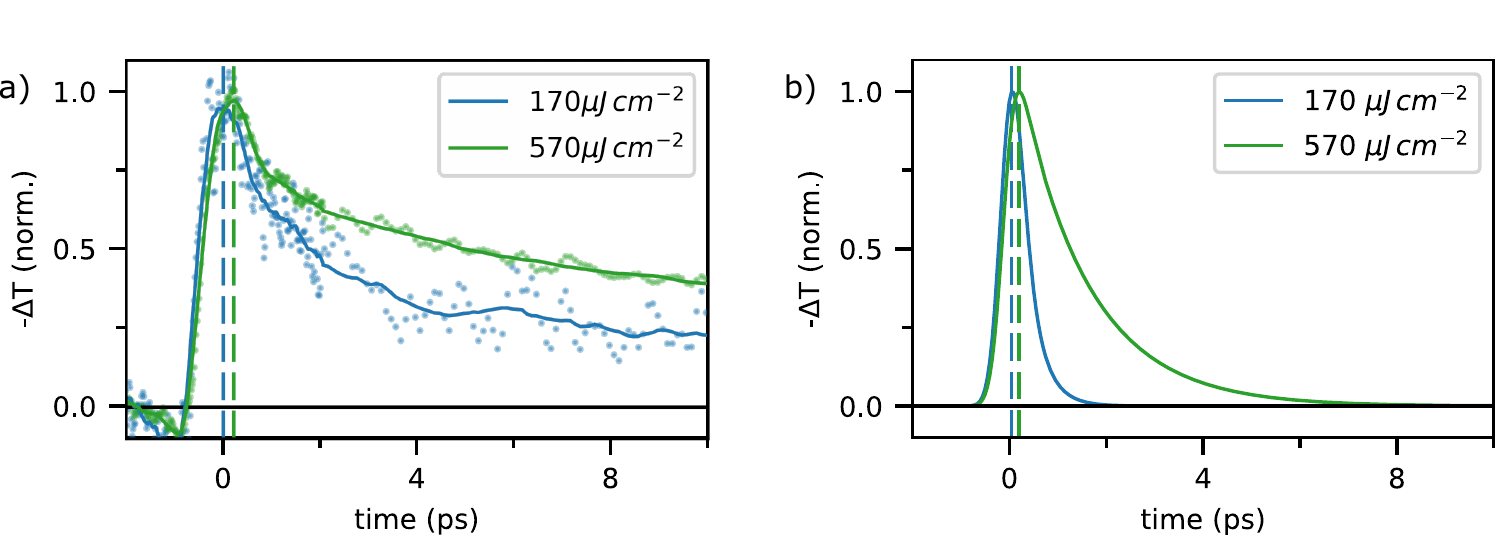}
		\caption{\textbf{Rate equation dynamics \textit{vs} experimentally observed dynamics. a} Experimentally measured $-\Delta$T at low and high fluence, showing a small increase in rise time at high fluence. The solid lines are results from a smoothing algorithm applied to the data. \textbf{b} Numerically calculated density as a function of time convoluted with the Gaussian response of the system, which also shows a slight increase in rise time at high fluence. The vertical dashed lines indicate the time delay of the peak signal. The rate k$_{10}$ is chosen to fit the fast lifetime of the 800 nm photoconductivity decay.}
		\label{rateSims}
	\end{center}
\end{figure}
To further study the behavior of this system of equations, we look at the fluence dependence of the calculated $-\Delta$T in comparison to experimental observation, shown in Supplementary Fig. \ref{rateSims}a, b. As stated previously, we are not able to reproduce the time-dependence of the photoconductive decay quantitatively with this model, as evidenced by the longer-lived component of the experimental curve at both low and high fluence. However, we do see the lifetime enhancement at high fluence and, additionally, we see a small increase in rise time at high fluence, which is consistent with experiment. We note that, especially in the low fluence case, the small signal makes extraction of the rise time quite difficult and it was necessary to use a smoothing algorithm (Savitsky-Golay filter in Scipy). While this makes it more difficult to draw conclusions from the rise time alone, it does provide circumstantial support for the trapping model alongside the excellent quantitative agreement seen in Fig. 5d of the main text.
\clearpage

\section{\supl\noteMatSum}
\begin{figure}[htb]
	\begin{center}
		\includegraphics[scale=1]{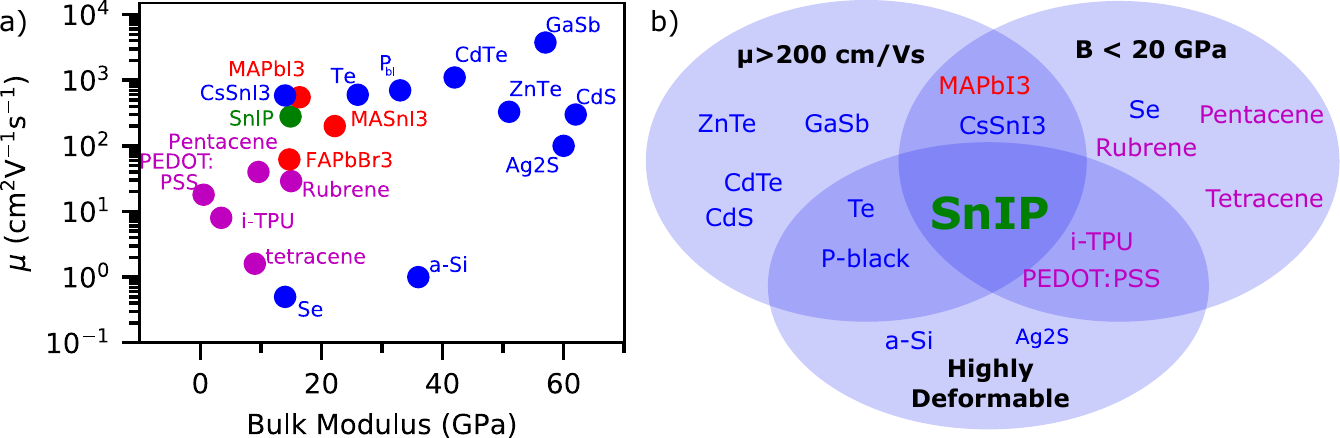}
		\caption{\textbf{Electronic and mechanical properties of common materials. a} Carrier mobilities of several common organic, inorganic, and hybrid organic-inorganic semiconductors with bulk modulus, B, less than 60 GPa. Blue: inorganic, purple: organic, red: hybrid organic/inorganic, green: SnIP (also inorganic). \textbf{b} Illustration of the unique combination of softness, high carrier mobility, and flexibility demonstrated by SnIP, in which individual needles can be bent beyond 90$^\cdot$ reversibly.}
		\label{matSummary}
	\end{center}
\end{figure}
Shown in Supplementary Fig. \ref{matSummary}a is a survey of the carrier mobility versus bulk modulus for a wide variety of semiconductor materials. We chose 60 GPa as an upper limit o the bulk modulus to include several III-V and II-VI semiconductors that are relatively soft compared to their counterparts. We see that SnIP is significantly softer than typical main-group semiconductors, as noted in reference \cite{ott_flexible_2019}. The 60 GPa limit excludes several technologically-relevant high-mobility semiconductors such as GaAs and Si, which are more than 5 times harder than SnIP. We have chosen to include only high mobility organic semiconductors, resulting in the omission of many that are significantly softer than SnIP. 
\par

The high-mobility single-crystal organic semiconductors and the inorganic or hybrid perovskites have a comparable bulk modulus to SnIP, however, these materials are brittle with a low fracture energy and therefore do not share the extreme elastic flexibility demonstrated by SnIP. The high-mobility elastomer i-TPU and the polymer PEDOT:PSS show similar flexibility to SnIP, however, the carrier mobility in these materials, as well as the single-crystal organic semiconductors, is many times lower than SnIP. One interesting material that is both flexibile and has reasonably high mobility is $\alpha$-AgS$_2$, which was recently reported as the first room-temperature ductile semiconductor\cite{shi_room-temperature_2018}. We note, however, that it is significantly harder than SnIP and has mobility 2 times lower.
\par

The most comparable material to SnIP is Tellurium, which, interestingly, is also (single) helical at the atomic scale. It has a high field effect mobility and, moreover, nanoscale tellurium chains are quite flexible. Recent work has led to a great deal of excitement towards applications in nanoelectronics\cite{qin_raman_2020}. Nevertheless, Te is considered a rare earth element and is mildly toxic, in contrast to SnIP which is composed of non-toxic and abundant elements. Furthermore, the bulk modulus in Te is nearly double that of SnIP. In Supplementary Fig. \ref{matSummary}b we qualitatively illustrate the unique combination of flexibility, softness, and high carrier mobility in SnIP that make it attractive for applications in flexible devices. Furthermore, as discussed in main text, there is reason to believe that SnIP could be just the first of many materials that share the nested double-helix structure. It is interesting to speculate that this new class of materials could share the unique combination of mechanical and electronic properties of SnIP. 
\par

Refs. - Bulk modulus: CdTe, GaSb, CdS, ZnTe, and CsSnI$_3$ \cite{kumar_elastic_2015}, MAPbI$_3$ and MASnI$_3$\cite{feng_mechanical_2014}, FAPbBr$_3$\cite{ferreira_elastic_2018}, Se\cite{keller_effect_1977}, for Te\cite{bandyopadhyay_pressure_1999}, pentacene\cite{oehzelt_crystal_2006}, rubrene\cite{wu_strain_2016}, tetracene\cite{oehzelt_crystal_2006}, $\alpha$-Ag$_2$S\cite{wang_effect_2017}, black phosphorus\cite{cartz_effect_1979}, amorphous silicon\cite{liu_mechanically_2019} and SnIP\cite{ott_flexible_2019}. It was more difficult to find reliable values for the bulk modulus of PEDOT:PSS\cite{lang_mechanical_2009} and i-TPU\cite{uddin_enhanced_2017}, which were estimated from the Young's modulus and Poisson ratio. Carrier mobility: GaSb, CdS, and ZnTe \cite{madelung_semiconductors_1996}, CdTe \cite{roessler_cdte_2009}, CsSnI$_3$, MAPbI$_3$, MASnI$_3$, and FAPbI$_3$ \cite{herz_charge-carrier_2017}, Se\cite{bhaskar_mobility_2017}, Te\cite{qin_raman_2020}, pentacene\cite{jurchescu_interface_controlled_2007}, rubrene\cite{yamagishi_high-mobility_2007}, tetracene\cite{reese_high-performance_2006}, $\alpha$-Ag$_2$S\cite{shi_room-temperature_2018}, black phosphorus\cite{rudenko_intrinsic_2016}, amorphous silicon\cite{gao_high_2015}, PEDOT:PSS\cite{wang_high_2018}, i-TPU\cite{park_highly_2019}, and SnIP (this work).
\clearpage 

\section{\supl\noteTemp}
\begin{figure}[htb]
	\begin{center}
		\includegraphics{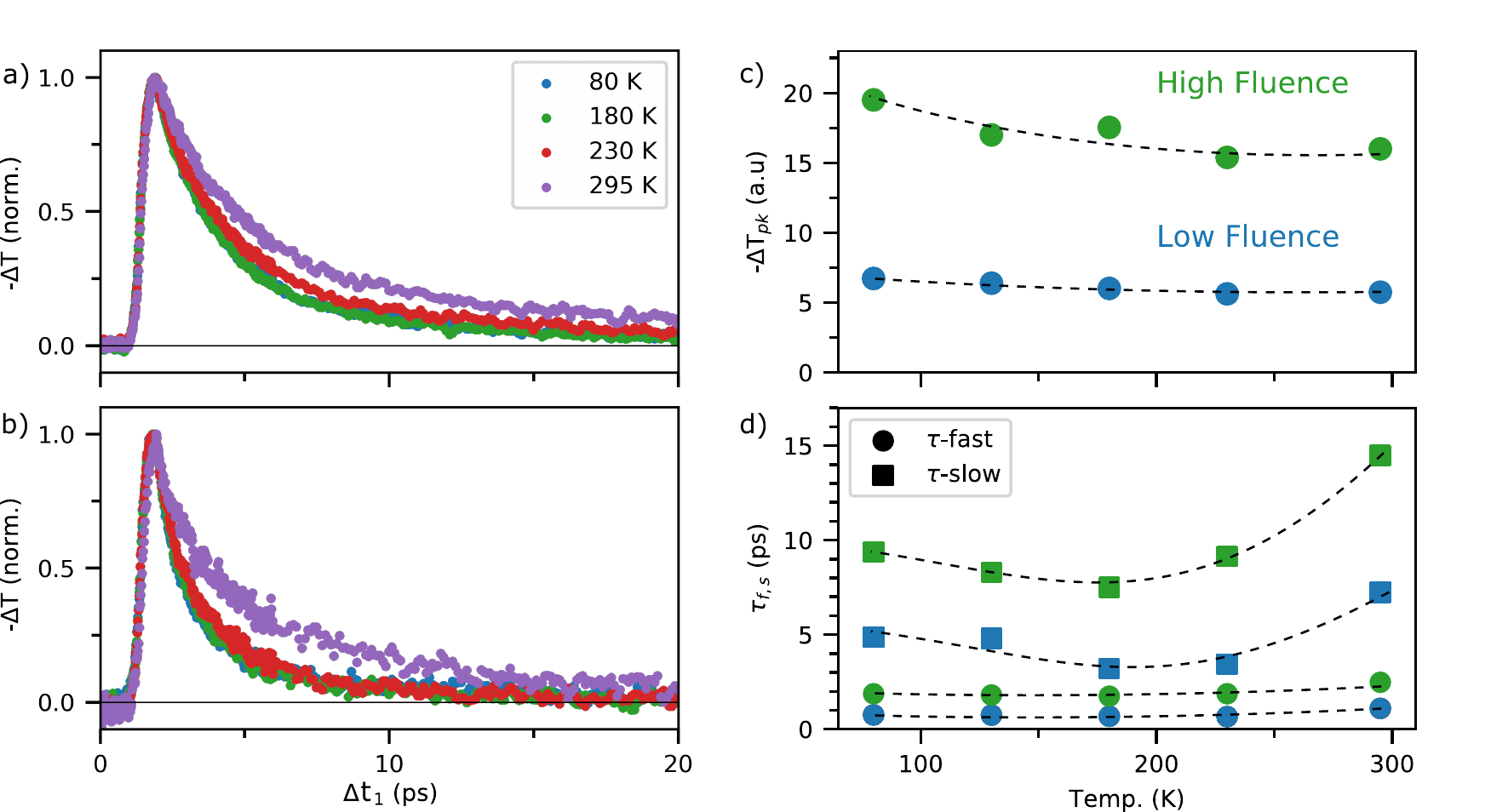}
		\caption{\textbf{Temperature-dependent THz photoconductivity and relaxation dynamics.} Relaxation dynamics with 400 nm photoexcitation for \textbf{a} high fluence ($\sim$440 $\mu J cm^{-2}$) and \textbf{b} low fluence ($\sim$95 $\mu J cm^{-2}$) at 4 temperatures varying from 295 K to 80 K. \textbf{c} Amplitude of differential signal, $-\Delta T_{pk}$, in the low (blue) and high (green) fluence cases. The dashed line is drawn as a guide to the eye. \textbf{d} Variation of the fast (circles) and slow (squares) lifetime for the low fluence (blue) and high fluence (green) cases. Again the dashed lines are drawn as a guide to the eye.}
		\label{temp1}
	\end{center}
\end{figure}
The normalized relaxation dynamics as a function of temperature is shown for high fluence and low fluence in Supplementary Fig. \ref{temp1}a, b, respectively. In general, the lifetime tends to decrease with decreasing temperature for both the high fluence and low fluence case. The largest change occurs between 295 and 230 K. Along with the decreasing limetime, there is in general a small increase in peak differential signal as the temperature is reduced, as seen in Supplementary Fig. \ref{temp1}c. The fast and slow lifetimes from the bi-exponential fits are shown in Supplementary Fig. \ref{temp1}d, where it can be seen that the largest change is in the slow lifetime.
\par

We suggest there are two different contributions to this behavior. For traps with energies comparable to the thermal energy, increasing temperature could lower the trap occupancy due to increased thermal ionization, which would lower the rate of recombination. Alternatively, it is possible that an increased mobility with low temperature could lead to more rapid diffusion to the surface, which would result in a faster lifetime. 
\par

\begin{figure}
	\begin{center}
		\includegraphics[scale=0.9]{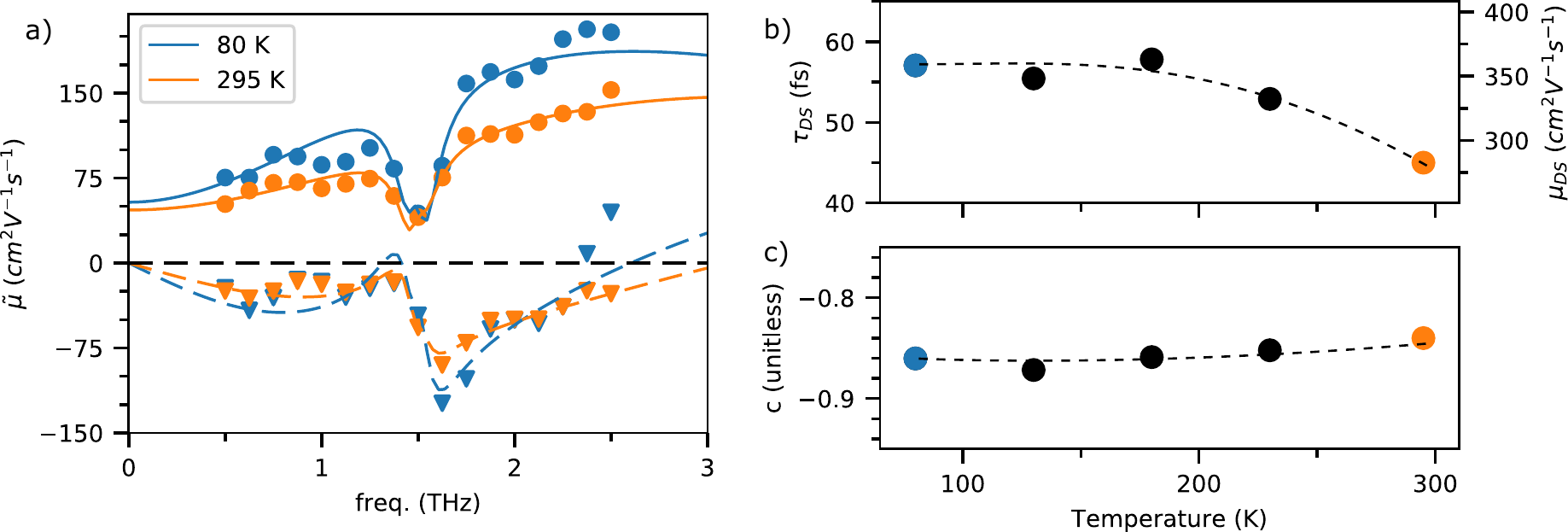}
		\caption{\textbf{Temperature dependent mobility spectrum. a} Mobility spectra with 400 nm excitation at 80 K and 295 K. \textbf{b} and \textbf{c} Temperature dependent scattering time and localization parameter. }
		\label{temp2}
	\end{center}
\end{figure}
Shown in Supplementary Fig. \ref{temp2}a is the mobility spectrum, defined as $\mu(\omega)=\sigma(\omega)\cdot d/N\cdot d$, at 80 K and 295 K. The 80 K spectra shows the same features as at room temperature, namely, the negative imaginary conductivity, increasing real conductivity with frequency, and suppression of the 1.5 THz vibrational mode. The Drude-Smith scattering time and localization parameter are plotted in Supplementary Fig. \ref{temp2}b, c, respectively. The scattering time increases with decreasing temperature with the most dramatic difference between room temperature and 230 K. This initially seems to support the conclusion that increased mobility is the mechanism underlying the reduction in lifetime at low temperature, however, the diffusion constant is also inversely proportional to temperature. The increase in scattering time, with $\tau_{DS,80K}\approx 1.3\tau_{DS,295K}$, is small compared to the reduction 3.7 times reduction in temperature, which points to the ionization of traps as the dominant mechanism for increased lifetime at room temperature relative to low temperature. 
\par

The width of the feature at 1.5 THz also appears narrower at low temperature, which could suggest a slight reduction in line width. Shown in Supplementary Fig. \ref{temp3} is the change in linewidth, $\Delta\gamma_2$, extracted from the Peak-shift fit as a function of temperature, which generally shows a more negative $\Delta\gamma_2$ at lower temperature. However, it appears that the fit quality is not as high at low temperature. We suspect that this is due in part to the changing oscillator parameters; the peak-shift model using the room-temperature parameters will no longer be able to accurately reproduce the differential lineshape if the unexcited lineshape changes with temperature. This could account for the qualitatively different behavior of $\Delta\gamma_2$ between 80 K and 130K. Future temperature-dependent studies will therefore need to perform both TRTS and TDS to obtain a more detailed picture of the dynamics.
\begin{figure}[h]
	\begin{center}
		\includegraphics{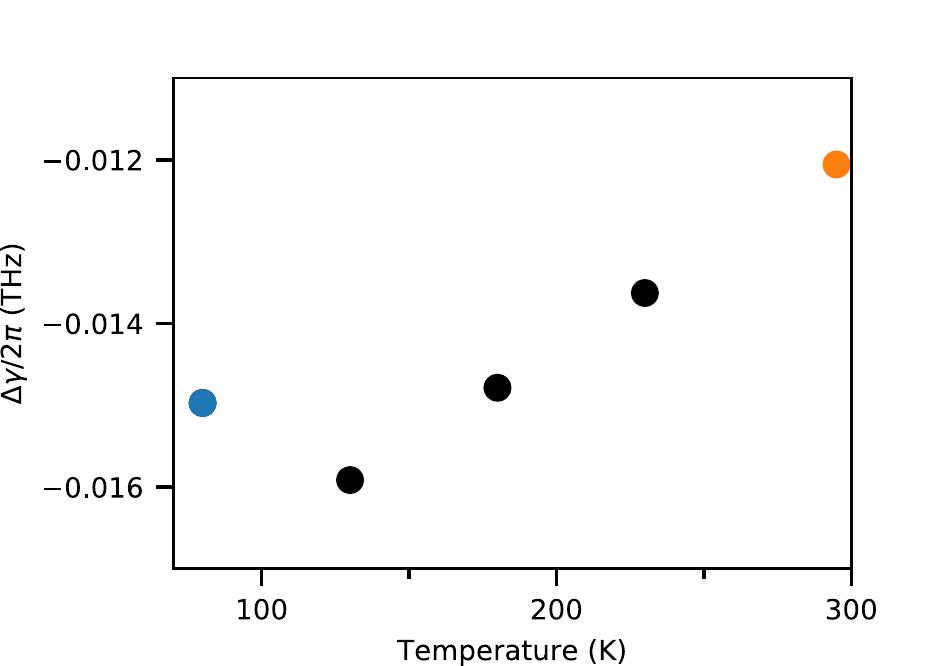}
		\caption{\textbf{Narrowing oscillator with temperature.} $\Delta\gamma_2$ extracted from the temperature dependent photoconductivity measurement. }
		\label{temp3}
	\end{center}
\end{figure}

\clearpage

\section{\supl\noteXRD}
\begin{figure}[htb]
	\begin{center}
		\includegraphics[scale=0.9]{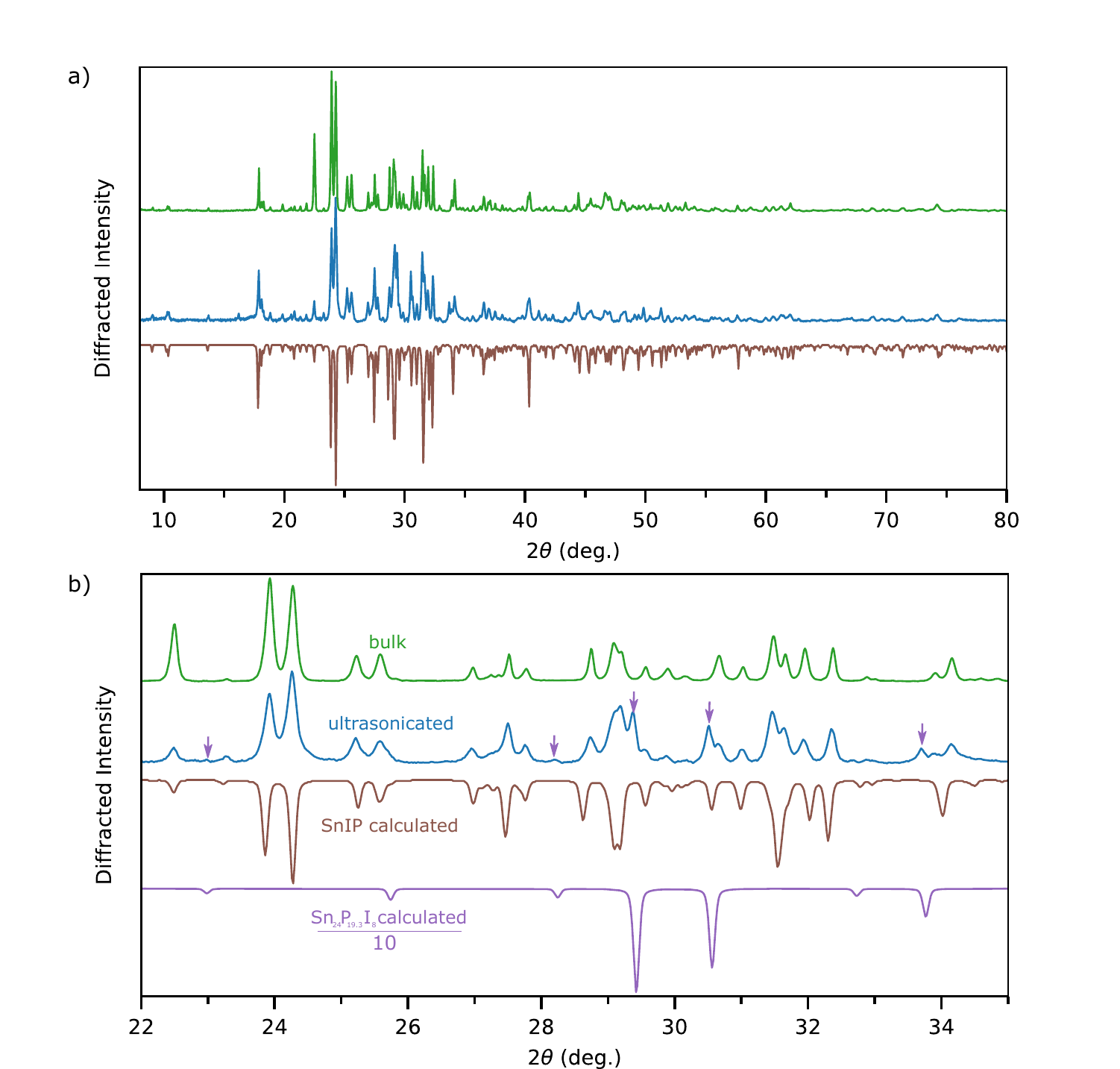}
		\caption{\textbf{X-Ray diffraction probe of crystal quality in ultrasonicated nanowires \textit{vs} bulk needles. a} Full range of $2\theta$ for bulk needles SnIP (green), ultrasonicated nanowires (blue), and the calculated powder XRD pattern (maroon). \textbf{b} Reduced range of $2\theta$ with the impurity peaks attributed to $Sn_{24}P_{19.3}I_8$ (calculated spectrum in purple) indicated with purple arrows for the ultrasonicated SnIP.}
		\label{xrd}
	\end{center}
\end{figure}
Supplementary Fig. \ref{xrd} contains a comparison of the x-ray diffraction (XRD) pattern for bulk needles SnIP \textit{vs} ultrasonicated SnIP nanowires, as defined \supl\noteRam. In Supplementary Fig. \ref{xrd}a, we see excellent agreement between the powder XRD pattern calculated from the solved crystal structure and both the bulk and ultrasonicated samples. In Supplementary Fig. \ref{xrd}b, we focus on a reduced window. In general, we see that the linewidth of the ultrasonicated SnIP shows only a small broadening compared to the bulk SnIP, indicating that the ultrasonicated material retains high crystalline quality. This is consistent with Raman spectroscopy (\supl\noteRam). However, the ultrasonicated SnIP contains several peaks not expected from the pure SnIP structure. These are an indication of a small amount of the clathrate compound $Sn_{24}P_{19.3}I_8$ \cite{shatruk_first_1999}. 
\par

A Rietveld-based phase analysis of the two existing phases after ultrasonication, SnIP and Sn$_{24}$P$_{19.8}$I$_4$, resulted in phase fractions of 0.909(6):0.091. For the Rietveld analysis the two structure models were taken from literature \cite{shatruk_first_1999,pfister_inorganic_2016} using isotropic displacement parameters for each atom and keeping all crystallographic parameters fixed, except the lattice parameters and phase fraction. The final R values after refinements of the profile, lattice and phase fraction parameters are (in \%): Profile R values: Rp 1.49, wRp 2.09; Sn$_{24}$P$_{19.8}$I$_4$: Robs(F) 3.37, wRall(F) 3.94; SnIP: Robs(F) 2.87, wRall(F) 3.62.

\clearpage

\section{\supl\noteRam}
\begin{figure}[htb]
	\begin{center}
		\includegraphics{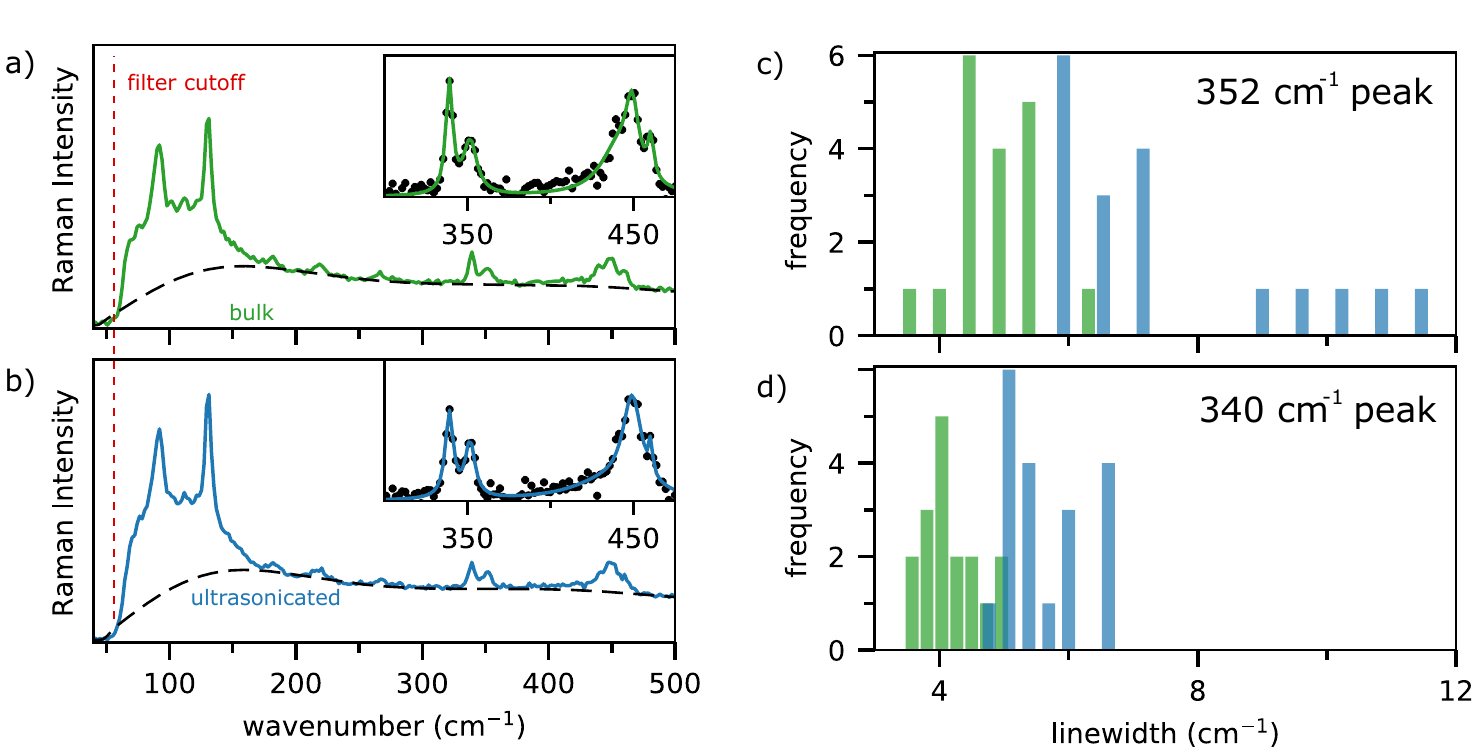}
		\caption{\textbf{Comparison of Raman spectra for SnIP needles and nanowires. a} Micro-Raman spectrum of a single bulk SnIP needle and \textbf{b} Raman spectrum of a single ultrasonicated SnIP nanowire. The insets show a magnified version of the high frequency peaks with the baseline subtracted. Histograms of the linewidth of the 340 and 352 $cm^{-1}$ peaks are shown in \textbf{c} and \textbf{d}, respectively. }
		\label{raman}
	\end{center}
\end{figure}
Supplementary Fig. \ref{raman}a, b contain a comparison of the Raman spectra for bulk SnIP needles and ultrasonicated SnIP nanowires, as defined in Supplemental Note 1. The spectra are qualitatively very similar, with all peaks in the bulk also present in the ultrasonicated SnIP. We note that the filter cutoff near 70 cm$^{-1}$ is at 2.1 THz, which is near the high-frequency edge of the THz spectrum in this work. We note that in SnIP, which is centrosymmetric, all modes are either Raman or infrared active, not both, so Raman and THz spectroscopy probe a different subset of modes.
\par

To be more quantitative, we have plotted a histogram of the linewidth of the 352 $cm^{-1}$ (Supplementary Fig. \ref{raman}c) and 340 $cm^{-1}$ peaks in bulk (green) and ultrasonicated (blue) peaks measured on a variety of individual needles and nanowires. These peaks were chosen for study as they are most easily distinguishable from the baseline and each other. Before fitting, the baseline was subtracted using a Python implementation of the ALS baseline subtraction routine.
\par

Even in the bulk material, the peak width fluctuates from needle to needle. In general, there is a slight increase in average peak width in the ultrasonicated materials. This indicates that the ultrasonication process may result in, at most, a small reduction in crystallinity, which is consistent with the XRD analysis. Alternatively, it is also possible that the peak-broadening is due to the increasing surface to volume ratio.
\clearpage

\section{\supl\noteTinkMod}
\begin{figure}[htb]
	\begin{center}
		\includegraphics{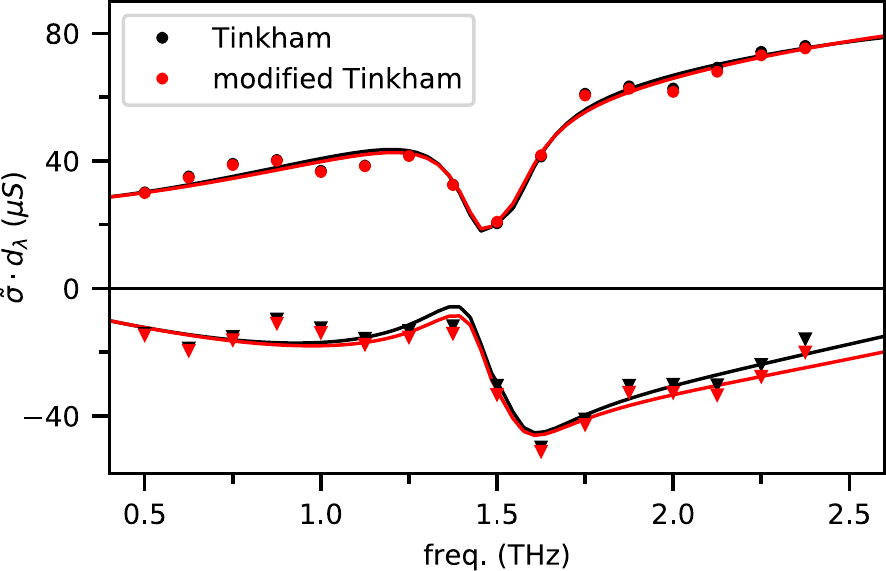}
		\caption{\textbf{Effect of the thin-film static dispersion on the differential conductivity.} Comparison of the conductivity extracted using the standard Tinkham formula and the modified Tinkham film.}
		\label{mod_tinkham}
	\end{center}
\end{figure}
The complex amplitude of an electromagnetic wave transmitted into a substrate with a thin conducting film on the surface is given by\cite{glover_conductivity_1957},
\begin{equation}
\tilde{E}_{ref}=\frac{2}{1+\tilde{n}_{sub}+Z_0\tilde{\sigma}d_{film}}\tilde{E}_{inc},
\label{eq:trans1}
\end{equation}
where $d$ is the film thickness, $\tilde{n}_{sub}$ is the substrate refractive index, $Z_0$ is the impedance of free space, and $\tilde{\sigma}$ is the conductivity of the film. The Tinkham formula in the main text is obtained from the ratio of transmitted field without the conductive film ($\tilde{\sigma}=0$) to the transmitted field with the conductive film and solving for the conductivity. We can see that this method is valid for the TDS measurements but does not take into account the static dispersion of the thin film for the TRTS measurements. To take this into account, we rewrite eq. \ref{eq:trans1} as,
\begin{equation}
\tilde{E}_{samp}=\frac{2}{1+\tilde{n}_{sub}+Z_0(\tilde{\sigma}d_{film}+\Delta\tilde{\sigma}d_{\lambda})}\tilde{E}_{inc},
\label{eq:trans2}
\end{equation}
where we have separated the contribution of the differential conductivity, $\Delta\sigma d_\lambda$, and the static conductivity, $\tilde{\sigma}d_{film}$, to the total areal conductivity. If we now take the ratio of fields, $\tilde{t}=\tilde{E}_{samp}/\tilde{E}_{ref}$, we can solve for the differential conductivity and find,
\begin{equation}
\Delta\tilde{\sigma}d_\lambda=\frac{1+\tilde{n}_{sub}+Z_0\tilde{\sigma}_0d_{film}}{Z_0}\left(\frac{1}{\tilde{t}}-1\right).
\label{eq:tinkMod}
\end{equation}
This modified Tinkham formula introduces small corrections to the extracted differential conductivity. 
\par

To see the difference, in Supplementary Fig. \ref{mod_tinkham} we plot the SnIP differential conductivity for 400 nm excitation extracted from the standard Tinkham formula, \textit{i.e.}, we compare $\Delta\tilde{\sigma}$ calculated with the modified Tinkham formula (eq. \ref{eq:tinkMod}) to the standard Tinkham equation from the Methods section in the main text. As can be seen, the modified Tinkham formula yields only a small correction to the differential conductivity, as suggested in the main text.
\par

\clearpage

\section{\supl\noteLST}
\begin{figure}[htb]
	\begin{center}
		\includegraphics[scale=0.8]{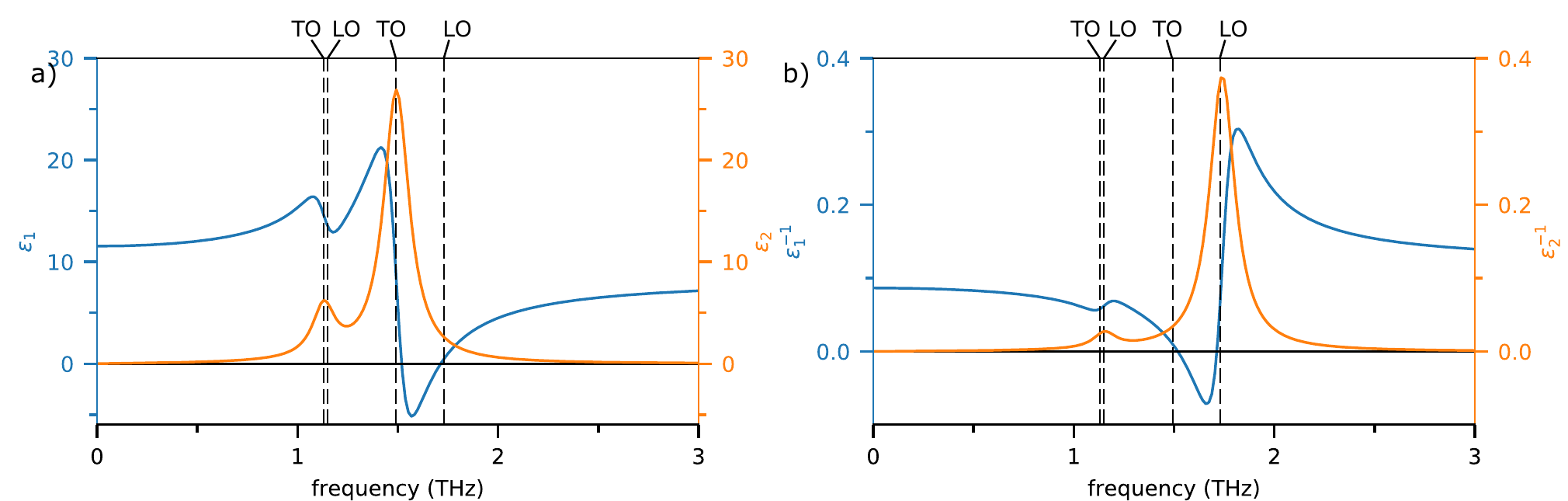}
		\caption{\textbf{TO and LO frequencies. a} Estimated real (blue) and imaginary (orange) dielectric function along the double-helix axis calculated from the fit parameters from the main text. The calculation assumes the average dielectric constant is a linear superposition of the air/nanowire dielectric constants weighted by their filling fraction and orientation fraction. \textbf{b} Real (blue) and imaginary (orange) part of the inverse of the dielectric function in \textbf{a}, showing peaks at the LO phonon frequencies. The vertical dashed lines indicate the TO phonon frequency extracted from THz spectroscopy and the LO phonon frequency calculated from the LST relationship.}
		\label{LST}
	\end{center}
\end{figure}
The Lyddane-Sachs-Teller (LST) relation applies strictly to cubic crystals with a single TO and LO phonon frequency in the case of zero damping \cite{lyddane_polar_1941}. It can be easily extended to the case of multiple resonances in cubic crystals using the idea that longitudinal resonances occur at frequencies where the dielectric constant is zero, \textit{i.e.}, at poles in the inverse dielectric function \cite{born1988dynamical}, which we note is the basis of the LST relation itself. In the zero-damping limit, the imaginary part of the inverse dielectric function consists of zero-width peaks at the LO frequencies and introducing damping results in broadening these peaks. Now, in cubic crystals with both multiple resonances and damping, we can associate longitudinal modes with peaks in the inverse dielectric function \cite{huber_femtosecond_2005}.
\par

Finally, to deal with lower-symmetry crystals in the zero-damping limit, the condition that the dielectric is equal to zero for longitudinal resonances is generalized to the condition that the determinant of the dielectric tensor is zero, \textit{i.e.}, $det(\epsilon_{ij})=0$. It is easy to recover from this expression the condition of $\epsilon=0$ for cubic crystals, which have no off-diagonal elements of $\epsilon_{ij}$ and $det(\epsilon_{ij})=\epsilon_{11}\epsilon_{22}\epsilon_{33}$. The zeros, and therefore LO frequencies, can then be calculated individually for each polarization. The off-diagonal components, which can be present in low-symmetry crystals, are what make the calculation of LO frequencies more complicated. 
\par

We now see that in using the LST relation to find the LO frequency for mode 2, we have made the approximations that the off-diagonal components are zero and that the dielectric constant in the vicinity of the LO mode due to the lower frequency oscillator at 1.1 THz is dispersionless. In \supl\noteEpsDFT, we note that the calculated off-diagonal components of the dielectric tensor in SnIP ($\epsilon_{xz}=0$ for monoclinic crystals) are significantly smaller than the diagonal components, which motivates this approximation. To show that including the 1.1 THz oscillator does not affect the calculation of the LO frequency, in Supplementary Fig. \ref{LST}a, b we plot the dielectric function after account for the anisotropy and filling fraction (see \enquote{Methods} in the main text) and it's inverse, respectively, with the LO frequencies calculated via the LST relation indicated by dashed lines. We see excellent agreement between the frequencies calculated from the LST relation and the peaks in the inverse dielectric function. 

\clearpage

\bibliographystyle{naturemag}
\bibliography{SnIP_supplementary}% Produces the bibliography via BibTeX.